\documentclass[
 amsmath,amssymb,
 aps,
 twocolumn,
]{revtex4-2}

\usepackage{graphicx}
\usepackage{dcolumn}
\usepackage{bm}

\usepackage{color}
\definecolor{orange}{rgb}{0.8, 0.3, 0}

\begin{document}


\title{Hyperfine spectroscopy in a quantum-limited spectrometer}

\author{S.~Probst$^{1}$, G.-L. Zhang$^{2}$, M. Ran\v{c}i\'{c}$^{1}$, V. Ranjan$^{1}$, M. Le Dantec$^{1}$, Z. Zhang$^{3}$ B. Albanese$^{1}$, A. Doll$^{4}$, R.-B. Liu$^{2}$, J.J.L. Morton$^{5}$, T. Chaneli{\`e}re$^{6}$, P. Goldner$^{3}$, D. Vion$^{1}$, D. Esteve$^{1}$, P. Bertet$^{1}$}

\affiliation{$^{1}$Quantronics group, SPEC, CEA, CNRS, Universit\'e Paris-Saclay, CEA Saclay 91191 Gif-sur-Yvette Cedex, France}

\affiliation{$^{2}$Department of Physics and The Hong Kong Institute of Quantum Information Science and Technology, The Chinese University of Hong Kong, Shatin, New Territories, Hong Kong, China}

\affiliation{$^{3}$Chimie ParisTech, PSL University, CNRS, Institut de Recherche de Chimie Paris, 75005 Paris, France}

\affiliation{$^{4}$Laboratory of nanomagnetism and oxides, SPEC, CEA, CNRS, Universit\'e Paris-Saclay, CEA Saclay 91191 Gif-sur-Yvette Cedex, France}

\affiliation{$^{5}$London Centre for Nanotechnology, University College London, London WC1H 0AH, United Kingdom}

\affiliation{$^{6}$Univ. Grenoble Alpes, CNRS, Grenoble INP, Institut N\'eel, 38000 Grenoble, France}

\date{\today}

\begin{abstract}
We report measurements of electron spin echo envelope modulation (ESEEM) performed at millikelvin temperatures in a custom-built high-sensitivity spectrometer based on superconducting micro-resonators. The high quality factor and small mode volume (down to 0.2pL) of the resonator allow to probe a small number of spins, down to $5\cdot 10^2$. We measure 2-pulse ESEEM on two systems: erbium ions coupled to $^{183}$W nuclei in a natural-abundance $\text{CaWO}_4$ crystal, and bismuth donors coupled to residual $^{29}$Si nuclei in a silicon substrate that was isotopically enriched in the $^{28}$Si isotope. We also measure 3- and 5-pulse ESEEM for the bismuth donors in silicon. Quantitative agreement is obtained for both the hyperfine coupling strength of proximal nuclei, and the nuclear spin concentration.
\end{abstract}


\maketitle

\section{\label{sec:intro}Introduction}

Electron paramagnetic resonance (EPR) spectroscopy provides a set of versatile tools to study the magnetic environment of unpaired electron spins~\cite{schweiger_principles_2001}. EPR spectrometers rely on the inductive detection of the spin signal by a three-dimensional microwave resonator tuned to the spin Larmor frequency. While concentration sensitivity is the main concern for dilute samples available in macroscopic volumes~\cite{song_toward_2016}, there are also cases in which the absolute spin detection sensitivity matters, motivating research towards alternative detection methods to measure smaller and smaller numbers of spins. Electrical~\cite{elzerman_single-shot_2004,veldhorst_addressable_2014,morello_single-shot_2010,pla_single-atom_2012}, optical~\cite{wrachtrup_optical_1993,jelezko_observation_2004}, and scanning-probe-based~\cite{rugar_single_2004,baumann_electron_2015} detection of magnetic resonance have reached sufficient sensitivity to detect individual electron spins.

In parallel, recent results have shown that the inductive detection method can also be pushed to much higher absolute sensitivity than previously achieved, using planar micro-resonators~\cite{narkowicz_scaling_2008,artzi_induction-detection_2015} and micro-helices~\cite{sidabras_extending_2019}. Superconducting resonators~\cite{wallace_microstrip_1991,benningshof_superconducting_2013,sigillito_fast_2014} are particularly useful in that context since they combine low mode volume and narrow linewidth $\kappa$. Inductive-detection spectrometers relying on a superconducting planar micro-resonator combined with a Josephson Parametric Amplifier (JPA), cooled down to millikelvin temperatures~\cite{bienfait_reaching_2015,eichler_electron_2017,probst_inductive-detection_2017}, have achieved a sensitivity of $10$ spin/$\sqrt{\mathrm{Hz}}$ for detecting Hahn echoes emitted by donors in silicon~\cite{ranjan_electron_2020}. A particular feature of these quantum-limited spectrometers is that quantum fluctuations of the microwave field play an important role. First, the system output noise is governed by these quantum fluctuations, with negligible thermal noise contribution. Second, quantum fluctuations also impact spin dynamics by triggering spontaneous emission of microwave photons at a rate $\Gamma_P = 4 g^2 / \kappa$, $g$ being the spin-photon coupling~\cite{bienfait_controlling_2016,eichler_electron_2017,ranjan_pulsed_2020}. This Purcell effect forbids $T_1$ to become prohibitively long since it is at most equal to $\Gamma_P^{-1}$, making spin detection with a reasonable repetition rate possible even at the lowest temperatures.

Hahn echoes are the simplest pulse sequence used in EPR spectroscopy, useful to determine the electron spin density as well as the spin Hamiltonian parameters and their distribution. The richness of EPR comes from the ability to characterize the local magnetic environment of the electron spins, often consisting of a set of nuclear spins or of other electron spins. For that, hyperfine spectroscopy is required, which uses more elaborate pulse sequences and requires larger detection bandwidth. Previous hyperfine spectroscopy measurements with superconducting micro-resonators include the electron-nuclear double resonance detection of donors in silicon~\cite{sigillito_electrically_2017} and the electron-spin-echo envelope modulation (ESEEM) of erbium ions by the nuclear spin of yttrium in a $\text{Y}_2\text{SiO}_5$ crystal~\cite{probst_microwave_2015}. 

Here, we demonstrate that hyperfine spectroscopy is compatible with quantum-limited EPR spectroscopy despite its additional requirements in terms of pulse complexity and bandwidth, by measuring ESEEM in two model electron spin systems. We measure the ESEEM of erbium ions coupled to $^{183}$W nuclei in a scheelite crystal ($\text{CaWO}_4$) with a simple two-pulse sequence, and get quantitative agreement with a simple dipolar interaction model. We also measure the ESEEM of bismuth donors in silicon caused by $^{29}$Si nuclei using 2, 3, and 5-pulse sequences ~\cite{schweiger_principles_2001,kasumaj_5-_2008}. Compared to other ESEEM measurements on donors in silicon~\cite{witzel_decoherence_2007,abe_electron_2010}, ours are performed in an isotopically purified sample having a $100$ times lower concentration in $^{29}$Si ($500$\,ppm) than natural abundance. As a result, the dominant hyperfine interactions in the ESEEM signal are very low (on the order of $100$\,Hz) and have to be detected at low magnetic fields (around $0.1$\,mT). These results bring quantum-limited EPR spectroscopy one step closer to real-world applications.

\section{ESEEM spectroscopy : theory}

\subsection{Phenomenology}

\begin{figure}[h!]%
\centering
\includegraphics[width=0.9\columnwidth]{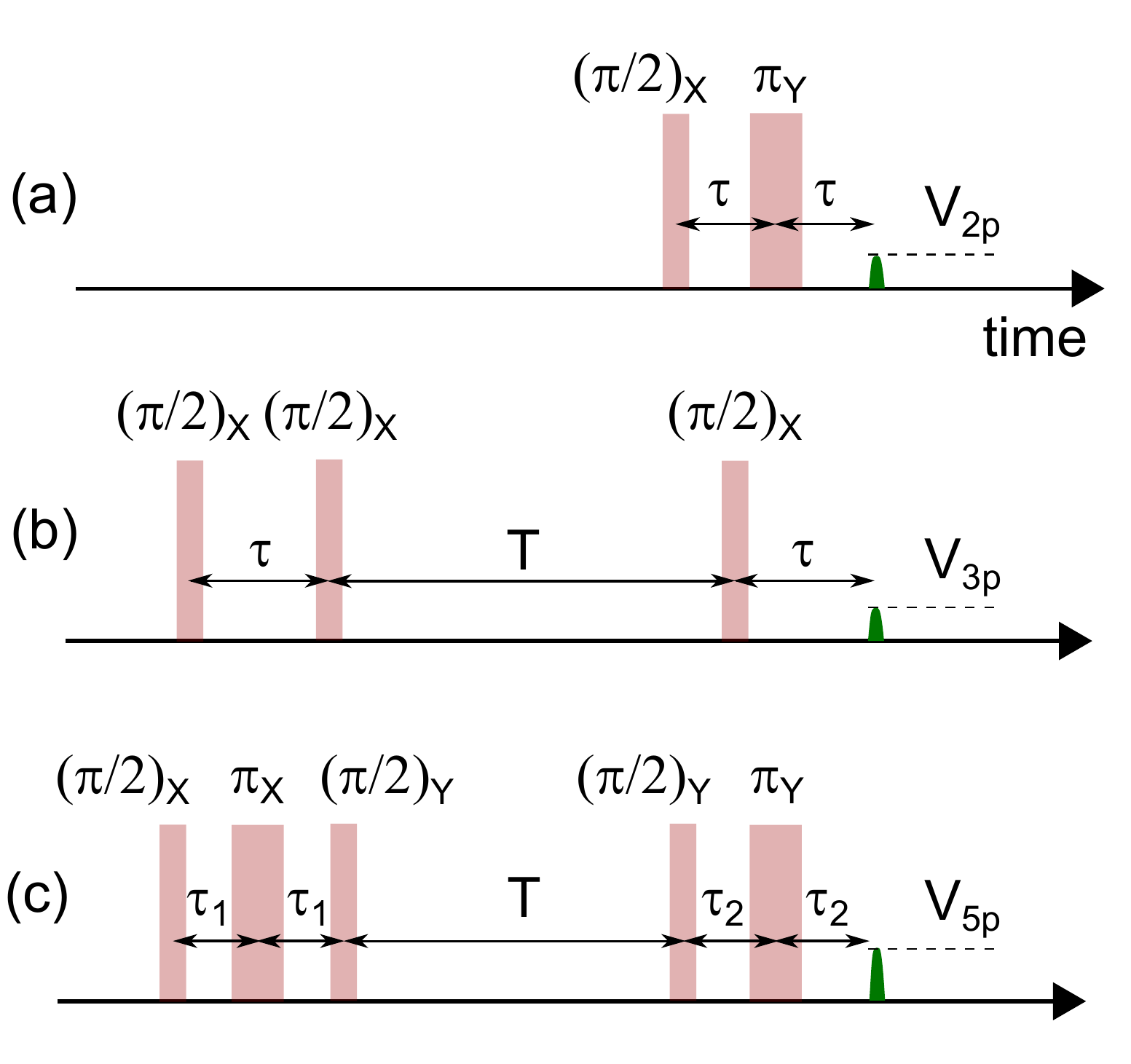}%
\caption{Sequences used for 2-pulse (a), 3-pulse (b), and 5-pulse (c) ESEEM measurements.}%
\label{fig:PulseSequences}%
\end{figure}

We start by briefly discussing the ESEEM phenomenon. Consider an ensemble of electron spins placed in a magnetic field $B_0$. The spin ensemble linewidth $\Gamma$ is broadened by a variety of mechanisms : spatial inhomogeneity of the applied field $B_0$, local magnetic fields generated by magnetic impurities throughout the sample, and spatially inhomogeneous strain or electric fields. One prominent way to mitigate the effect of this inhomogeneous broadening is the spin-echo sequence (also called Hahn echo, or two-pulse echo). It consists of a $\pi/2$ pulse at time $t=0$ and a $\pi$ pulse after a delay $\tau$ (see Fig.\ref{fig:PulseSequences}a). This $\pi$ pulse reverses the evolution of the phase of the precessing magnetic dipoles, which leads at a later time $2 \tau$ to their refocussing and the emission of a microwave pulse (the echo) of amplitude  $V_{\rm 2p}(\tau)$.

In general, $V_{\rm 2p}(\tau)$ decays monotonically; it can however also display oscillations. Such ESEEM was first observed by Mims and co-workers~\cite{mims_spectral_1961,rowan_electron-spin-echo_1965} for $\text{Ce}^{3+}$ ions in a $\text{CaWO}_4$ crystal, and was interpreted as being caused by the dipolar interaction of the electronic spin of the $\text{Ce}^{3+}$ ions with the $^{183}\text{W}$ nuclear spins of the crystal. The oscillation frequencies appearing in the ESEEM pattern are related to the nuclear spin Larmor frequencies and to their coupling to the electron spin. As such, ESEEM measurements provide spectroscopic information on the nature of the nuclear spin bath and its density, and ESEEM spectroscopy has become an essential tool in advanced EPR \citep{schweiger_principles_2001,mims_exchange_1990}. ESEEM has also been observed for individual spins measured optically, in particular for individual NV centers in diamond coupled to a bath of $^{13}\text{C}$ nuclear spins~\cite{childress_coherent_2006}. A more complete theory of ESEEM is presented in~\cite{mims_envelope_1972}. Our goal here is to provide a simple picture of the physics involved, as well as to introduce useful formulas and notations. 

\subsection{Two-spin-1/2 model}

\begin{figure}[h!]%
\centering
\includegraphics[width=0.9\columnwidth]{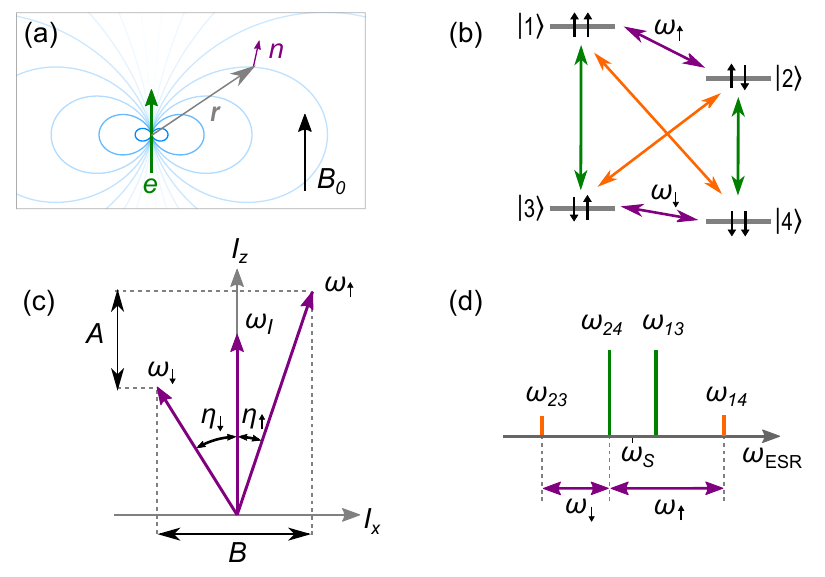}%
\caption{ESEEM model system for electron spin $S = 1/2$ and nuclear spin $I = 1/2$ with $\omega_I, A, B > 0$. (a) Nuclear spin (purple) subject to external field $B_0$ and dipole field (blue) of a nearby electron spin (green) located at relative position $\boldsymbol{r}$. (b) Energy diagram showing the electron transitions (green), the nuclear transitions (purple), and the (normally forbidden) electro-nuclear transitions (orange). The energy levels $|1\rangle$, ..., $|4\rangle$ are labeled according to the eigenstates of the Zeeman basis. (c) Quantization axes $\omega_\uparrow$ and $\omega_\downarrow$ due to mixing of the nuclear states, which results in inclination of the quantization axis from $z$ by the angles $\eta_\uparrow$ and $\eta_\downarrow$, respectively. (d) EPR spectrum showing the electron transitions (green) and the electro-nuclear transitions (orange) as well as the relation of these ESR transitions to the nuclear frequencies $\omega_\uparrow$ and $\omega_\downarrow$ (purple).}%
\label{figESEEMtheo}%
\end{figure}

We follow the analysis in Ref.\citep{schweiger_principles_2001} of the model case depicted in Fig.\ref{figESEEMtheo}a. An electron spin $S=1/2$, with an isotropic g-tensor, is coupled to a proximal nuclear spin $I=1/2$. Both are subject to a magnetic field $B_0$ applied along $z$. The system Hamiltonian is

\begin{equation}
    H_0 = H_\text{e} + H_\text{n} + H_\text{hf},
\end{equation}

\noindent where $H_\text{e} = \omega_S S_z$ ($H_\text{n} = \omega_I I_z$) is the Zeeman Hamiltonian of the electron (nuclear) spin with Larmor frequency $\omega_S$ ($\omega_I$), and $H_\text{hf}$ is the electron-nuclear hyperfine interaction, which includes their dipole-dipole coupling and may include a Fermi contact term as well. We assume that $\omega_S $ is much larger than the hyperfine interaction strength, in which case terms proportional to the $S_x$ and $S_y$ operators can be neglected. This secular approximation leads to a hyperfine Hamiltonian of the form $H_\text{hf} = A S_z I_z + B S_z I_x$, with the expressions for $A$ and $B$ depending on the details of the hyperfine interaction\citep{schweiger_principles_2001}.

Overall, the system Hamiltonian is

\begin{equation}\label{eq:spinH}
H_0 = \omega_S S_z + \omega_I I_z + A S_z I_z + B S_z I_x.
\end{equation}

Because of the $B S_z I_x$ term, the nuclear spin is subjected to an effective magnetic field whose direction (and magnitude) depend on the electron spin state $|\uparrow_{\rm e}\rangle$ or $|\downarrow_{\rm e}\rangle$. Its eigenstates therefore depend on the electron spin state, so that transitions become allowed between all the spin system energy levels $|1\rangle - |4 \rangle$, leading to the ESEEM phenomenon. Relevant parameters are the electron-spin-state-dependent angles between the effective magnetic field seen by the nuclear spin and the quantization axis $z$

\begin{eqnarray}
\label{eq:etatheo}
\eta_\uparrow = \arctan\frac{B}{A+2\omega_I} \notag \\
\eta_\downarrow = \arctan\frac{B}{A-2\omega_I}.
\end{eqnarray}

and the electron-spin-dependent nuclear-spin frequencies

\begin{eqnarray*}
\omega_\uparrow & = & (\omega_I + \frac{A}{2})\cos \eta_\uparrow - \frac{B}{2}\sin \eta_\uparrow \notag \\
\omega_\downarrow & = & (\omega_I - \frac{A}{2})\cos \eta_\downarrow - \frac{B}{2}\sin \eta_\downarrow.
\end{eqnarray*}

When $\eta_\uparrow,\eta_\downarrow$ are close to equal, only the nuclear-spin preserving transitions are allowed; this occurs either when $B = 0$ (due to a specific orientation of the dipolar field, or to a purely isotropic hyperfine coupling), or when $B\neq0$ but $\omega_I \gg A$ (very weak coupling limit) or $\omega_I \ll A$ (very strong coupling limit). On the contrary, when the direction of the effective magnetic field seen by the nuclear spin is electron-spin-dependent, all transitions become allowed. This occurs when $B \neq 0$ and $\omega_I \simeq \pm A/2$.

\subsection{Multi-pulse ESEEM}

Because of the level structure shown in Fig.\ref{figESEEMtheo}, and assuming for simplicity microwave pulses so short that their bandwidth is much larger than $\omega_{\uparrow,\downarrow}$, microwave pulses at the electron spin frequency $\omega_S$ excite the allowed transitions $|1\rangle \leftrightarrow |3\rangle$ and $|2\rangle \leftrightarrow |4\rangle$, but also the normally forbidden $|1\rangle \leftrightarrow |4\rangle$ and $|2\rangle \leftrightarrow |3\rangle$, leading to coherence transfer between the levels and to beatings. Note that for simplicity we assume that the microwave pulses are ideal and so short that their bandwidth is much larger than $\omega_{12}$ and $\omega_{34}$.  

It is then possible to compute analytically the effect of a two-pulse echo sequence consisting of an instantaneous ideal $\pi/2$ pulse and an instantaneous ideal $\pi$ pulse (see Fig.\ref{fig:PulseSequences}), disregarding any decoherence. The resulting echo amplitude \citep{schweiger_principles_2001} is given by  

\begin{eqnarray}
\label{eq:V2p}
V_{\rm 2p}(\tau) & = & 1 - \frac{k}{4} [2 - 2 \cos(\omega_\uparrow \tau) - 2 \cos(\omega_\downarrow \tau) \notag \\
& & + \cos ((\omega_\uparrow-\omega_\downarrow) \tau) + \cos ((\omega_\uparrow+\omega_\downarrow) \tau)],
\end{eqnarray}

with 

\begin{equation}
k = \left[\frac{B \omega_I}{\omega_\uparrow \omega_\downarrow} \right]^2.
\end{equation}

The spin-echo amplitude is modulated by a function whose frequency spectrum and amplitude contain information about the nuclear spin Larmor frequency $\omega_I$ as well as its hyperfine coupling $(A,B)$ to the electron spin. The modulation contrast $0\leq k \leq 1$ is maximal when transitions $|1\rangle-|4\rangle$ and $|2\rangle-|3\rangle$ are maximally allowed, corresponding to $\omega_I \simeq A/2$.

The above results are exact, as long as the secular approximation is valid and the pulses are ideal. In the weak-coupling limit $A,B \ll \omega_I$, $\omega_\uparrow \simeq \omega_\downarrow \simeq \omega_I$ so that $V_{\rm 2p}(\tau) = 1 - \frac{k}{4} [3 - 4 \cos(\omega_I \tau) +  \cos (2 \omega_I \tau) ]$, with $k = (B/\omega_I)^2 \ll 1$. In this limit, the echo modulation spectrum directly yields the nuclear spin Larmor frequency, and also contains components at twice this frequency. Note however that in practice, the $\pi$ pulse bandwidth is always finite, because of the resonator bandwidth or limited pulse power; this sets a limit to the range of detectable modulation frequencies.

The electron spin is often coupled to $N$ nuclear spins, with $N>1$. Since all nuclear spin subspaces can be diagonalized separately, the total ESEEM modulation is simply given by the product of each nuclear spin modulation $V_{2p,l}(\tau)$, $l$ being the nuclear spin index. Taking also into account that the electron spin is also subject to decoherence processes, modelled for instance by an exponential decay with time constant $T_2$, the echo envelope is

\begin{equation}
\label{eq:2PEProductRule}
V'_{\rm 2p}(\tau) = \exp{(-2 \tau / T_2)}  \prod_{l=1}^N V_{2p,l}(\tau).
\end{equation}

The modulation pattern $V'_{\rm 2p}(\tau)$ yields quantitative information about the nature and coupling of the nuclear spins surrounding the electron spin whose echo is measured, and is therefore a useful tool in EPR spectroscopy. When the environmental nuclei have a certain probability $p$ to be of a given isotope with a nuclear spin $I=1/2$, and a probability $1-p$ to be of an isotope with $I=0$, the above formulas are straightforwardly modified~\cite{rowan_electron-spin-echo_1965} by writing 

\begin{eqnarray}
V_{\rm 2p,l}(\tau) & = & 1 - \frac{p k_l}{4} [2 - 2 \cos(\omega_{\uparrow,l} \tau) - 2 \cos(\omega_{\downarrow,l} \tau) \notag \\
& & + \cos ((\omega_{\uparrow,l}-\omega_{\downarrow,l}) \tau) + \cos ((\omega_{\uparrow,l}+\omega_{\downarrow,l}) \tau)].
\label{eq:2pESEEMConcentrationDependence}
\end{eqnarray}

The echo signal $V'_{\rm 2p}(\tau)$ is the sum of terms that have the general form $p^L \prod_{l=1}^{l=L} k_l \cos(\omega_{\mu,l} \tau)$, where $l$ runs over a subset of $L$ nuclei and $\mu = \uparrow,\downarrow$. If $p \ll 1$, this expression is well approximated by keeping only the $L=1$ terms, which then yields

\begin{eqnarray}
V_{\rm 2p}(\tau) & \simeq & 1 - \sum_{l=1}^{l=N} \frac{p k_l}{4} [2 - 2 \cos(\omega_{\uparrow,l} \tau) - 2 \cos(\omega_{\downarrow,l} \tau) \notag \\
& & + \cos ((\omega_{\uparrow,l}-\omega_{\downarrow,l}) \tau) + \cos ((\omega_{\uparrow,l}+\omega_{\downarrow,l}) \tau)].
\label{eq:2pESEEMSmallConcentration}
\end{eqnarray}

One limitation of the previous pulse sequence is that the modulation envelope can only be measured up to a time of order $T_2$ due to electron spin decoherence, which may be too short for appreciable spectral resolution. This limitation can be overcome by the three-pulse echo sequence shown in Fig.~\ref{fig:PulseSequences}b. It consists of a $\pi/2$ pulse applied at $t=0$ followed, after a time $\tau$ chosen such that $\tau < T_2$, by a second $\pi/2$ pulse. After a variable delay $T$, a third $\pi/2$ pulse is applied, leading to the emission of a stimulated echo at time $t=T+2\tau$. The interest of this sequence is that the first pair of $\pi/2$ pulses generates nuclear spin coherence that can survive up to the nuclear spin coherence time $T_{2,\rm{n}}$ which is in general much longer than $T_2$ (and close to the electron energy spin relaxation time $T_1$). An analytical formula can be derived for the three-pulse echo amplitude in the ideal pulse approximation~\cite{schweiger_principles_2001}

\begin{eqnarray}
\label{eq:V3ptheo}
V_{\rm 3p} (T) & = & \exp(-T/T_{2,\rm{n}}) \exp (-2\tau /T_2) \notag \\
& &  \{1 - \frac{k}{4} [[1 - \cos \omega_\downarrow \tau][1 - \cos \omega_\uparrow (T+\tau)] \notag \\
&  & + [1 - \cos \omega_\uparrow \tau][1 - \cos \omega_\downarrow (T+\tau)]]\}.
\end{eqnarray}

Contrary to two-pulse ESEEM, three-pulse echo modulation as a function of $T$ only contains the $\omega_\downarrow,\omega_\uparrow$ frequency components, and not their sum or difference; that is, in the weak-coupling limit $A,B \ll \omega_I$, only the nuclear spin Larmor frequency $\omega_I$ appears in the spectrum. Another difference is that the modulation pattern and amplitude depend on $\tau$; in particular, its amplitude is zero whenever $\omega_{\downarrow,\uparrow} \tau = 2 \pi n$ with $n$ integer ({\it blind spots}).

For weakly coupled nuclei, the modulation amplitude of 3-pulse ESEEM can be enhanced by up to one order of magnitude by using a more complex pulse sequence known as 5-pulse ESEEM~\cite{schweiger_principles_2001,kasumaj_5-_2008}, and shown in Fig.\ref{fig:PulseSequences}. The analytical formula for the five-pulse echo amplitude $V_{\rm 5p}$ is given in the Supplementary Information.

Equation \ref{eq:2PEProductRule}, with proper modification to take into account contributions of different pathways, can be applied to the 3- and 5-pulse ESEEM to treat coupling to multiple nuclear spins. The details are shown in Section 3C of the Supplementary Information.  

\subsection{Fictitious spin model}

The electronic spins that we consider in this work involve an unpaired electron with spin $S_0=1/2$ either located around or trapped by an ionic defect, which itself can possess a non-zero nuclear spin $I_0$. These two spins of the defect are strongly coupled and form therefore a multi-level system, which can nevertheless be mapped to an effective, fictitious, spin-1/2 model as explained below~\cite{schweiger_principles_2001}, to which the model of Section 2.3 can be applied.

The system spin Hamiltonian writes

\begin{equation}
\label{eq:Hetheo}
H_\text{ion} = \beta_\text{e}\boldsymbol{B_{0}}\cdot\bar{\bold{g}}_\text{e}\cdot\boldsymbol{S_0}+\boldsymbol{S}_{0}\cdot\bar{\bold{A}}_0\cdot\boldsymbol{I}_{0},
\end{equation}
Here, $\beta_\text{e}$ is the electron Bohr magneton, $\bar{\bold{g}}_\text{e}$ is the (possibly anisotropic) gyromagnetic tensor, and $\bar{\bold{A}}_0$ the hyperfine tensor. The nuclear Zeeman interaction of the defect system, being small compared to the hyperfine interaction in the range of magnetic fields explored here, is neglected from the Hamiltonian.

This multi-level electron-spin system is coupled to other nuclear spins in the lattice, giving rise to ESEEM. Consider a nuclear spin at a lattice site $j$, defined by its location $\boldsymbol{r}_j$ with respect to the electron spin. The nuclear Zeeman Hamiltonian is $H_j = \omega_I I_{j,z}$, with $\omega_I=g_n \beta_n B_0$, $g_n$ being the nuclear g-factor and $\beta_n$ the nuclear magneton. Its hyperfine coupling to the electron spin system is described by the Hamiltonian

\begin{equation}
\label{eq:Hhftheo}
  H_\text{j,hf}=\boldsymbol{S_0} \cdot \bar{\bold{A}}_\text{j} \cdot \boldsymbol{I_j},
\end{equation}

with 

\begin{equation}
\label{eq:Aj}
\bar{\bold{A}}_\text{j} = \bar{\bold{A}}_\text{j,cf}+\bar{\bold{A}}_\text{j,dd}.
\end{equation}

This hyperfine tensor consists of a Fermi contact term $\bar{\bold{A}}_\text{j,cf} =\frac{2}{3} \mu_0 \beta_e g_n \beta_n \bar{\bold{g}}_\text{e} |\psi(\bold{r}_\text{j})|^2 $ and a dipole-dipole term $\bar{\bold{A}}_\text{j,dd} =\frac{3 \mu_0}{4\pi |\boldsymbol{r}_j|^5} \beta_e \beta_n g_n [r_j^2 \boldsymbol{g_\text{e}} - 3 (\boldsymbol{g_\text{e}} \cdot \boldsymbol{r}_j) \boldsymbol{r}_j]$, $\psi(\bold{r}_\text{j})$ being the electron wavefunction at the nuclear spin location.

The Hamiltonian $H_\text{ion}$ (Eq.\ref{eq:Hetheo}) can be diagonalized, yielding $4I_0+2$ energy levels. It is in general possible to isolate two levels $|\alpha \rangle$ and $|\beta \rangle$ that are coupled by an ESR-allowed transition and are resonant or quasi-resonant with the microwave cavity, with a transition frequency $\omega_S$. If these two levels are sufficiently separated in energy from other levels of $H_\text{ion}$, they define a fictitious $S=1/2$ system. Writing the total Hamiltonian $H_\text{ion}+H_\text{j}+H_\text{hf,j}$ restricted to this two-dimensional subspace yields

\begin{eqnarray}\label{eq:spinHFict}
H_0 & = & \omega_S S_z + (\omega_I + \frac{m_S^\alpha + m_S^\beta}{2} A_{j,zz}) I_{j,z} \notag \\
& + & \frac{m_S^\alpha + m_S^\beta}{2} A_{j,zx} I_{j,x} \notag \\
& + & (m_S^\alpha - m_S^\beta) (A_{j,zz} S_z I_{j,z} + A_{j,zx} S_z I_{j,x})
\end{eqnarray}

where $m_S^{\alpha,\beta} = \langle \alpha,\beta | S_{0,z} | \alpha, \beta \rangle $.

Equation \ref{eq:spinHFict} maps the more complex system to the simple model of section 2.2. Compared to Eq.\,(\ref{eq:spinH}), two differences appear. First, the hyperfine interaction parameters $A,B$ are rescaled by the effective longitudinal magnetization difference $(m_S^\alpha - m_S^\beta)$ which depends on the two levels considered. Second, when the average longitudinal magnetization of the two levels $(m_S^\alpha + m_S^\beta)$ is non-zero, the nuclear spin sees an extra Zeeman contribution which may be tilted with respect to the $z$ axis. Once taken into account these corrections, the analysis and formulas of Section 2.3 remain valid.

\section{Spin systems}

\subsection{Erbium-doped $\text{CaWO}_{4}$}

\begin{figure}[t!]
\centering
\includegraphics[width=0.96\columnwidth]{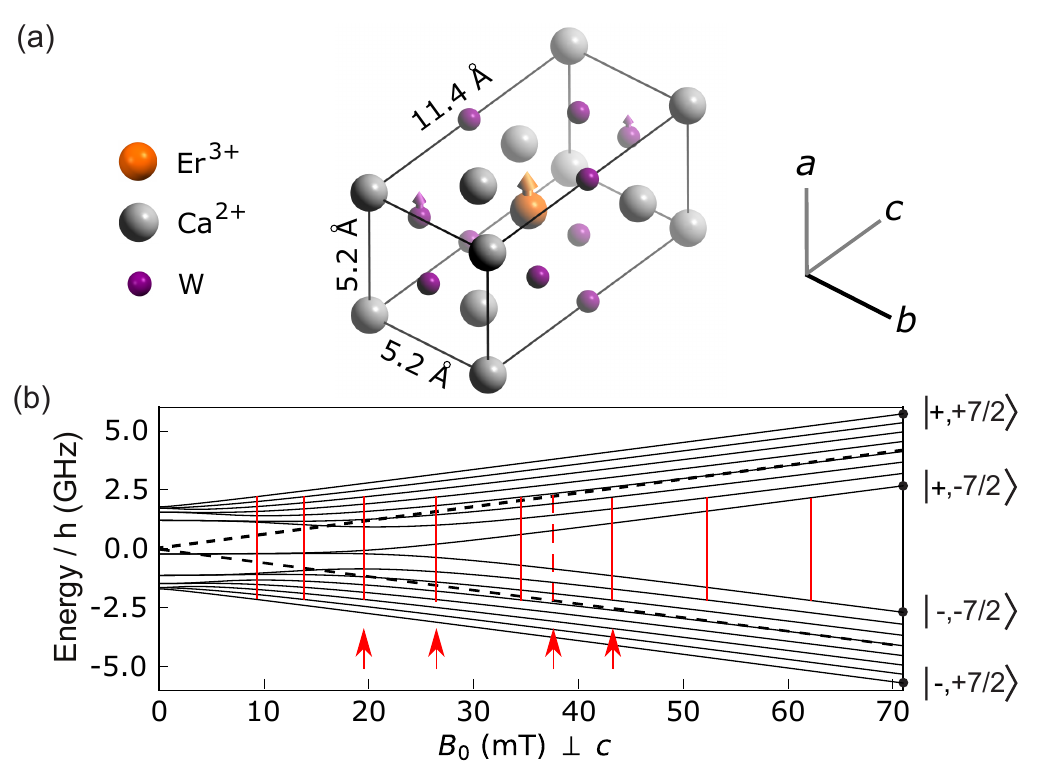}%
\caption{Structure and energy diagram of erbium ions in $\mathrm{CaWO}_4$. (a) Crystal structure with oxygen atoms hidden for clarity. Erbium atoms are in substitution of the Calcium. The crystal has a rotational symmetry around the $c$ axis. A fraction $p=0.13$ of the $W$ atoms are of the $^{183}$W isotope, with a nuclear spin $1/2$. (b) Energy level diagram of the $I=0$ erbium isotopes (black dashed line) and of the $^{167}$Er isotope (black solid lines) with $I=7/2$, for $B_0$ applied perpendicular to the $c$ axis. Red vertical lines indicate the value of $B_0$ for which an allowed EPR transition becomes resonant with the $4.372$ GHz frequency of our detection resonator (see main text, Section 4). Four red arrows indicate the values of $B_0$ at which ESEEM data were measured. 
}%
\label{fig:ErbiumTransitions}%
\end{figure}

The first system investigated consists of erbium Er$^{3+}$ ions doped into a $\text{CaWO}_4$ matrix, substituting $\mathrm{Ca}^{2+}$. The crystal has a tetragonal body-centered structure (see  Fig.\,\ref{fig:ErbiumTransitions}) with lattice constants $a=b=0.524$\,nm and $c=1.137$\,nm. Rare-earth ions with an odd number of electrons such as Er$^{3+}$ have a ground state consisting of two levels that are degenerate in zero magnetic field, and separated from other levels by an energy scale equivalent to several tens of Kelvin due to the crystalline electric field and the spin-orbit interaction. This pair of electronic levels is known as a Kramers doublet, and forms an effective $S_0 = 1/2$ electron spin system, with a spin Hamiltonian $H_\text{Er}$\citep{abragam_electron_2012} whose form is given by Eq.(\ref{eq:Hetheo}). 

Due to the S4 site symmetry in which rare earth ions are found in $\text{CaWO}_{4}$, the g-tensor is diagonal in the crystallographic frame with $g_{xx}=g_{yy}=8.38$ and $g_{zz}=1.247$~\cite{antipin_a._paramagnetic_1968} (${x,y,z}$ corresponding to ${a,b,c}$). Of all erbium atoms, 77\% are from an isotope that has nuclear spin $I_0 = 0$ and therefore no contribution from the hyperfine term in Eq.(\ref{eq:Hetheo}). Their energy levels are shown in Fig.\ref{fig:ErbiumTransitions} for $B_0$ applied in the $(a,b)$ plane. 

The remaining 23\% are from the $^{167}\text{Er}$ isotope with $I_0=7/2$. Its hyperfine coupling tensor to the $\text{Er}^{3+}$ electron spin is diagonal, with coefficients $A_{xx}=A_{yy}=873$ MHz and $A_{zz}=130$ MHz. The $16$ eigenfrequencies of the $^{167}\text{Er}$ spin Hamiltonian are also shown in Fig.\ref{fig:ErbiumTransitions}, again for $B_0$ applied in the $(a,b)$ plane. In the high-magnetic field limit $B_0 \gg A_\text{Er}/(g_\text{Er} \beta_e)$, the eigenstates are simply described by $|\pm,m_I \rangle$, $\pm$ describing the electron spin quantum number $m_S = \pm 1/2$ and $m_I$ the nuclear spin quantum number. For $B_0<100$\, mT as is the case in the measurements described here, this limit is only approximate, but we will use nevertheless the high-field state vectors as labels for the lower-field eigenstates. The strongest EPR-allowed transitions are the $m_I$-preserving transitions. In the following we will apply the fictitious spin model with $|\alpha,\beta\rangle$ = $|\pm, m_I \rangle$.

The $\text{CaWO}_4$ matrix also contains nuclear spins. Indeed, the $^{183}$W isotope has a spin $I=1/2$ with nuclear g-factor $g_n = 0.235$ (corresponding to a gyromagnetic ratio of $1.8$ MHz T$^{-1}$), and is present in a $p=0.13$ abundance, whereas the other tungsten isotopes are nucelar-spin-free. The interaction of the $^{183}$W atoms with the erbium ions gives rise to the ESEEM studied below. Because the 4f electron wavefunction is mainly located on the $\text{Er}^{3+}$ ion, the contact hyperfine with the nuclear spins of the lattice is expected to be negligibly small. We therefore model the hyperfine interaction with $^{183}$W by the dipole-dipole term in Eq.(\ref{eq:Aj}).

\subsection{Bismuth donors in Silicon}

\begin{figure*}[t!]%
\centering
\includegraphics[width=1.4\columnwidth]{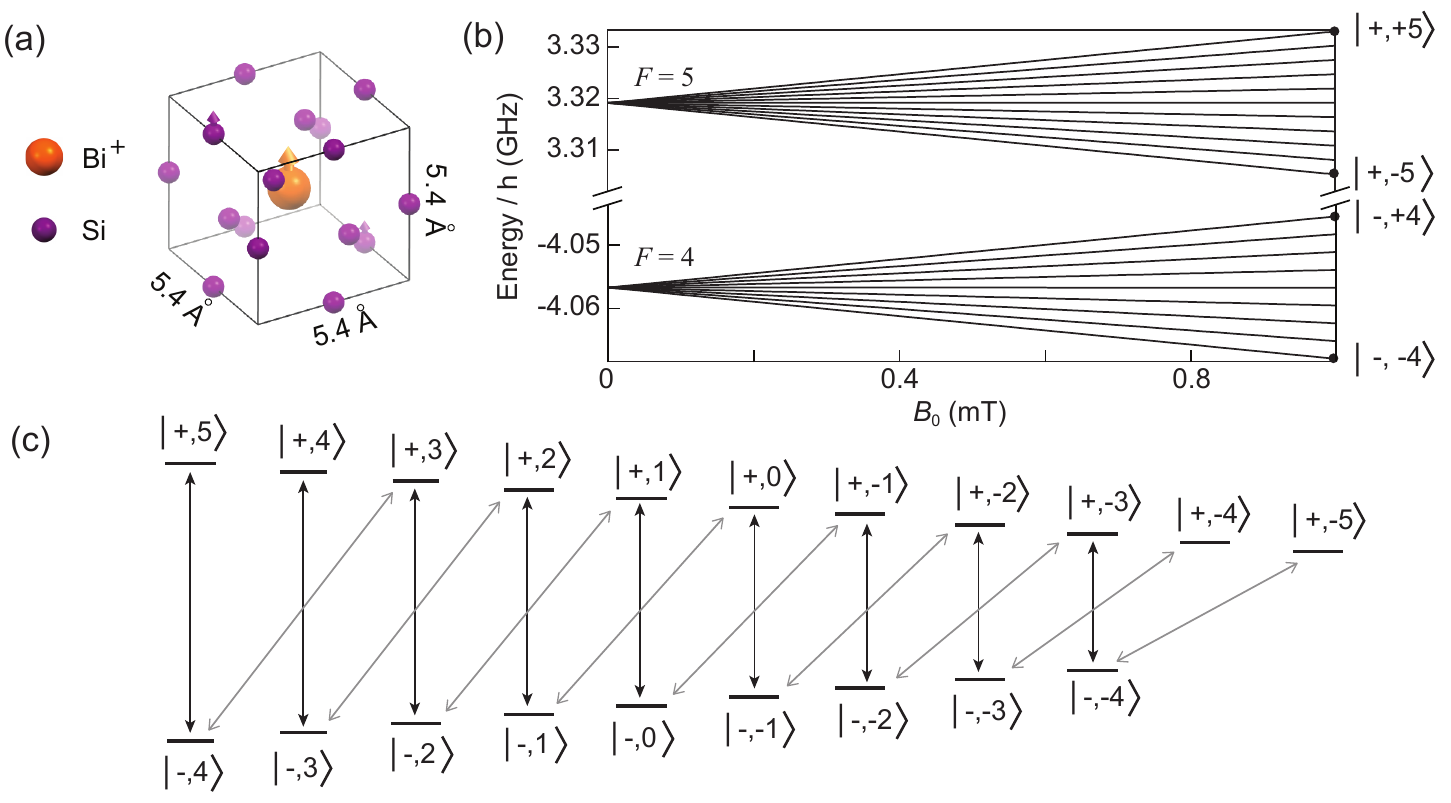}%
\caption{Structure and energy diagram of bismuth donors in silicon. (a) Silicon crystal structure, showing a substitutional bismuth atom coupled to nearby $^{29}$Si nuclear spins. The donor electron is trapped around the $\mathrm{Bi}^+$ ion and its wavefunction covers many lattice sites. (b) Energy levels of the bismuth donor, for $B_0 < 1$ mT. (c) Schematic representation of the allowed transitions (black and grey arrows) between the bismuth donor energy levels in the low field limit.}%
\label{fig:BismuthTransitions}%
\end{figure*}

The other system considered is the bismuth donor in silicon. Bismuth, as an element of the 5th column, substitutes in the silicon lattice by making 4 covalent bonds with neighboring atoms, leaving one unpaired electron that can be weakly trapped by the hydrogenic potential generated by the $\text{Bi}^+$ ion, whose spin gives rise to the resonance signal (see Fig.\ref{fig:BismuthTransitions}a). The donor wavefunction $\psi(\boldsymbol{r})$ has a complex structure that extends over $\approx 1.5$\,nm in the silicon lattice~\cite{kohn_theory_1955,feher_electron_1959} (see Supp. Info). As for $\mathrm{Er:CaWO}_4$, the donor spin Hamiltonian $H_\text{Bi}$ is given by Eq.(\ref{eq:Hetheo}). However in this case the g-tensor $ g_\text{e} \boldsymbol{1} $ is isotropic with $g_\text{e} = 2$, and the hyperfine tensor $A_\text{Bi} \boldsymbol{1} $ with the nuclear spin $I_0 = 9/2$ of the Bismuth atom is also isotropic, with $A_\text{Bi} /2\pi = 1.4754$ GHz. 

The eigenstates of $H_\text{Bi}$ have simple properties because of its isotropic character. Denoting $m_S$ ($m_I$) the eigenvalue of $S_{z,0}$ ($I_{z,0}$), we note that $m = m_I + m_S$ is a good quantum number since $H_\text{Bi}$ commutes with $S_{z,0} + I_{z,0}$ \citep{mohammady_bismuth_2010}, $z$ being the direction of $\boldsymbol{B_0}$. States with equal $m$ are hybridized by $H_\text{Bi}$. States $|m=5 \rangle$ and $|m=-5 \rangle$, corresponding to $|m_S=+1/2,m_I=9/2\rangle$ and $|m_S=-1/2,m_I=-9/2\rangle$, are non-degenerate and are thus also eigenstates of $H_\text{Bi}$. States with $|m| \leq 4$ belong to $9$ two-dimensional subspaces spanned by ${|m_S=+1/2,m_I=m-1/2\rangle,|m_S=-1/2,m_I=m+1/2\rangle}$ within which the $2$ eigenstates of $H_\text{Bi}$ are given by $|\pm,m\rangle = a_m^\pm |\pm \frac{1}{2},m \mp \frac{1}{2} \rangle + b_m^\pm |\mp \frac{1}{2},m \pm \frac{1}{2} \rangle$, with values of $a_m^\pm,b_m^\pm$ that can be determined analytically \citep{mohammady_bismuth_2010}.

Contrary to the erbium case, the measurements of bismuth donor spins are performed in the low-field limit $\left|g_{\rm e} \beta_{\rm e} B_0 \right|\ll \left|A_{\rm Bi}\right|$, in which the eigenstates are fully hybridized. In this limit, a useful approximate expression for the eigenenergy of level $|\pm,m \rangle$ is

\begin{equation}
   \label{eq:BiSiEnergyLevels}
E_m^{\pm} \approx -\frac{A_{\rm Bi}}{2} \pm \frac{5A_{\rm Bi}}{2} \pm\frac{m g_{\rm e} \beta_{\rm e} B_0}{10}.
\end{equation}

The magnetic-field dependence of the $|\pm,m \rangle$ energy levels is shown in Fig.~\ref{fig:BismuthTransitions}(b) for $B_0 < 1$\,mT. Note in particular that the separation between neighboring hyperfine levels is given by $E_m^{\pm}-E_{m-1}^{\pm} \approx  \pm\frac{g_{\rm e} \beta_{\rm e} B_0}{10} = \pm 2\pi \times 2.8\,\mathrm{B}_0\,\mathrm{GHz}$.

Because of the hybridization, all transitions that satisfy $|\Delta m| = 1$ are to some extent EPR-allowed at low field i.e., have a non-zero matrix element of operator $S_{0,x}$. In this work, we particularly focus on the $18$ $|\Delta m| = 1$ transitions that are in the $\simeq 7$GHz frequency range at low magnetic fields $|+,m\rangle \leftrightarrow |-,m-1\rangle$ and $|-,m\rangle \leftrightarrow |+,m-1\rangle$, as shown in Fig.\ref{fig:BismuthTransitions}c. The $|-,m\rangle \leftrightarrow |+,m+1\rangle$ and $|-,m+1\rangle \leftrightarrow |+,m-1\rangle$ transitions are degenerate in frequency for $-4 \leq m < 4$ as seen from Eq.(\ref{eq:BiSiEnergyLevels}), which results in only $10$ different transition frequencies (see Figs.~\ref{fig:BismuthTransitions}b,c, and \ref{fig:lineshape}a).

The most abundant isotope of silicon is $^{28}\text{Si}$, which is nuclear-spin-free. The lattice also contains a small percentage $p$ of $^{29}\text{Si}$ atoms that have a nuclear spin $I=1/2$ and give rise to the ESEEM. The g-factor of $^{29}\text{Si}$ is $g_\text{n} = -1.11$, yielding a gyromagnetic ratio of $8.46$ MHz T$^{-1}$.

The donor-$^{29}\text{Si}$ hyperfine interaction is given by Eq.(\ref{eq:Aj}). Due to the spatial extent of the electron wavefunction, the Fermi contact term is not negligible and needs to be taken into account together with the dipole-dipole coupling~\cite{hale_shallow_1969}; more details can be found in the Supplementary Information.

The restriction of the total system Hamiltonian to each of the $18$ ESR-allowed transitions of the Bismuth donor manifold can be mapped onto the fictitious spin-1/2 model of Section 2.4. Note however that the hyperfine term $|A_j|$ can take values up to $\sim 1$ MHz for proximal nuclear spins, which is comparable to or larger than the frequency difference between hyperfine states of the Bismuth donor manifold at low field as explained above. The validity of the fictitious spin-1/2 model in this context will be discussed in Section 5.

\section{Experimental setup and samples}

\begin{figure}[t!]%
\centering
\includegraphics[width=0.9\columnwidth]{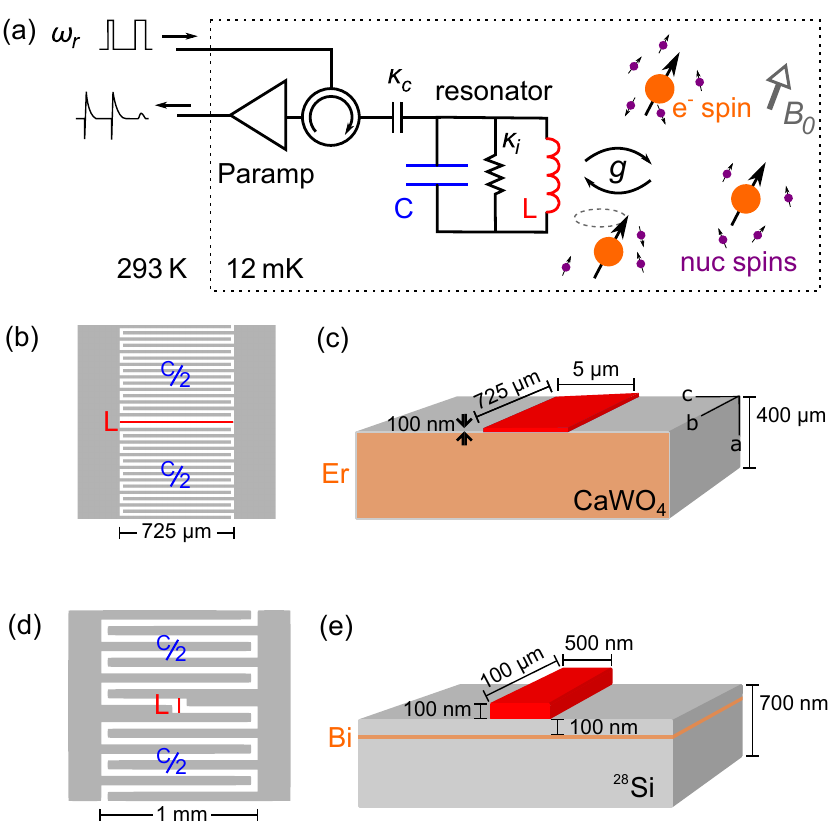}%
\caption{Experimental setup and samples. (a) Schematic of the low-temperature EPR spectrometer. The LC resonator is inductively coupled to electron spins, which are coupled to a nuclear spin bath that causes the ESEEM. The spins are probed by sequences of microwave pulses at the resonator frequency $\omega_\text{r}=1/\sqrt{LC}$. Reflected pulses as well as the echo signal are routed to a parametric amplifier, and are further amplified at $4$\,K, and finally demodulated and digitized at room temperature. (b,c) Design of the LC resonator used for the detection of erbium ion spins, with a $725\,\mu \mathrm{m}$-long, $5\,\mu \mathrm{m}$-wide inductor. It is patterned out of a $100$ nm thick niobium film deposited on top of a $\mathrm{CaWO}_4$ substrate bulk-doped with Er$^{3+}$ ions. (d,e) Design of the LC resonator used for the detection of bismuth donor spins, with a $100\,\mu \mathrm{m}$-long, $0.5\,\mu \mathrm{m}$-wide inductor. It is patterned out of a $100$ nm thick aluminum film deposited on top of a silicon substrate isotopically enriched in $^{28}$Si, in which bismuth ions were implanted at a 50-100 nm depth.}%
\label{fig:Setup}%
\end{figure}

The EPR spectrometer has been described in detail in refs.~\cite{bienfait_reaching_2015,probst_inductive-detection_2017} and is shown schematically in Fig.\ref{fig:Setup}a. It is built around a superconducting micro-resonator of frequency $\omega_\mathrm{r}$ consisting of a planar interdigitated capacitor shunted by an inductor, directly patterned on the crystal. We detect the spins that are located in the immediate vicinity of the resonator inductance. Note that the microwave $B_1$ field generated by the inductance is spatially inhomogeneous. If the spin location is broadly distributed, this can make the application of control pulses with a well-defined Rabi angle problematic\citep{ranjan_pulsed_2020}. As explained below, the resonator is more strongly coupled to the measurement line than in Ref.\,\citep{bienfait_reaching_2015} to increase the measurement bandwidth as requested for ESEEM spectroscopy.

The sample is mounted in a copper sample holder thermally anchored at the mixing chamber of a dilution refrigerator. A DC magnetic field $B_0$ is applied parallel to the sample surface and along the resonator inductance. The resonator is coupled capacitively to an antenna, which is itself connected to a microwave measurement setup in reflection. To minimize heat load, the coaxial cables between 4 K and 10 mK are in superconducting NbTi. To suppress thermal noise, the input line is heavily attenuated at low temperatures. Microwave pulses for driving the spins are sent to the resonator input, and their reflection or transmission, together with the echo signal emitted by the spins, is fed into a superconducting Josephson Parametric Amplifier, either of the flux-pumped type \citep{zhou_high-gain_2014} or of the Josephson Traveling-Wave Parametric Amplifier (JTWPA) type~\cite{macklin_nearquantum-limited_2015}.
Further microwave amplification takes place at 4K with a High-Electron-Mobility-Transistor (HEMT) from Low-Noise Factory, and then at room-temperature, before homodyne demodulation which yields the two signal quadratures $[I(t),Q(t)]$. The echo-containing quadrature signal is integrated to yield the echo amplitude $A_{\rm e}$. Such a setup was shown to reach sensitivities of order $10^2 - 10^3$ spin/$\sqrt{Hz}$ \citep{bienfait_reaching_2015,eichler_electron_2017,probst_inductive-detection_2017}.

Because of the small resonator mode volume and high quality factor, little microwave power is needed to drive the spins. The exact amount depends on the resonator geometry, as conveniently expressed by the power-to-field conversion factor $\alpha = B_1 / \sqrt{P_{in}}$. In the experiments reported here, the maximum microwave power used to drive the spins is on the order of $10$ nW. At this power, the superconducting pre-amplifiers saturate; however they recover rapidly enough (within a few microseconds) to amplify the much weaker subsequent spin-echoes. Flux-pumped JPAs are moreover switched off during the control pulses by pulsing the pump tone, whereas the JTWPA was kept on all the time. All microwave powers reaching the $4$ K HEMT are low enough that neither saturation nor damage are to be expected at this stage.

The erbium-doped sample (from Scientific Materials) was prepared by mixing erbium oxide with calcium and tungsten oxides before crystal growth, yielding a uniform Er concentration of $6\cdot10^{17}\,\mathrm{cm}^{-3}$ ($50$ ppm) throughout the sample. For resonator fabrication, the bulk crystal was cut and polished to a thin rectangular sample with dimensions $0.4$ mm $\times$ 3 mm $\times$ 6 mm parallel to $a\times b\times c$ axes. The resonator was patterned out of a $100$ nm thick (sputtered) Nb layer, using a design similar to that shown in Ref~\cite{bienfait_reaching_2015}. More specifically, 15 interdigitated fingers on either side of a $720$ $\mathrm{\mu m}\,\times 5 \mathrm{\mu m}$ inductive wire form an LC resonator, corresponding to a detection volume of $V_\text{Er}\sim 20$ pL. In the absence of magnetic field, the resonance frequency is $\omega_\text{\rm r}/2\pi = 4.323$ GHz. Its total quality factor of $8 \cdot 10^3$ is set both by the internal losses, characterized by the energy loss rate $\kappa_{i} = 5\cdot10^5$ s$^{-1}$, and by its coupling to the measurement line $\kappa_{C}=3\cdot10^{6}$ s$^{-1}$. For this geometry, the power-to-field factor is $\alpha = 1.7$ T W$^{-1/2}$.

The bismuth donors have been implanted at $\approx 100$ nm depth with a peak concentration of $8\cdot 10^{16}$ cm$^{-3}$ in a silicon sample. They lie in a 700 nm-thick silicon epilayer enriched in the nuclear-spin-free $^{28}$Si isotope (nominal concentration of 99.95\%), grown on top of a natural-abundance silicon sample. The resonator is patterned out of a 50 nm-thick aluminum film. It has the same geometry as reported in~\cite{probst_inductive-detection_2017}, with a $100$ $\mu$m-long, $500$ nm-wide inductor, and a detection volume of $0.2$ pL. Its frequency $\omega_\text{\rm r}/2\pi=7.370$ GHz is only slightly below the zero-field splitting of unperturbed Bi:Si donors $5A_{\rm Bi}/(2\pi) = 7.37585$ GHz~\cite{wolfowicz_atomic_2013}. The resonator internal loss is given by $\kappa_i = 3 \cdot 10^5$ s$^{-1}$. The coupling to the measurement line can be tuned at will by modifying the length of a microwave antenna that capacitively couples the measurement waveguide to the on-chip resonator via the copper sample holder~\cite{bienfait_reaching_2015,probst_inductive-detection_2017}. For the experiments reported below we used two settings : one for which the resonator was over-coupled ($\kappa_{C1} = 10^7$  s$^{-1}$), corresponding to a loaded quality factor $Q_1 = 4 \cdot 10^3$, and one for which the coupling was closer to critical ($\kappa_{C2} = 10^6$ s$^{-1}$), corresponding to a loaded quality factor $Q_2=3.4 \cdot 10^4$. In the low-Q case, square microwave pulses were used, of duration $\simeq 100$ ns similar to the cavity field damping time. In the high-Q case, shaped pulses were used~\cite{probst_shaped_2019} so that the intra-cavity field was a square pulse of $1\,\mu$s without any ringing. In some experiments, we additionally used a train of $\pi$ pulses (CPMG sequence), which generated extra echoes for significant gain in signal-to-noise ratio. More details on the pulse sequences used, the phase cycling scheme, and the repetition time, will be given in the following sections, together with experimental results. For this geometry, the power-to-field factor is $\alpha = 9$ T W$^{-1/2}$ for the low-Q case, and $\alpha = 21$ T W$^{-1/2}$ for the high-Q case.

\section{Results}

\subsection{Erbium-doped $\text{CaWO}_{4}$}

\subsubsection{Spectroscopy}

\begin{figure}[t!]%
\centering
\includegraphics[width=0.9\columnwidth]{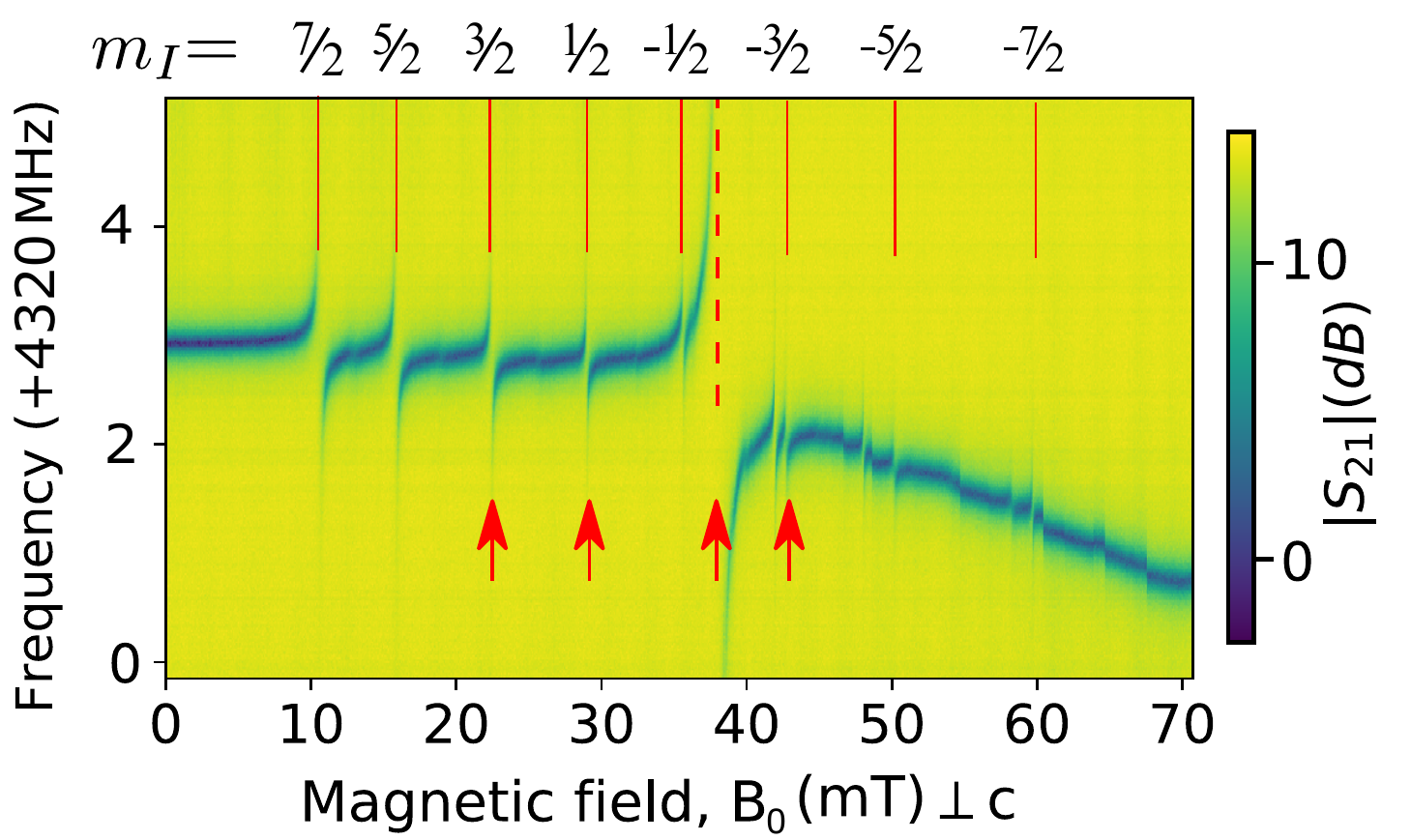}%
\caption{Spectroscopy of $\mathrm{Er}^{3+}$:$\mathrm{CaWO}_4$. Transmission coefficient $|S_{21}|(\omega)$ at $100$ mK as a function of the magnetic field $B_0$ applied along to the $a$ crystalline axis, around $4.323$ GHz. Red vertical lines indicate the expected Erbium transitions either for the $I=0$ isotopes (dashed) or the $I=7/2$ isotope (solid). Red arrows indicate the field at which the ESEEM data are measured.}%
\label{fig:ErSpectro}%
\end{figure}

Figure \ref{fig:ErSpectro} shows a spectrum comprising a series of microwave transmission measurements recorded on a vector network analyser, measured at $100$ mK, as a function of the magnetic field $B_0$ applied along the $b$ crystal axis ~\cite{data_repo}. Note that compared to Fig.\ref{fig:Setup}a, the resonator is coupled to the measurement line in a hanger geometry \citep{day_broadband_2003}, so that its resonance appears as a dip in the amplitude transmission coefficient $|S_{21}|$ (see Fig.\ref{fig:ErSpectro}). The 9 red lines indicate the values of $B_0$ at which the calculated $\text{Er}^{3+}$ ion transitions are equal to $\omega_\text{\rm r}$ (see Fig.~\ref{fig:ErbiumTransitions}b). Avoided level crossings are observed, which indicate a strong coupling of the resonator to the erbium transitions. Several additional anti-crossings and discontinuities are visible above $40$ mT. These are attributed to ytterbium impurities $^{171}$Yb and $^{173}$Yb and magnetic flux vortices penetrating the resonator.

Noticeable in the spectrum at $37$ mT is the large anti-crossing attributed to the highly concentrated $I=0$ erbium isotopes. Here the high-cooperativity regime $(C>30)$ is reached between the electronic spins and the resonator~\cite{kubo_strong_2010,probst_anisotropic_2013}. Typical linewidths $\Gamma/2\pi \sim 20$ MHz is observed. The coupling strength is also observed to be different for the eight $^{167}\mathrm{Er}$ transitions, which are labeled according to their corresponding nuclear spin projections $m_{I}$. This is explained by the partial polarisation of the ground-state hyperfine levels of $^{167}\text{Er}^{3+}$ at millikelvin temperatures (see Fig.~\ref{fig:ErbiumTransitions}b). 

\subsubsection{Two-Pulse ESEEM}

Four values of $B_0$ were selected for investigating ESEEM, indicated by the arrows in Fig.~\ref{fig:ErSpectro}; the first, second, and fourth corresponding to electronic-spin transitions of $^{167}\mathrm{Er}$, and the third one to the $I=0$ isotopes. The two-pulse echo sequence of Fig.\ref{fig:PulseSequences}a was implemented with square pulses of $1\,\mu$s duration applied at the resonator input, with double amplitude for the second pulse. Note that due to the $B_1$ spatial inhomogeneity combined with the homogeneous spin distribution throughout the crystal, the spread of Rabi frequency is too large to observe a well-defined nutation signal. The Rabi angle is therefore not well defined, and the echo is the average of different rotation angles. 

The control pulses driving the spins are filtered by the resonator bandwidth $\kappa / 2\pi \simeq 600$ kHz, corresponding to a field decay time $2 \kappa^{-1} = 3.3\,\mu \mathrm{s}$. The repetition time between echo sequences was 1 second, close to the spin relaxation time $T_1 \sim 1-2$ s measured by saturation recovery on the transitions studied. The echo signal was averaged 10 times with phase-cycling of the $\pi$-pulse to improve signal-to-noise and to remove signal offsets.

\begin{figure}[t!]%
\centering
\includegraphics[width=0.9\columnwidth]{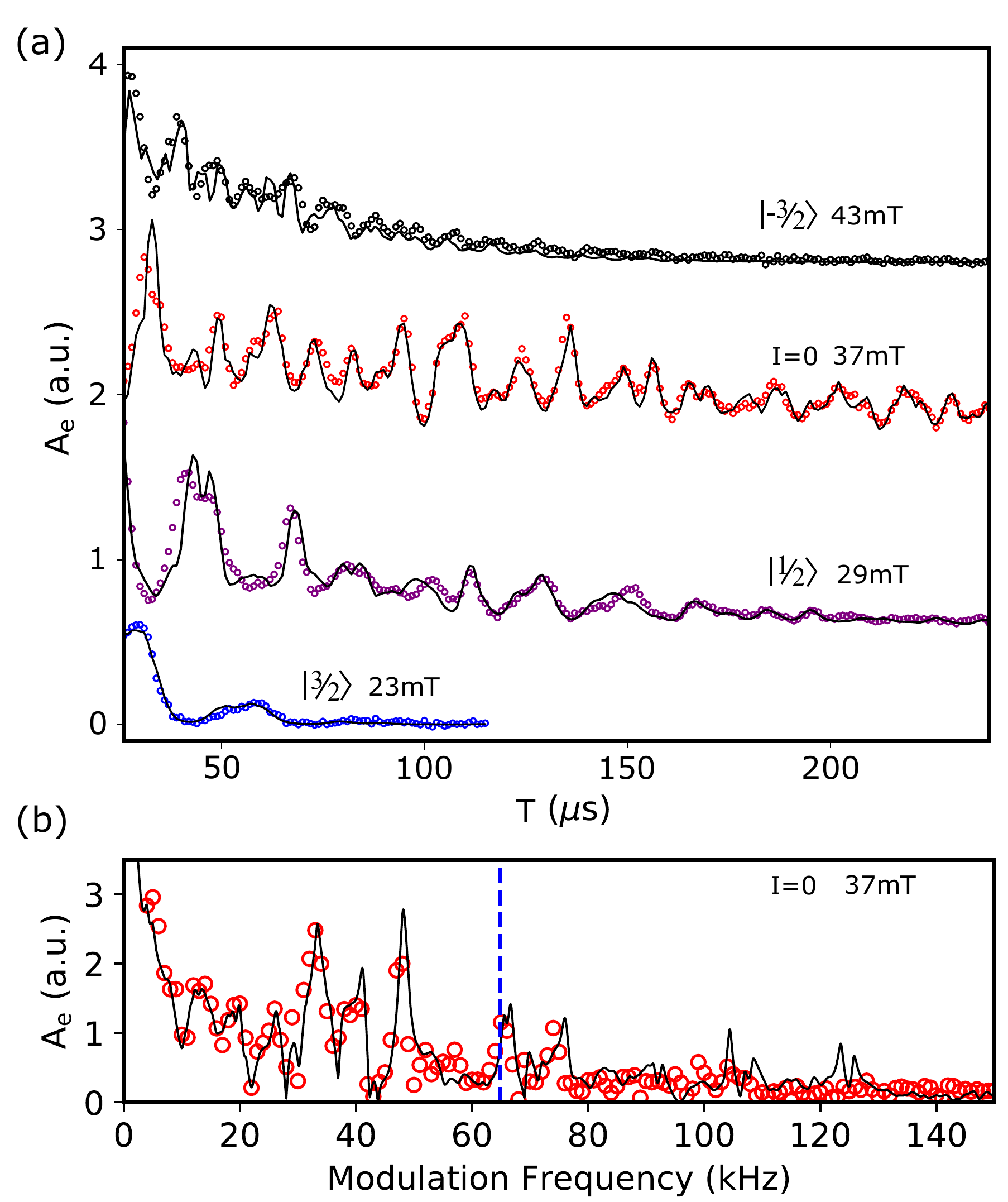}%
\caption{Two-pulse ESEEM on Er:CaWO4. (a) Integrated echo area $A_e$ as a function of the inter-pulse delay $\tau$, for four values of $B_0$ corresponding to different transitions. Open circles are measurements, and solid lines are the results of the ESEEM calculations as explained in the text Sec. V.A.3. (b) Measured (open red circles) and computed (solid line) fast Fourier transform of the $I=0$ data. The blue dashed line shows the Larmor frequency of $^{183}$W nuclei in free space.}%
\label{fig:Er2pESEEM}%
\end{figure}

Figure \ref{fig:Er2pESEEM} shows the two-pulse echo integrated amplitude $A_e$ as a function of $\tau$ for each of the four Er transitions investigated ~\cite{data_repo}. A clear envelope modulation signal is observed, together with an overall damping. Here we are interested only in the modulation pattern; a detailed study of the coherence time $T_2$ will be provided elsewhere. Qualitatively, we observe that the modulation frequency increases with $B_0$ and the modulation amplitude overall decreases with $B_0$, as expected from the discussion in Section 2. A Fourier transform of the $I=0$ data (see Fig.~\ref{fig:Er2pESEEM}b) shows the ESEEM spectrum. Well resolved peaks are observed in the $5 - 100$ kHz range, distributed around the $^{183}$W bare Larmor frequency $\omega_\text{W}$.

A very rough estimate of the number of erbium ions contributing to the signal is $[\mathrm{Er}]V_\text{Er}\kappa/\Gamma$, which is $ 2.5\cdot 10^8$ for the $I=0$ data, and $10^7$ for each $^{167}\text{Er}$ transition.

\subsubsection{Comparison with the model}

We compute the echo envelope $V'_{\mathrm{2p}}(\tau)$ described in Section 2.3, with the nearest 1000 coupled tungsten nuclei $(N=1000)$ and a natural $^{183}\mathrm{W}$ abundance of 14.4\% $(p=0.144)$. The hyperfine interaction is taken to be purely dipolar, as already explained \citep{guillot-noel_direct_2007,car_selective_2018}. The fitting proceeds by assigning an initial `guess' to six free parameters, then minimising using the L-BFGS-B algorithm \citep{byrd_limited_1995}. Three of these parameters $\left(|B_0|,\phi,\theta\right)$ describe the applied magnetic field:

\begin{align*}
B_{0} & =|B_0| \left[\sin\theta\cos\phi\,\hat{x}+\sin\theta\sin\phi\,\hat{y}+\cos\theta\,\hat{z}\right]
\end{align*}

Here $\theta$ is the angle of the field relative to the crystal \textit{c}-axis $(\hat{z})$ and $\phi$ is the angle relative to the \textit{a}-axis $(\hat{x})$ in the \textit{a}-$b$ plane ($\hat{x}$-$\hat{y}$ plane). The other three parameters ($C,T_{2},n)$ account for the echo envelope decay

\begin{alignat*}{1}
\mathrm{A_{e}}(\tau) & =V_{\mathrm{2p}}(\tau)\cdot C\exp\left(-\frac{2\tau}{T_{2}}\right)^{n},
\end{alignat*}

\noindent where $C$ represents the signal magnitude, $T_{2}$ the coherence time and $n\in[1,2]$ accounts for non-exponential decay. To determine the global minimum of the fit, the minimisation is repeated 200 times with randomly seeded initial values for the six parameters, bounded within the known uncertainty of the applied magnetic field $B_0$, signal strength $C$ and coherence time $T_2$. This approach reveals single local minima for each fitted parameter within the bounded range, with the variance of the 200 outcomes determining the uncertainty for each parameter. In particular, it yields precise values for the angles $\theta=91.47\pm 0.01^{0}$ and $\phi=90.50\pm 0.01^{0}$. The result of this fitting is presented in Fig.7(a), overlaid on the data for the $I=0$ transition at $37$ mT. Only the decay parameters ($C,T_{2},n)$ and magnetic field magnitude $|B_{0}|$ are left free when fitting the other three transitions in Fig.~\ref{fig:Er2pESEEM}(a). This was done for consistency between data sets, and because the $I=0$ data yields the most accurate values for $\phi$ and $\theta$ due to the low decoherence rate. The fits yield coherence times $T_2$ varying between $40\,\mu \mathrm{s}$ and $400\,\mu \mathrm{s}$, depending on the transition considered. Good agreement was also reached between the fitted and expected (pre-calibrated) field magnitudes.

Note that good fits to the data are also achieved by including only the nearest 100 tungsten nuclei, although noticeable deviations between the data and fit are observed with any less. The dimensionless `anisotropic hyperfine interaction parameter' $\rho$ described in the seminal publication on ESEEM \citep{rowan_electron-spin-echo_1965} is not required here. This parameter was introduced with the earliest attempts of ESEEM fitting, likely to compensate for the low number of simulated nuclear spins (typically 10 nearest nuclei or less), and was interpreted as an account for a potential distortion of the local environment caused by dopant insertion. Finally, a consideration of the spectral components presented in Fig.7(b) helps to more clearly identify the difference between the fit and the data. In particular, the high frequency components of the fitted model are not present experimentally due to the filtering effect of the superconducting resonance (260 kHz HWHM). This high-Q resonator greatly reduces the bandwidth of the RF field absorbed by the coupled Er-$^{183}\text{W}$ system and further limits the bandwidth of the detected echo signal.

\subsection{Bismuth donors sample}

\subsubsection{Spectroscopy}

\begin{figure}[t!]%
\centering
\includegraphics[width=0.9\columnwidth]{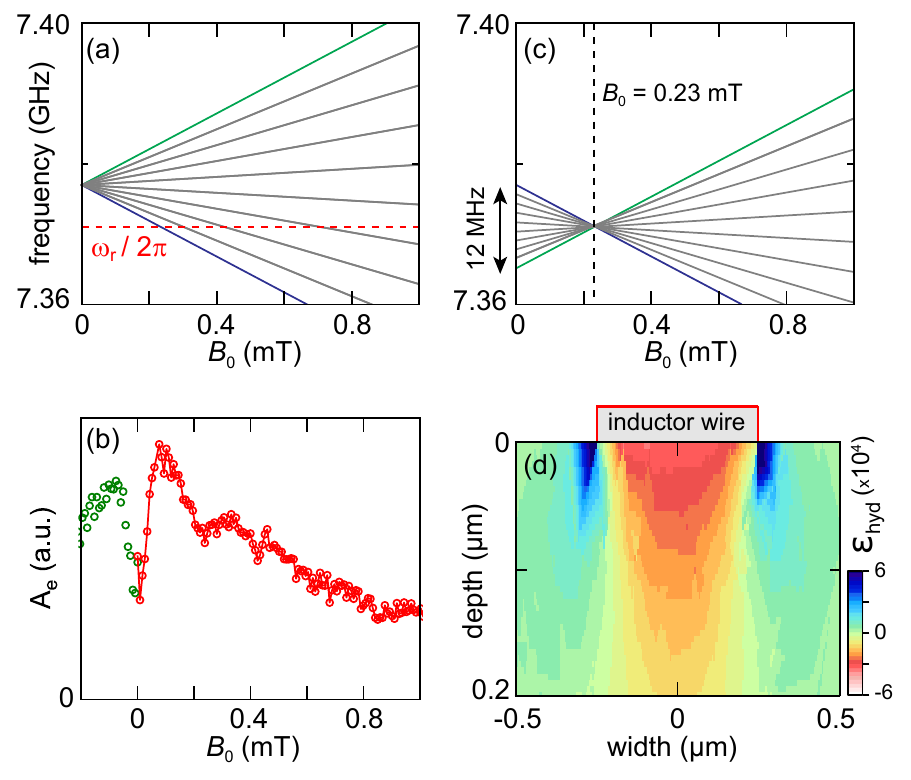}%
\caption{(a) EPR-allowed transitions of a bismuth donor in silicon for $0<B_0<1$ mT. The red dashed line denotes the resonator frequency $\omega_\text{\rm r}$. The spectrum is for an unstrained donor, for which the frequency at $B_0 = 0$ is $5A_\text{Bi}/(2\pi)$. (b) Echo-detected field sweep. The echo integral $A_e$ is plotted versus $B_0$. (c) Frequency of all $18$ Bismuth donor transitions that may contribute to the echo signal at a given field (here, $B_0 = 0.23$ mT). This is made possible by the strain-induced spread in $A_\text{Bi}$ between different donors. (d) Hydrostatic component of strain in silicon simulated using COMSOL. }%
\label{fig:lineshape}%
\end{figure}

Given the resonator frequency $\omega_{\rm r}$, four bismuth donor resonances should be observed when varying $B_0$ between $0$ and $1$  mT, as seen in Fig.\ref{fig:lineshape}a. Figure ~\ref{fig:lineshape}(b) shows an echo-detected field sweep, measured at $12$ mK: the integrated area $A_e$ of echoes obtained with a sequence shown in Fig.~\ref{fig:PulseSequences}a with $\tau = 50\,\mu$s pulse separation is plotted as a function of $B_0$ ~\cite{data_repo}. Instead of showing well-separated peaks as in the Erbium case, echoes are observed for all fields below $1$ mT, with a maximum close to $0.1$ mT, and extends in particular down to $B_0=0$  mT . This is the sign that each of the expected peaks is broadened and overlaps with neighboring transitions. Close to zero field, the echo amplitude goes down by a factor $2$ on a scale of $\sim 0.1$ mT, before showing a sharp increase at exactly zero field. These zero-field features are not currently understood, but they are reproducible as confirmed by the measurements at $B_0<0$, which are approximately symmetric to the $B_0>0$ data as they should be. 

Line broadening was reported previously for bismuth donors in silicon in related experiments \citep{bienfait_reaching_2015,probst_inductive-detection_2017}, and was attributed to the mechanical strain exerted by the aluminum resonator onto the silicon substrate due to differential thermal contractions between the metal and the substrate. At low strain, $A_{\rm Bi}$ depends linearly on the hydrostatic component of the strain tensor $\epsilon_{\rm hs}=(\epsilon_{xx}+\epsilon_{yy}+\epsilon_{zz})/3$ with a coefficient $dA_{\rm Bi}/d\epsilon_{\rm hs} / (2\pi)= 28$ GHz\citep{mansir_linear_2018}. Quantitative understanding of the lineshape was achieved in a given sample geometry based on this mechanism \citep{pla_strain-induced_2018}, using a finite-element modelling to estimate the strain profile induced upon sample cooldown. A similar modelling was performed for the Bi sample reported here (see Fig.~\ref{fig:lineshape}(d)). Based on the typical strain distribution $|\epsilon_{hyd}| \sim 3 \cdot 10^{-4}$ and on the hyperfine to strain coefficient $dA_{\rm Bi}/d\epsilon_{\rm hs}/(2\pi) = 28$  GHz, we expect the zero-field splitting $5A_{\rm Bi} / (2 \pi)$ to have a spread of $\sim 50$ MHz, which would indeed result in complete peak overlap in the $B_0 < 1$ mT region, as observed in Fig.~\ref{fig:lineshape}(b).

This broadening has two consequences worth highlighting. First, the bismuth donor echo signals can be measured down to $B_0=0$ mT, which otherwise is generally impossible in X-band spectroscopy. Here, this is enabled by the large hyperfine coupling of the Bi:Si donor, combined with strain-induced broadening. This makes it possible to detect ESEEM caused by very-weakly-coupled nuclear spins, which requires low magnetic fields as explained in Section 2. Second, at a given magnetic field, the spin-echo signal contains contributions from several overlapping EPR transitions. This last point is best understood from Fig.~\ref{fig:lineshape}(c), which shows how several classes of Bismuth donors, each with different hyperfine coupling $A_{\rm Bi}$, may have transitions resonant with $\omega_{\rm r}$. We will assume in the following that the inhomogeneous distribution of $A_{\rm Bi}$ is so broad that each of the $10$ $A_{\rm Bi}$ values for which one bismuth donor transition is resonant with $\omega_{\rm r}$ at fixed $B_0$ is equally probable, which is likely to be valid for $B_0 < 1$ mT. 

\subsubsection{Two-Pulse ESEEM}

\begin{figure}[t!]%
\centering
\includegraphics[width=0.9\columnwidth]{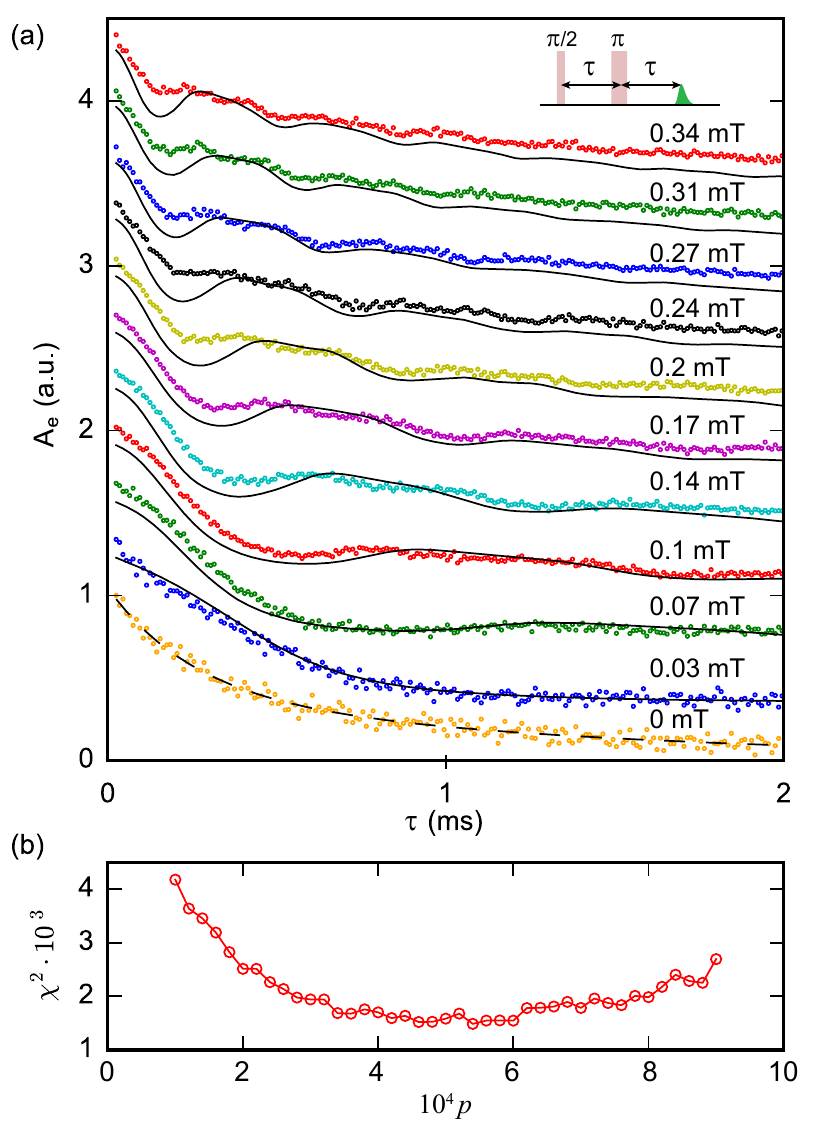}%
\caption{Two-pulse ESEEM of Bi:Si donors. (a) Echo integral $A_e$ versus inter-pulse delay $\tau$ for a 2-pulse echo sequence, for varying magnetic field $B_0$. Dots are experimental data, lines are results of the model (see text), assuming a concentration in $^{29}$Si of $p=4.4\cdot 10^{-4}$. The curves are vertically shifted, for clarity (b) Fit residue $\chi^2$ for different $^{29}$Si relative abundance $p$. The best fit is obtained for $p=4.4 \pm 1 \cdot 10^{-4}$, in agreement with the specified value.}%
\label{fig:Bi2pESEEM}%
\end{figure}

Two-pulse echoes are measured with the pulse sequence shown in Fig.~\ref{fig:PulseSequences}, which consists of a square $\pi/2_\text{X}$ pulse of duration $50$ ns followed by a square $\pi_\text{Y}$ pulse of duration $100$ ns after a delay $\tau$. Note that due to the donor spatial location in a shallow layer below the surface and to the strain shifting of their Larmor frequency~\cite{pla_strain-induced_2018}, the Rabi frequency is more homogeneous than in the erbium-doped sample, and Rabi rotations with a well-defined angle can be applied~\cite{pla_strain-induced_2018,probst_inductive-detection_2017}. To increase the signal-to-noise ratio, a CPMG sequence of $198$ $\pi$ pulses separated by $10\,\mu$s are used following the echo sequence~\cite{probst_inductive-detection_2017}. The curves are repeated $20$ times, with a delay of $2$ s in-between to enable spin relaxation of the donors. All the resulting echoes are then averaged. Phase cycling is performed by alternating sequences with opposite phases for the $\pi/2$ pulses and subtracting the resulting echoes. The data are obtained in the low-Q configuration (see section 4). 

Figure \ref{fig:Bi2pESEEM} shows the integral of the averaged echoes $A_{\rm e} (\tau)$ as a function of $\tau$, for various values of $B_0$ ~\cite{data_repo}. At non-zero field, $A_{\rm e} (\tau)$ shows $B_0$-dependent oscillations on top of an exponential decay with time constant $ T_2 = 2.6$ ms. Similar decay times were measured on the same chip with another resonator \citep{probst_inductive-detection_2017}, and are attributed to a combination of donor-donor dipolar interactions and magnetic noise from defects at the sample surface.

\begin{figure}[t!]%
\centering
\includegraphics[width=0.9\columnwidth]{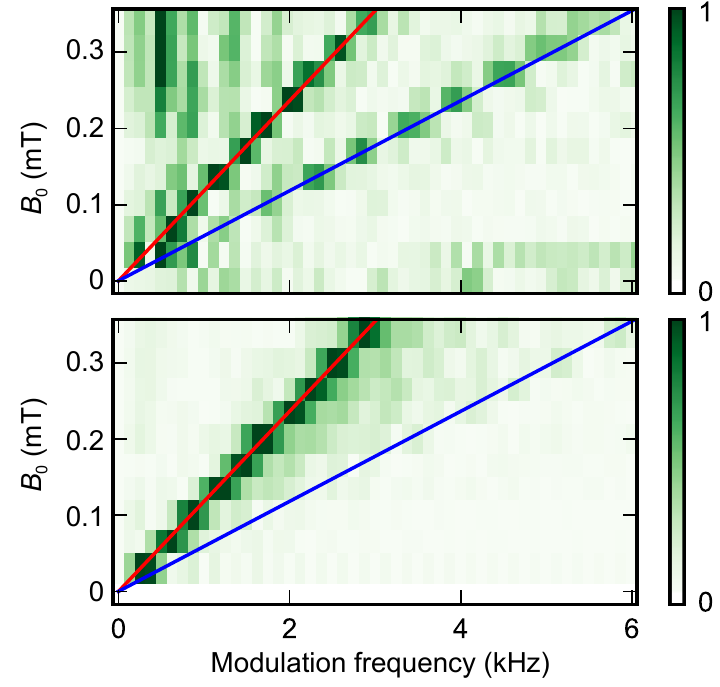}%
\caption{Amplitude of the Fourier transform of the experimental (top panel) and theoretical (bottom panel) 2-pulse Bi:Si donors ESEEM data.}%
\label{fig:BiFFT2PESEEM}%
\end{figure}

In the subsequent discussion, we concentrate on the ESEEM pattern. To analyze the data, each curve was divided by a constant exponential decay with $2.6$ ms time constant, mirrored at $t=0$, and Fourier transformed (see Fig.~\ref{fig:BiFFT2PESEEM}). Only two peaks are observed. Their frequencies vary linearly with $B_0$, and are found to be approximately $8$ kHz mT$^{-1}$ and $16$ kHz mT$^{-1}$. This is in good agreement with the gyromagnetic ratio of $^{29}$Si ($8.46$ kHz mT$^{-1}$); the presence of the second peak at twice this value is expected as explained in Section 2 for the two-pulse ESEEM in the weak-coupling limit. The oscillation amplitude goes down with $B_0$, again as expected from the model put forward in Section 2. 

A rough estimate of the number of donors contributing to the measurements shown in Fig.~\ref{fig:Bi2pESEEM} can be obtained by comparison with~\cite{probst_inductive-detection_2017}. Given the nearly identical resonator geometry, and assuming identical strain broadening in both samples, the ratio of the number of donors involved in both measurements is simply given by the ratio of resonator bandwidths. For the low-Q configuration, such as the two-pulse-echo of Fig.~\ref{fig:Bi2pESEEM}, this corresponds to $\simeq 5 \cdot 10^3$ dopants; in the high-Q configuration (see the 3- and 5-pulse data in the next paragraph), this number is reduced to $\simeq 5\cdot 10^2$ dopants.

\subsubsection{Three- and Five-Pulse ESEEM}

The spectral resolution provided by the measurement protocol is limited because of the finite electron coherence time $T_2$. As discussed in Section 2.3, this can be overcome by 3- or 5-pulse ESEEM.

\begin{figure*}[t!]%
\centering
\includegraphics[width=1.8\columnwidth]{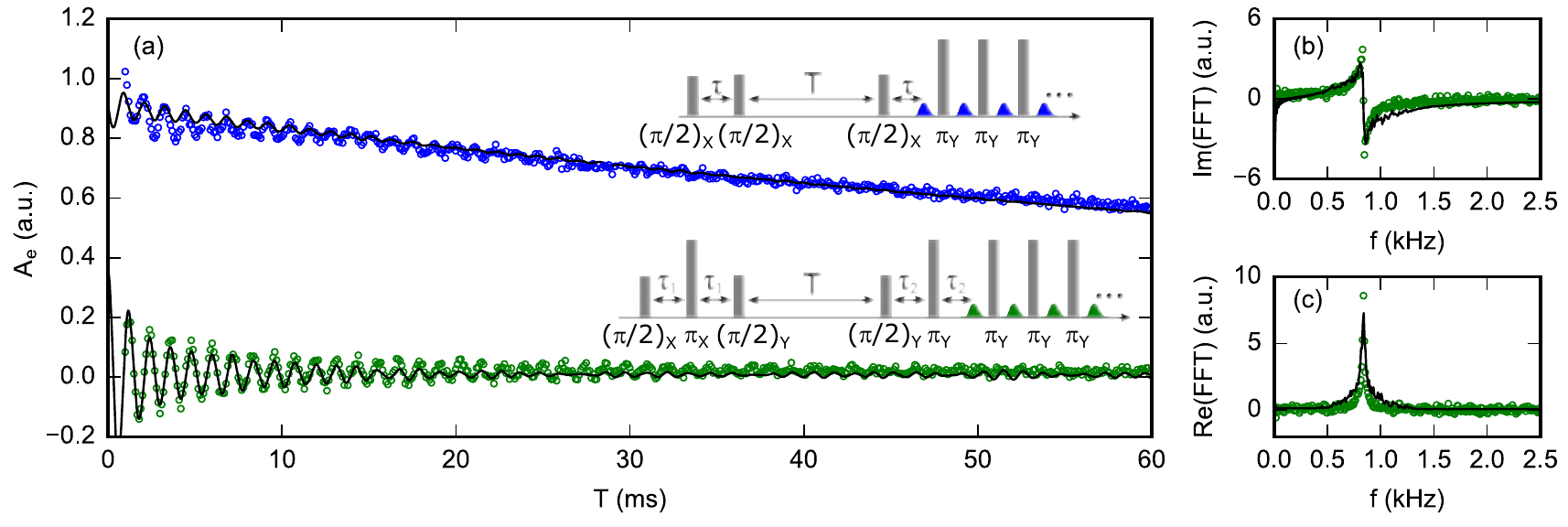}%
\caption{(a) 3-pulse (blue circles) and 5-pulse (green circles) ESEEM signals of Bi:Si donors at $B_0 = 0.1$ mT. Black lines are simulations assuming a $^{29}$Si concentration $p=4.4\cdot 10^{-4}$. (b) Imaginary and (c) real part of the Fourier transform of the 5-pulse ESEEM data. The spectrum only contains a peak at $850$ Hz, which is the $^{29}\mathrm{Si}$ nuclei Larmor frequency at this field.}%
\label{fig:Bi3p5pESEEM}%
\end{figure*}

We measure 3- and 5- pulse ESEEM with the pulse sequence shown in Fig.~\ref{fig:Bi3p5pESEEM}. The high-Q configuration is chosen, for which $T_1=120$ ms is measured (see Supplementary Information); shaped pulses generate an intra-cavity field in the form of a rectangular pulse of $1\, \mu$s duration with sharp rise and fall~\cite{probst_shaped_2019} despite the high resonator quality factor. The data are acquired at $B_0=0.1$ mT, so that $\omega_\text{\rm I}/2\pi \simeq 850$ Hz. The first blind spot for 3-pulse ESEEM is thus at $2 \pi / \omega_\text{\rm I} = 1.2$ ms; we chose $\tau=290\,\mu$s for the 3-pulse echo, and $\tau_1 = \tau_2 = 290\,\mu$s for the 5-pulse sequence. A sequence of $19$ CPMG $\pi$ pulses, separated by $50\,\mu$s, was used to enhance the signal-to-noise ratio. The sequences were repeated after a fixed waiting time of $100$ ms between the last $\pi$ pulse of one sequence and the first $\pi/2$ pulse of the following, to enable spin relaxation. Phase-cycling is used to suppress unwanted echoes (see Supplementary Information for the schemes~\cite{schweiger_principles_2001,kasumaj_5-_2008}). Each point is averaged over $2.5\cdot 10^4$ sequences, with a total acquisition time of two weeks for each curve ~\cite{data_repo}. 

The results are shown in Fig.~\ref{fig:Bi3p5pESEEM}, together with their fast Fourier transform ~\cite{data_repo}. Both the 3-pulse ESEEM (3PE) and 5-pulse ESEEM (5PE) curves show oscillations that last one order of magnitude longer than the electron spin $T_2$ (up to $20$  ms), enabling higher spectral resolution of the ESEEM signal. The 5PE curve has a higher oscillation amplitude than the 3PE by a factor 2-3, as expected. The decay of the oscillations occurs in $\sim 10$ ms, one order of magnitude faster than the stimulated echo amplitude (see the $3$PE curve), suggesting that it is an intrinsic feature of the ESEEM signal, as discussed below. 

The spectrum shows only one peak at the $^{29}$Si frequency. This is consistent with the expression provided in Section 2 and the Supplementary Information for the 3- and 5-pulse ESEEM, in which the terms oscillating at the sum and difference frequency are absent in contrast to the 2-pulse ESEEM. The peak width is $\simeq 100$ Hz, which indicates that the nuclei contributing to the ESEEM signal have hyperfine coupling strengths $A,B$ of at most $100$ Hz. Neglecting the contact interaction term, this corresponds to $^{29}$Si nuclei that are located at least $\sim 5$ nm away from the donor spin. 

The measured ESEEM spectrum of the bismuth donor sample qualitatively differs from the erbium sample, since it only contains a peak at the unperturbed silicon nuclei Larmor frequency (and at twice this frequency for the 2-pulse ESEEM), instead of the many peaks observed in Fig.\ref{fig:Er2pESEEM} indicating nuclear spin contribution with vastly different hyperfine strengths. This can be qualitatively understood by examining Eq.\ref{eq:2pESEEMSmallConcentration}. Defining $N_l$ as the number of lattice sites with approximately the same hyperfine parameters $A_l,B_l$ and modulation frequency $\omega_{\downarrow/\uparrow,l}$, the component at $\omega_{\downarrow/\uparrow,l}$ is visible in the spectrum if $N_l k_l p \sim 1$, which can only be achieved if $N_l p \sim 1$. In the case of erbium, $p=0.144$ so that even the sites closest to the ion (for which $N_l$ is of order unity) may satisfy this condition for well-chosen $B_0$. In the bismuth donor sample where $p=4.4 \cdot 10^{-4}$, this condition can only be met for $N_l \sim 10^3$, and therefore for crystal sites $l$ that are far from the donor, for which the hyperfine coupling is small, so that $\omega_{\downarrow/\uparrow,l} \simeq \omega_\text{\rm I}$. This is confirmed by the more quantitative modelling below.

\subsubsection{Comparison with the model}\label{sec:theo}

As explained above, the measured echo signal results from the contribution of all $18$ Bi:Si transitions because of strain broadening. To model the data, we therefore apply the fictitious spin-1/2 model to each transition, and sum the resulting echo amplitudes weighted by their relative contribution, which we determine using numerical simulations described in the Supplementary Information. 

Moreover, as discussed in Section 3.2, and in contrast to the erbium case, the fictitious spin model for a given transition needs to be validated in the low-$B_0$ regime because the energy difference between neighboring hyperfine levels of the bismuth donor manifold ($E_m^{\pm}-E_{m-1}^{\pm}) / h  \simeq 0.3$ MHz for $B_0=0.1$ mT is comparable to or even lower than the hyperfine coupling to some $^{29}$Si nuclei. In that case, the hyperfine interaction induces significant mixing between the bismuth donor and the $^{29}$Si eigenstates, and we should describe the coupled electron spin ${\mathbf S}_0$- $^{209}$Bi nuclear spin-${\mathbf I}_0$+$^{29}$Si nuclear spin ${\mathbf I}$ as a single $40$-level quantum system.

This study is described in the Supplementary Information Sec.IV for a $^{29}$Si with strong hyperfine coupling ($\geq 200$ kHz). The state mixing makes many transitions EPR-allowed, and the interference between these transitions causes fast oscillations in the spin echo signal, as seen in Fig.~S7 in the Supplementary Information. The frequencies of these oscillations depend greatly on the local Overhauser field on the donor electron spin. Since the latter has a large inhomogeneous broadening ($\sim 0.5$ MHz), the ensemble average leads to a rapid decay of the signal ($<1\,\mu$s). Given the $^{29}$Si concentration, about $10$\% of the donors have one or more $^{29}$Si with coupling $>300$ kHz in the proximity, which therefore leads to a rapid decay of the total echo signal within $\sim 1\,\mu$s by about $10$\%. In the experimental data, this fast decay is not visible because the echo signal is measured at longer times, and therefore the ESEEM signals presented in Fig.S5 in the Supplementary Information are those from $^{29}$Si with couplings $<200$ kHz. 

As for spins with a coupling strength between $20$ kHz and $200$ kHz, they lead to ESEEM amplitude much less than $1$\% as shown in Figs.~S7-S9 of the SI. For nuclear spins with a  hyperfine coupling $<100$ kHz, the fictitious spin model produces results with negligible errors of the modulation frequencies from the exact solution (Figs.~S5 and S6 in the Supplementary Information). Furthermore, the systematic numerical studies (Figs.S9 in the Supplementary Information) show that a nearby Si nuclear spin with coupling $<100$ kHz has little effects on the ESEEM due to other distant nuclear spins.
 
Considering these different contributions of Si nuclear spins of different hyperfine couplings, as discussed in the paragraph above and in more details in the Supplementary Information, we apply the fictitious spin-1/2 model to each EPR-allowed transition of the bismuth donor manifold, considering only Si nuclear spins that have a hyperfine coupling weaker than a certain cut-off which we choose as $20$ kHz, and discarding all the others. 

For each transition, we compute the hyperfine parameters that enter the fictitious spin-1/2 model for all sites of the silicon lattice. We then generate a large number of random configurations of nuclear spins. We compute the corresponding 2-, 3-, or 5- pulse ESEEM signal using the analytical formulas of section 2.4 after discarding all nuclei whose hyperfine coupling is larger than $20$ kHz. We average the signal for one configuration over all bismuth donor transitions using the weights determined by simulation, and then average the results over all the configurations computed. In this way, we obtain the curves shown in Fig.\ref{fig:Bi2pESEEM}. 

We use the two-pulse-Echo dataset to determine the most likely sample concentration in $^{29}$Si, using $p$ as a fitting parameter. As seen in Fig.~\ref{fig:Bi2pESEEM}b, the best fit is obtained for $p=4.4 \pm 1 \times 10^{-4}$, which is compatible with the specified $5 \times 10^{-4}$. The agreement is satisfactory but not perfect, as seen for instance in the amplitude of the short-time ESEEM oscillations which are lower in the measurements than in the simulations, particularly at larger field. Also, the peak at $2 \omega_\text{\rm I}$ is notably broader and has a lower amplitude than in the experiment.  

For the fitted value of $p$, the 3- and 5-pulse theoretical signals are also computed, and found to be in overall agreement with the data, even though the decay of the ESEEM signal predicted by the model is faster than in the experiment, and correspondingly the predicted ESEEM spectrum broader than the data. 

\section{Discussion and Conclusion}

We have reported 2-, 3- and 5-pulse ESEEM measurements using a quantum-limited EPR spectrometer on two model systems: erbium ions in a $\mathrm{CaWO}_4$ matrix, and bismuth donors in silicon. Whereas the erbium measurements are done in a commonly used regime of high field, the bismuth donor measurements are performed in an unusual regime of low nuclear spin density, low hyperfine coupling, and almost zero magnetic field. Good agreement is found with the simplest analytical ESEEM models. 

Having demonstrated that ESEEM is feasible in a millikelvin quantum-limited EPR spectrometer setup on two model spin systems, it is worth speculating in broader terms about its potential for real-world hyperfine spectroscopy. First, high magnetic fields are desirable for a better spectral resolution. Superconducting resonators in Nb, NbN or NbTiN can retain a high quality factor up to $\sim 1\,\mathrm{T}$~\cite{graaf_magnetic_2012,samkharadze_high-kinetic-inductance_2016,mahashabde_fast_2020}, so that quantum-limited EPR spectroscopy at Q-band can in principle be envisioned. Resonator bandwidths larger than demonstrated here are also desirable. Given, increasing $\kappa$ in the Purcell regime leads to longer relaxation times $T_1$, this should be done with care. One option is to increase also the coupling constant $g$, by further reduction of the resonator mode volume~\cite{ranjan_electron_2020}. Interestingly, this provides another motivation to apply higher magnetic fields, since $g$ is proportional to $\omega_{\rm r}$. Overall, a resonator at $\omega_\text{\rm r}/2\pi \simeq 30$ GHz, in a magnetic field $B_0 \simeq 1$ T, and with a $\kappa/2\pi \sim 10$ MHz bandwidth seems within reach, while keeping the Purcell $T_1$ well below $1$ s. Such high-bandwidth, high-sensitivity EPR spectrometer would be ideally suited for studying surface defects. One potential concern, however, is the power-handling capability of the resonator, as the kinetic inductance causes a non-linear response at high power. 

\bigskip
\paragraph*{Code and data availability}
All code and data necessary for generating figures 6-11 can be found at \\ https://doi.org/10.7910/DVN/ZJ2EEX. The analysis and plotting code is written in Python (.py) and Igor (.pxp). These files are sorted according to figure number, with the relevant files for each figure compressed into a single 7zip file (.7z). \bigskip

\paragraph*{Author Contributions}
S.P., M.R., M.L.D, A.D., and P.B. planned and designed the experiment. Z.Z., P.G. prepared the Er:$\mathrm{CaWO}_4$ crystal. J.M. prepared and provided the bismuth-donor-implanted silicon sample. S.P., M. R., and M.L.D. fabricated the devices, set up the experiment, and acquired the data. S.P., G.L.Z, M.R., V.R., M.L.D., B.A., A.D. , R.B.L., T.C., P.G., P.B. worked on the data analysis. The project was supervised by R.B.L. and P.B. All authors contributed to manuscript preparation.

\paragraph*{Competing interests}The authors declare that they have no conflict of interest.

\paragraph*{Acknowledgements}
We thank P. S\'enat, D. Duet and J.-C. Tack for the technical support, and are grateful for fruitful discussions within the Quantronics group. We acknowledge IARPA and Lincoln Labs for providing a Josephson Traveling-Wave Parametric Amplifier used in some of the measurements. We acknowledge support of the European Research Council under the European Community’s Seventh Framework Programme (FP7/2007-2013) through grant agreement No. 615767 (CIRQUSS) and under the European Union’s Horizon 2020 research and innovation programme [Grant Agreement No. 771493 (LOQO-MOTIONS)], of the Agence Nationale de la Recherche under the Chaire Industrielle NASNIQ (grant number ANR-17-CHIN-0001) supported by Atos, the project QIPSE (Hong Kong RGC – French ANR Joint Scheme Fund Project A-CUHK403/15), and the project MIRESPIN, and of Region Ile-de-France Domaine d'Interet Majeur SIRTEQ under grant REIMIC. This project has received funding from the European Union’s Horizon 2020 research and innovation programme under the Marie Skłodowska-Curie grant agreement No 792727. AD acknowledges a SNSF mobility fellowship (177732).


\begin{thebibliography}{54}%
\makeatletter
\providecommand \@ifxundefined [1]{%
 \@ifx{#1\undefined}
}%
\providecommand \@ifnum [1]{%
 \ifnum #1\expandafter \@firstoftwo
 \else \expandafter \@secondoftwo
 \fi
}%
\providecommand \@ifx [1]{%
 \ifx #1\expandafter \@firstoftwo
 \else \expandafter \@secondoftwo
 \fi
}%
\providecommand \natexlab [1]{#1}%
\providecommand \enquote  [1]{``#1''}%
\providecommand \bibnamefont  [1]{#1}%
\providecommand \bibfnamefont [1]{#1}%
\providecommand \citenamefont [1]{#1}%
\providecommand \href@noop [0]{\@secondoftwo}%
\providecommand \href [0]{\begingroup \@sanitize@url \@href}%
\providecommand \@href[1]{\@@startlink{#1}\@@href}%
\providecommand \@@href[1]{\endgroup#1\@@endlink}%
\providecommand \@sanitize@url [0]{\catcode `\\12\catcode `\$12\catcode
  `\&12\catcode `\#12\catcode `\^12\catcode `\_12\catcode `\%12\relax}%
\providecommand \@@startlink[1]{}%
\providecommand \@@endlink[0]{}%
\providecommand \url  [0]{\begingroup\@sanitize@url \@url }%
\providecommand \@url [1]{\endgroup\@href {#1}{\urlprefix }}%
\providecommand \urlprefix  [0]{URL }%
\providecommand \Eprint [0]{\href }%
\providecommand \doibase [0]{https://doi.org/}%
\providecommand \selectlanguage [0]{\@gobble}%
\providecommand \bibinfo  [0]{\@secondoftwo}%
\providecommand \bibfield  [0]{\@secondoftwo}%
\providecommand \translation [1]{[#1]}%
\providecommand \BibitemOpen [0]{}%
\providecommand \bibitemStop [0]{}%
\providecommand \bibitemNoStop [0]{.\EOS\space}%
\providecommand \EOS [0]{\spacefactor3000\relax}%
\providecommand \BibitemShut  [1]{\csname bibitem#1\endcsname}%
\let\auto@bib@innerbib\@empty
\bibitem [{\citenamefont {Schweiger}\ and\ \citenamefont
  {Jeschke}(2001)}]{schweiger_principles_2001}%
  \BibitemOpen
  \bibfield  {author} {\bibinfo {author} {\bibfnamefont {A.}~\bibnamefont
  {Schweiger}}\ and\ \bibinfo {author} {\bibfnamefont {G.}~\bibnamefont
  {Jeschke}},\ }\href@noop {} {\emph {\bibinfo {title} {Principles of pulse
  electron paramagnetic resonance}}}\ (\bibinfo  {publisher} {Oxford University
  Press},\ \bibinfo {year} {2001})\BibitemShut {NoStop}%
\bibitem [{\citenamefont {Song}\ \emph {et~al.}(2016)\citenamefont {Song},
  \citenamefont {Liu}, \citenamefont {Kaur}, \citenamefont {Esquiaqui},
  \citenamefont {Hunter}, \citenamefont {Hill}, \citenamefont {Smith},\ and\
  \citenamefont {Fanucci}}]{song_toward_2016}%
  \BibitemOpen
  \bibfield  {author} {\bibinfo {author} {\bibfnamefont {L.}~\bibnamefont
  {Song}}, \bibinfo {author} {\bibfnamefont {Z.}~\bibnamefont {Liu}}, \bibinfo
  {author} {\bibfnamefont {P.}~\bibnamefont {Kaur}}, \bibinfo {author}
  {\bibfnamefont {J.~M.}\ \bibnamefont {Esquiaqui}}, \bibinfo {author}
  {\bibfnamefont {R.~I.}\ \bibnamefont {Hunter}}, \bibinfo {author}
  {\bibfnamefont {S.}~\bibnamefont {Hill}}, \bibinfo {author} {\bibfnamefont
  {G.~M.}\ \bibnamefont {Smith}},\ and\ \bibinfo {author} {\bibfnamefont
  {G.~E.}\ \bibnamefont {Fanucci}},\ }\bibfield  {title} \bibinfo {title} {Toward increased concentration sensitivity for
  continuous wave {EPR} investigations of spin-labeled biological
  macromolecules at high fields} {\bibfield  {journal} {\bibinfo
  {journal} {Journal of Magnetic Resonance}\ }\textbf {\bibinfo {volume}
  {265}},\ \bibinfo {pages} {188} (\bibinfo {year} {2016})}\BibitemShut
  {NoStop}%
\bibitem [{\citenamefont {Elzerman}\ \emph {et~al.}(2004)\citenamefont
  {Elzerman}, \citenamefont {Hanson}, \citenamefont {Beveren}, \citenamefont
  {Witkamp}, \citenamefont {Vandersypen},\ and\ \citenamefont
  {Kouwenhoven}}]{elzerman_single-shot_2004}%
  \BibitemOpen
  \bibfield  {author} {\bibinfo {author} {\bibfnamefont {J.~M.}\ \bibnamefont
  {Elzerman}}, \bibinfo {author} {\bibfnamefont {R.}~\bibnamefont {Hanson}},
  \bibinfo {author} {\bibfnamefont {L.~H. W.~v.}\ \bibnamefont {Beveren}},
  \bibinfo {author} {\bibfnamefont {B.}~\bibnamefont {Witkamp}}, \bibinfo
  {author} {\bibfnamefont {L.~M.~K.}\ \bibnamefont {Vandersypen}},\ and\
  \bibinfo {author} {\bibfnamefont {L.~P.}\ \bibnamefont {Kouwenhoven}},\
  }\bibfield  {title} {{\bibinfo {title} {Single-shot
  read-out of an individual electron spin in a quantum dot}},\ }\href
  {https://doi.org/10.1038/nature02693} {\bibfield  {journal} {\bibinfo
  {journal} {Nature}\ }\textbf {\bibinfo {volume} {430}},\ \bibinfo {pages}
  {431} (\bibinfo {year} {2004})}\BibitemShut {NoStop}%
\bibitem [{\citenamefont {Veldhorst}\ \emph {et~al.}(2014)\citenamefont
  {Veldhorst}, \citenamefont {Hwang}, \citenamefont {Yang}, \citenamefont
  {Leenstra}, \citenamefont {de~Ronde}, \citenamefont {Dehollain},
  \citenamefont {Muhonen}, \citenamefont {Hudson}, \citenamefont {Itoh},
  \citenamefont {Morello},\ and\ \citenamefont
  {Dzurak}}]{veldhorst_addressable_2014}%
  \BibitemOpen
  \bibfield  {author} {\bibinfo {author} {\bibfnamefont {M.}~\bibnamefont
  {Veldhorst}}, \bibinfo {author} {\bibfnamefont {J.~C.~C.}\ \bibnamefont
  {Hwang}}, \bibinfo {author} {\bibfnamefont {C.~H.}\ \bibnamefont {Yang}},
  \bibinfo {author} {\bibfnamefont {A.~W.}\ \bibnamefont {Leenstra}}, \bibinfo
  {author} {\bibfnamefont {B.}~\bibnamefont {de~Ronde}}, \bibinfo {author}
  {\bibfnamefont {J.~P.}\ \bibnamefont {Dehollain}}, \bibinfo {author}
  {\bibfnamefont {J.~T.}\ \bibnamefont {Muhonen}}, \bibinfo {author}
  {\bibfnamefont {F.~E.}\ \bibnamefont {Hudson}}, \bibinfo {author}
  {\bibfnamefont {K.~M.}\ \bibnamefont {Itoh}}, \bibinfo {author}
  {\bibfnamefont {A.}~\bibnamefont {Morello}},\ and\ \bibinfo {author}
  {\bibfnamefont {A.~S.}\ \bibnamefont {Dzurak}},\ }\bibfield  {title}
  {{\bibinfo {title} {An addressable quantum dot qubit with
  fault-tolerant control-fidelity}},\ }\href
  {https://doi.org/10.1038/nnano.2014.216} {\bibfield  {journal} {\bibinfo
  {journal} {Nature Nanotechnology}\ }\textbf {\bibinfo {volume} {9}},\
  \bibinfo {pages} {981} (\bibinfo {year} {2014})}\BibitemShut {NoStop}%
\bibitem [{\citenamefont {Morello}\ \emph {et~al.}(2010)\citenamefont
  {Morello}, \citenamefont {Pla}, \citenamefont {Zwanenburg}, \citenamefont
  {Chan}, \citenamefont {Tan}, \citenamefont {Huebl}, \citenamefont {Mottonen},
  \citenamefont {Nugroho}, \citenamefont {Yang}, \citenamefont {van
  Donkelaar},\ and\ \citenamefont {{others}}}]{morello_single-shot_2010}%
  \BibitemOpen
  \bibfield  {author} {\bibinfo {author} {\bibfnamefont {A.}~\bibnamefont
  {Morello}}, \bibinfo {author} {\bibfnamefont {J.~J.}\ \bibnamefont {Pla}},
  \bibinfo {author} {\bibfnamefont {F.~A.}\ \bibnamefont {Zwanenburg}},
  \bibinfo {author} {\bibfnamefont {K.~W.}\ \bibnamefont {Chan}}, \bibinfo
  {author} {\bibfnamefont {K.~Y.}\ \bibnamefont {Tan}}, \bibinfo {author}
  {\bibfnamefont {H.}~\bibnamefont {Huebl}}, \bibinfo {author} {\bibfnamefont
  {M.}~\bibnamefont {Mottonen}}, \bibinfo {author} {\bibfnamefont {C.~D.}\
  \bibnamefont {Nugroho}}, \bibinfo {author} {\bibfnamefont {C.}~\bibnamefont
  {Yang}}, \bibinfo {author} {\bibfnamefont {J.~A.}\ \bibnamefont {van
  Donkelaar}},\ and\ \bibinfo {author} {\bibnamefont {{others}}},\ }\bibfield
  {title} {\bibinfo {title} {Single-shot readout of an electron spin in
  silicon},\ }\href {https://doi.org/doi:10.1038/nature09392} {\bibfield
  {journal} {\bibinfo  {journal} {Nature}\ }\textbf {\bibinfo {volume} {467}},\
  \bibinfo {pages} {687} (\bibinfo {year} {2010})}\BibitemShut {NoStop}%
\bibitem [{\citenamefont {Pla}\ \emph {et~al.}(2012)\citenamefont {Pla},
  \citenamefont {Tan}, \citenamefont {Dehollain}, \citenamefont {Lim},
  \citenamefont {Morton}, \citenamefont {Jamieson}, \citenamefont {Dzurak},\
  and\ \citenamefont {Morello}}]{pla_single-atom_2012}%
  \BibitemOpen
  \bibfield  {author} {\bibinfo {author} {\bibfnamefont {J.~J.}\ \bibnamefont
  {Pla}}, \bibinfo {author} {\bibfnamefont {K.~Y.}\ \bibnamefont {Tan}},
  \bibinfo {author} {\bibfnamefont {J.~P.}\ \bibnamefont {Dehollain}}, \bibinfo
  {author} {\bibfnamefont {W.~H.}\ \bibnamefont {Lim}}, \bibinfo {author}
  {\bibfnamefont {J.~J.~L.}\ \bibnamefont {Morton}}, \bibinfo {author}
  {\bibfnamefont {D.~N.}\ \bibnamefont {Jamieson}}, \bibinfo {author}
  {\bibfnamefont {A.~S.}\ \bibnamefont {Dzurak}},\ and\ \bibinfo {author}
  {\bibfnamefont {A.}~\bibnamefont {Morello}},\ }\bibfield  {title}
  {{\bibinfo {title} {A single-atom electron spin qubit in
  silicon}},\ }\href {https://doi.org/10.1038/nature11449} {\bibfield
  {journal} {\bibinfo  {journal} {Nature}\ }\textbf {\bibinfo {volume} {489}},\
  \bibinfo {pages} {541} (\bibinfo {year} {2012})}\BibitemShut {NoStop}%
\bibitem [{\citenamefont {Wrachtrup}\ \emph {et~al.}(1993)\citenamefont
  {Wrachtrup}, \citenamefont {Von~Borczyskowski}, \citenamefont {Bernard},
  \citenamefont {Orritt},\ and\ \citenamefont
  {Brown}}]{wrachtrup_optical_1993}%
  \BibitemOpen
  \bibfield  {author} {\bibinfo {author} {\bibfnamefont {J.}~\bibnamefont
  {Wrachtrup}}, \bibinfo {author} {\bibfnamefont {C.}~\bibnamefont
  {Von~Borczyskowski}}, \bibinfo {author} {\bibfnamefont {J.}~\bibnamefont
  {Bernard}}, \bibinfo {author} {\bibfnamefont {M.}~\bibnamefont {Orritt}},\
  and\ \bibinfo {author} {\bibfnamefont {R.}~\bibnamefont {Brown}},\ }\bibfield
   {title} {\bibinfo {title} {Optical detection of magnetic resonance in a
  single molecule},\ }\href {https://doi.org/10.1038/363244a0} {\bibfield
  {journal} {\bibinfo  {journal} {Nature}\ }\textbf {\bibinfo {volume} {363}},\
  \bibinfo {pages} {244} (\bibinfo {year} {1993})}\BibitemShut {NoStop}%
\bibitem [{\citenamefont {Jelezko}\ \emph {et~al.}(2004)\citenamefont
  {Jelezko}, \citenamefont {Gaebel}, \citenamefont {Popa}, \citenamefont
  {Gruber},\ and\ \citenamefont {Wrachtrup}}]{jelezko_observation_2004}%
  \BibitemOpen
  \bibfield  {author} {\bibinfo {author} {\bibfnamefont {F.}~\bibnamefont
  {Jelezko}}, \bibinfo {author} {\bibfnamefont {T.}~\bibnamefont {Gaebel}},
  \bibinfo {author} {\bibfnamefont {I.}~\bibnamefont {Popa}}, \bibinfo {author}
  {\bibfnamefont {A.}~\bibnamefont {Gruber}},\ and\ \bibinfo {author}
  {\bibfnamefont {J.}~\bibnamefont {Wrachtrup}},\ }\bibfield  {title} {\bibinfo
  {title} {Observation of {Coherent} {Oscillations} in a {Single} {Electron}
  {Spin}},\ }\href {https://doi.org/10.1103/PhysRevLett.92.076401} {\bibfield
  {journal} {\bibinfo  {journal} {Physical Review Letters}\ }\textbf {\bibinfo
  {volume} {92}},\ \bibinfo {pages} {076401} (\bibinfo {year}
  {2004})}\BibitemShut {NoStop}%
\bibitem [{\citenamefont {Rugar}\ \emph {et~al.}(2004)\citenamefont {Rugar},
  \citenamefont {Budakian}, \citenamefont {Mamin},\ and\ \citenamefont
  {Chui}}]{rugar_single_2004}%
  \BibitemOpen
  \bibfield  {author} {\bibinfo {author} {\bibfnamefont {D.}~\bibnamefont
  {Rugar}}, \bibinfo {author} {\bibfnamefont {R.}~\bibnamefont {Budakian}},
  \bibinfo {author} {\bibfnamefont {H.}~\bibnamefont {Mamin}},\ and\ \bibinfo
  {author} {\bibfnamefont {B.}~\bibnamefont {Chui}},\ }\bibfield  {title}
  {\bibinfo {title} {Single spin detection by magnetic resonance force
  microscopy},\ }\href {https://doi.org/10.1038/nature02658} {\bibfield
  {journal} {\bibinfo  {journal} {Nature}\ }\textbf {\bibinfo {volume} {430}},\
  \bibinfo {pages} {329} (\bibinfo {year} {2004})}\BibitemShut {NoStop}%
\bibitem [{\citenamefont {Baumann}\ \emph {et~al.}(2015)\citenamefont
  {Baumann}, \citenamefont {Paul}, \citenamefont {Choi}, \citenamefont {Lutz},
  \citenamefont {Ardavan},\ and\ \citenamefont
  {Heinrich}}]{baumann_electron_2015}%
  \BibitemOpen
  \bibfield  {author} {\bibinfo {author} {\bibfnamefont {S.}~\bibnamefont
  {Baumann}}, \bibinfo {author} {\bibfnamefont {W.}~\bibnamefont {Paul}},
  \bibinfo {author} {\bibfnamefont {T.}~\bibnamefont {Choi}}, \bibinfo {author}
  {\bibfnamefont {C.~P.}\ \bibnamefont {Lutz}}, \bibinfo {author}
  {\bibfnamefont {A.}~\bibnamefont {Ardavan}},\ and\ \bibinfo {author}
  {\bibfnamefont {A.~J.}\ \bibnamefont {Heinrich}},\ }\bibfield  {title}
  {\bibinfo {title} {Electron paramagnetic resonance of individual atoms on a
  surface},\ }\href {https://doi.org/10.1126/science.aac8703} {\bibfield
  {journal} {\bibinfo  {journal} {Science}\ }\textbf {\bibinfo {volume}
  {350}},\ \bibinfo {pages} {417} (\bibinfo {year} {2015})}\BibitemShut
  {NoStop}%
\bibitem [{\citenamefont {Narkowicz}\ \emph {et~al.}(2008)\citenamefont
  {Narkowicz}, \citenamefont {Suter},\ and\ \citenamefont
  {Niemeyer}}]{narkowicz_scaling_2008}%
  \BibitemOpen
  \bibfield  {author} {\bibinfo {author} {\bibfnamefont {R.}~\bibnamefont
  {Narkowicz}}, \bibinfo {author} {\bibfnamefont {D.}~\bibnamefont {Suter}},\
  and\ \bibinfo {author} {\bibfnamefont {I.}~\bibnamefont {Niemeyer}},\
  }\bibfield  {title} {\bibinfo {title} {Scaling of sensitivity and efficiency
  in planar microresonators for electron spin resonance},\ }\href
  {https://doi.org/10.1063/1.2964926} {\bibfield  {journal} {\bibinfo
  {journal} {Review of Scientific Instruments}\ }\textbf {\bibinfo {volume}
  {79}},\ \bibinfo {pages} {084702} (\bibinfo {year} {2008})}\BibitemShut
  {NoStop}%
\bibitem [{\citenamefont {Artzi}\ \emph {et~al.}(2015)\citenamefont {Artzi},
  \citenamefont {Twig},\ and\ \citenamefont
  {Blank}}]{artzi_induction-detection_2015}%
  \BibitemOpen
  \bibfield  {author} {\bibinfo {author} {\bibfnamefont {Y.}~\bibnamefont
  {Artzi}}, \bibinfo {author} {\bibfnamefont {Y.}~\bibnamefont {Twig}},\ and\
  \bibinfo {author} {\bibfnamefont {A.}~\bibnamefont {Blank}},\ }\bibfield
  {title} {\bibinfo {title} {Induction-detection electron spin resonance with
  spin sensitivity of a few tens of spins},\ }\href
  {https://doi.org/10.1063/1.4913806} {\bibfield  {journal} {\bibinfo
  {journal} {Applied Physics Letters}\ }\textbf {\bibinfo {volume} {106}},\
  \bibinfo {pages} {084104} (\bibinfo {year} {2015})}\BibitemShut {NoStop}%
\bibitem [{\citenamefont {Sidabras}\ \emph {et~al.}(2019)\citenamefont
  {Sidabras}, \citenamefont {Duan}, \citenamefont {Winkler}, \citenamefont
  {Happe}, \citenamefont {Hussein}, \citenamefont {Zouni}, \citenamefont
  {Suter}, \citenamefont {Schnegg}, \citenamefont {Lubitz},\ and\ \citenamefont
  {Reijerse}}]{sidabras_extending_2019}%
  \BibitemOpen
  \bibfield  {author} {\bibinfo {author} {\bibfnamefont {J.~W.}\ \bibnamefont
  {Sidabras}}, \bibinfo {author} {\bibfnamefont {J.}~\bibnamefont {Duan}},
  \bibinfo {author} {\bibfnamefont {M.}~\bibnamefont {Winkler}}, \bibinfo
  {author} {\bibfnamefont {T.}~\bibnamefont {Happe}}, \bibinfo {author}
  {\bibfnamefont {R.}~\bibnamefont {Hussein}}, \bibinfo {author} {\bibfnamefont
  {A.}~\bibnamefont {Zouni}}, \bibinfo {author} {\bibfnamefont
  {D.}~\bibnamefont {Suter}}, \bibinfo {author} {\bibfnamefont
  {A.}~\bibnamefont {Schnegg}}, \bibinfo {author} {\bibfnamefont
  {W.}~\bibnamefont {Lubitz}},\ and\ \bibinfo {author} {\bibfnamefont {E.~J.}\
  \bibnamefont {Reijerse}},\ }\bibfield  {title} {{\bibinfo
  {title} {Extending electron paramagnetic resonance to nanoliter volume
  protein single crystals using a self-resonant microhelix}},\ }\href
  {https://doi.org/10.1126/sciadv.aay1394} {\bibfield  {journal} {\bibinfo
  {journal} {Science Advances}\ }\textbf {\bibinfo {volume} {5}},\ \bibinfo
  {pages} {eaay1394} (\bibinfo {year} {2019})}\BibitemShut {NoStop}%
\bibitem [{\citenamefont {Wallace}\ and\ \citenamefont
  {Silsbee}(1991)}]{wallace_microstrip_1991}%
  \BibitemOpen
  \bibfield  {author} {\bibinfo {author} {\bibfnamefont {W.~J.}\ \bibnamefont
  {Wallace}}\ and\ \bibinfo {author} {\bibfnamefont {R.~H.}\ \bibnamefont
  {Silsbee}},\ }\bibfield  {title} {\bibinfo {title} {Microstrip resonators for
  electron‐spin resonance},\ }\href {https://doi.org/10.1063/1.1142418}
  {\bibfield  {journal} {\bibinfo  {journal} {Review of Scientific
  Instruments}\ }\textbf {\bibinfo {volume} {62}},\ \bibinfo {pages} {1754}
  (\bibinfo {year} {1991})},\ \bibinfo {note} {publisher: American Institute of
  Physics}\BibitemShut {NoStop}%
\bibitem [{\citenamefont {Benningshof}\ \emph {et~al.}(2013)\citenamefont
  {Benningshof}, \citenamefont {Mohebbi}, \citenamefont {Taminiau},
  \citenamefont {Miao},\ and\ \citenamefont
  {Cory}}]{benningshof_superconducting_2013}%
  \BibitemOpen
  \bibfield  {author} {\bibinfo {author} {\bibfnamefont {O.~W.~B.}\
  \bibnamefont {Benningshof}}, \bibinfo {author} {\bibfnamefont {H.~R.}\
  \bibnamefont {Mohebbi}}, \bibinfo {author} {\bibfnamefont {I.~A.~J.}\
  \bibnamefont {Taminiau}}, \bibinfo {author} {\bibfnamefont {G.~X.}\
  \bibnamefont {Miao}},\ and\ \bibinfo {author} {\bibfnamefont {D.~G.}\
  \bibnamefont {Cory}},\ }\bibfield  {title} {{\bibinfo
  {title} {Superconducting microstrip resonator for pulsed {ESR} of thin
  films}},\ }\href {https://doi.org/10.1016/j.jmr.2013.01.010} {\bibfield
  {journal} {\bibinfo  {journal} {Journal of Magnetic Resonance}\ }\textbf
  {\bibinfo {volume} {230}},\ \bibinfo {pages} {84} (\bibinfo {year}
  {2013})}\BibitemShut {NoStop}%
\bibitem [{\citenamefont {Sigillito}\ \emph {et~al.}(2014)\citenamefont
  {Sigillito}, \citenamefont {Malissa}, \citenamefont {Tyryshkin},
  \citenamefont {Riemann}, \citenamefont {Abrosimov}, \citenamefont {Becker},
  \citenamefont {Pohl}, \citenamefont {Thewalt}, \citenamefont {Itoh},
  \citenamefont {Morton}, \citenamefont {Houck}, \citenamefont {Schuster},\
  and\ \citenamefont {Lyon}}]{sigillito_fast_2014}%
  \BibitemOpen
  \bibfield  {author} {\bibinfo {author} {\bibfnamefont {A.~J.}\ \bibnamefont
  {Sigillito}}, \bibinfo {author} {\bibfnamefont {H.}~\bibnamefont {Malissa}},
  \bibinfo {author} {\bibfnamefont {A.~M.}\ \bibnamefont {Tyryshkin}}, \bibinfo
  {author} {\bibfnamefont {H.}~\bibnamefont {Riemann}}, \bibinfo {author}
  {\bibfnamefont {N.~V.}\ \bibnamefont {Abrosimov}}, \bibinfo {author}
  {\bibfnamefont {P.}~\bibnamefont {Becker}}, \bibinfo {author} {\bibfnamefont
  {H.-J.}\ \bibnamefont {Pohl}}, \bibinfo {author} {\bibfnamefont {M.~L.~W.}\
  \bibnamefont {Thewalt}}, \bibinfo {author} {\bibfnamefont {K.~M.}\
  \bibnamefont {Itoh}}, \bibinfo {author} {\bibfnamefont {J.~J.~L.}\
  \bibnamefont {Morton}}, \bibinfo {author} {\bibfnamefont {A.~A.}\
  \bibnamefont {Houck}}, \bibinfo {author} {\bibfnamefont {D.~I.}\ \bibnamefont
  {Schuster}},\ and\ \bibinfo {author} {\bibfnamefont {S.~A.}\ \bibnamefont
  {Lyon}},\ }\bibfield  {title} {\bibinfo {title} {Fast, low-power manipulation
  of spin ensembles in superconducting microresonators},\ }\href
  {https://doi.org/10.1063/1.4881613} {\bibfield  {journal} {\bibinfo
  {journal} {Applied Physics Letters}\ }\textbf {\bibinfo {volume} {104}},\
  (\bibinfo {year} {2014})}\BibitemShut {NoStop}%
\bibitem [{\citenamefont {Bienfait}\ \emph {et~al.}(2015)\citenamefont
  {Bienfait}, \citenamefont {Pla}, \citenamefont {Kubo}, \citenamefont {Stern},
  \citenamefont {Zhou}, \citenamefont {Lo}, \citenamefont {Weis}, \citenamefont
  {Schenkel}, \citenamefont {Thewalt}, \citenamefont {Vion}, \citenamefont
  {Esteve}, \citenamefont {Julsgaard}, \citenamefont {Moelmer}, \citenamefont
  {Morton},\ and\ \citenamefont {Bertet}}]{bienfait_reaching_2015}%
  \BibitemOpen
  \bibfield  {author} {\bibinfo {author} {\bibfnamefont {A.}~\bibnamefont
  {Bienfait}}, \bibinfo {author} {\bibfnamefont {J.}~\bibnamefont {Pla}},
  \bibinfo {author} {\bibfnamefont {Y.}~\bibnamefont {Kubo}}, \bibinfo {author}
  {\bibfnamefont {M.}~\bibnamefont {Stern}}, \bibinfo {author} {\bibfnamefont
  {X.}~\bibnamefont {Zhou}}, \bibinfo {author} {\bibfnamefont {C.-C.}\
  \bibnamefont {Lo}}, \bibinfo {author} {\bibfnamefont {C.}~\bibnamefont
  {Weis}}, \bibinfo {author} {\bibfnamefont {T.}~\bibnamefont {Schenkel}},
  \bibinfo {author} {\bibfnamefont {M.}~\bibnamefont {Thewalt}}, \bibinfo
  {author} {\bibfnamefont {D.}~\bibnamefont {Vion}}, \bibinfo {author}
  {\bibfnamefont {D.}~\bibnamefont {Esteve}}, \bibinfo {author} {\bibfnamefont
  {B.}~\bibnamefont {Julsgaard}}, \bibinfo {author} {\bibfnamefont
  {K.}~\bibnamefont {Moelmer}}, \bibinfo {author} {\bibfnamefont
  {J.}~\bibnamefont {Morton}},\ and\ \bibinfo {author} {\bibfnamefont
  {P.}~\bibnamefont {Bertet}},\ }\bibfield  {title} {\bibinfo {title} {Reaching
  the quantum limit of sensitivity in electron spin resonance},\ }\href
  {https://doi.org/10.1038/nnano.2015.282} {\bibfield  {journal} {\bibinfo
  {journal} {Nature Nanotechnology}\ }\textbf {\bibinfo {volume} {11}},\
  \bibinfo {pages} {253 } (\bibinfo {year} {2015})}\BibitemShut {NoStop}%
\bibitem [{\citenamefont {Eichler}\ \emph {et~al.}(2017)\citenamefont
  {Eichler}, \citenamefont {Sigillito}, \citenamefont {Lyon},\ and\
  \citenamefont {Petta}}]{eichler_electron_2017}%
  \BibitemOpen
  \bibfield  {author} {\bibinfo {author} {\bibfnamefont {C.}~\bibnamefont
  {Eichler}}, \bibinfo {author} {\bibfnamefont {A.~J.}\ \bibnamefont
  {Sigillito}}, \bibinfo {author} {\bibfnamefont {S.~A.}\ \bibnamefont
  {Lyon}},\ and\ \bibinfo {author} {\bibfnamefont {J.~R.}\ \bibnamefont
  {Petta}},\ }\bibfield  {title} {\bibinfo {title} {Electron {Spin} {Resonance}
  at the {Level} of \$10{\textasciicircum}4\$ {Spins} {Using} {Low} {Impedance}
  {Superconducting} {Resonators}},\ }\href
  {https://doi.org/10.1103/PhysRevLett.118.037701} {\bibfield  {journal}
  {\bibinfo  {journal} {Phys. Rev. Lett.}\ }\textbf {\bibinfo {volume} {118}},\
  \bibinfo {pages} {037701} (\bibinfo {year} {2017})}\BibitemShut {NoStop}%
\bibitem [{\citenamefont {Probst}\ \emph {et~al.}(2017)\citenamefont {Probst},
  \citenamefont {Bienfait}, \citenamefont {Campagne-Ibarcq}, \citenamefont
  {Pla}, \citenamefont {Albanese}, \citenamefont {Barbosa}, \citenamefont
  {Schenkel}, \citenamefont {Vion}, \citenamefont {Esteve}, \citenamefont
  {Moelmer}, \citenamefont {Morton}, \citenamefont {Heeres},\ and\
  \citenamefont {Bertet}}]{probst_inductive-detection_2017}%
  \BibitemOpen
  \bibfield  {author} {\bibinfo {author} {\bibfnamefont {S.}~\bibnamefont
  {Probst}}, \bibinfo {author} {\bibfnamefont {A.}~\bibnamefont {Bienfait}},
  \bibinfo {author} {\bibfnamefont {P.}~\bibnamefont {Campagne-Ibarcq}},
  \bibinfo {author} {\bibfnamefont {J.~J.}\ \bibnamefont {Pla}}, \bibinfo
  {author} {\bibfnamefont {B.}~\bibnamefont {Albanese}}, \bibinfo {author}
  {\bibfnamefont {J.~F. D.~S.}\ \bibnamefont {Barbosa}}, \bibinfo {author}
  {\bibfnamefont {T.}~\bibnamefont {Schenkel}}, \bibinfo {author}
  {\bibfnamefont {D.}~\bibnamefont {Vion}}, \bibinfo {author} {\bibfnamefont
  {D.}~\bibnamefont {Esteve}}, \bibinfo {author} {\bibfnamefont
  {K.}~\bibnamefont {Moelmer}}, \bibinfo {author} {\bibfnamefont {J.~J.~L.}\
  \bibnamefont {Morton}}, \bibinfo {author} {\bibfnamefont {R.}~\bibnamefont
  {Heeres}},\ and\ \bibinfo {author} {\bibfnamefont {P.}~\bibnamefont
  {Bertet}},\ }\bibfield  {title} {\bibinfo {title} {Inductive-detection
  electron-spin resonance spectroscopy with 65
  spins/{Hz}{\textasciicircum}(1/2) sensitivity},\ }\href
  {https://doi.org/10.1063/1.5002540} {\bibfield  {journal} {\bibinfo
  {journal} {Applied Physics Letters}\ }\textbf {\bibinfo {volume} {111}},\
  \bibinfo {pages} {202604} (\bibinfo {year} {2017})}\BibitemShut {NoStop}%
\bibitem [{\citenamefont {Ranjan}\ \emph
  {et~al.}(2020{\natexlab{a}})\citenamefont {Ranjan}, \citenamefont {Probst},
  \citenamefont {Albanese}, \citenamefont {Schenkel}, \citenamefont {Vion},
  \citenamefont {Esteve}, \citenamefont {Morton},\ and\ \citenamefont
  {Bertet}}]{ranjan_electron_2020}%
  \BibitemOpen
  \bibfield  {author} {\bibinfo {author} {\bibfnamefont {V.}~\bibnamefont
  {Ranjan}}, \bibinfo {author} {\bibfnamefont {S.}~\bibnamefont {Probst}},
  \bibinfo {author} {\bibfnamefont {B.}~\bibnamefont {Albanese}}, \bibinfo
  {author} {\bibfnamefont {T.}~\bibnamefont {Schenkel}}, \bibinfo {author}
  {\bibfnamefont {D.}~\bibnamefont {Vion}}, \bibinfo {author} {\bibfnamefont
  {D.}~\bibnamefont {Esteve}}, \bibinfo {author} {\bibfnamefont {J.~J.~L.}\
  \bibnamefont {Morton}},\ and\ \bibinfo {author} {\bibfnamefont
  {P.}~\bibnamefont {Bertet}},\ }\bibfield  {title} {\bibinfo {title} {Electron
  spin resonance spectroscopy with femtoliter detection volume},\ }\href
  {https://doi.org/10.1063/5.0004322} {\bibfield  {journal} {\bibinfo
  {journal} {Applied Physics Letters}\ }\textbf {\bibinfo {volume} {116}},\
  \bibinfo {pages} {184002} (\bibinfo {year} {2020}{\natexlab{a}})},\ \bibinfo
  {note} {publisher: American Institute of Physics}\BibitemShut {NoStop}%
\bibitem [{\citenamefont {Bienfait}\ \emph {et~al.}(2016)\citenamefont
  {Bienfait}, \citenamefont {Pla}, \citenamefont {Kubo}, \citenamefont {Zhou},
  \citenamefont {Stern}, \citenamefont {Lo}, \citenamefont {Weis},
  \citenamefont {Schenkel}, \citenamefont {Vion}, \citenamefont {Esteve},
  \citenamefont {Morton},\ and\ \citenamefont
  {Bertet}}]{bienfait_controlling_2016}%
  \BibitemOpen
  \bibfield  {author} {\bibinfo {author} {\bibfnamefont {A.}~\bibnamefont
  {Bienfait}}, \bibinfo {author} {\bibfnamefont {J.}~\bibnamefont {Pla}},
  \bibinfo {author} {\bibfnamefont {Y.}~\bibnamefont {Kubo}}, \bibinfo {author}
  {\bibfnamefont {X.}~\bibnamefont {Zhou}}, \bibinfo {author} {\bibfnamefont
  {M.}~\bibnamefont {Stern}}, \bibinfo {author} {\bibfnamefont {C.-C.}\
  \bibnamefont {Lo}}, \bibinfo {author} {\bibfnamefont {C.}~\bibnamefont
  {Weis}}, \bibinfo {author} {\bibfnamefont {T.}~\bibnamefont {Schenkel}},
  \bibinfo {author} {\bibfnamefont {D.}~\bibnamefont {Vion}}, \bibinfo {author}
  {\bibfnamefont {D.}~\bibnamefont {Esteve}}, \bibinfo {author} {\bibfnamefont
  {J.}~\bibnamefont {Morton}},\ and\ \bibinfo {author} {\bibfnamefont
  {P.}~\bibnamefont {Bertet}},\ }\bibfield  {title} {\bibinfo {title}
  {Controlling {Spin} {Relaxation} with a {Cavity}},\ }\href
  {https://doi.org/doi:10.1038/nature16944} {\bibfield  {journal} {\bibinfo
  {journal} {Nature}\ }\textbf {\bibinfo {volume} {531}},\ \bibinfo {pages} {74
  } (\bibinfo {year} {2016})}\BibitemShut {NoStop}%
\bibitem [{\citenamefont {Ranjan}\ \emph
  {et~al.}(2020{\natexlab{b}})\citenamefont {Ranjan}, \citenamefont {Probst},
  \citenamefont {Albanese}, \citenamefont {Doll}, \citenamefont {Jacquot},
  \citenamefont {Flurin}, \citenamefont {Heeres}, \citenamefont {Vion},
  \citenamefont {Esteve}, \citenamefont {Morton},\ and\ \citenamefont
  {Bertet}}]{ranjan_pulsed_2020}%
  \BibitemOpen
  \bibfield  {author} {\bibinfo {author} {\bibfnamefont {V.}~\bibnamefont
  {Ranjan}}, \bibinfo {author} {\bibfnamefont {S.}~\bibnamefont {Probst}},
  \bibinfo {author} {\bibfnamefont {B.}~\bibnamefont {Albanese}}, \bibinfo
  {author} {\bibfnamefont {A.}~\bibnamefont {Doll}}, \bibinfo {author}
  {\bibfnamefont {O.}~\bibnamefont {Jacquot}}, \bibinfo {author} {\bibfnamefont
  {E.}~\bibnamefont {Flurin}}, \bibinfo {author} {\bibfnamefont
  {R.}~\bibnamefont {Heeres}}, \bibinfo {author} {\bibfnamefont
  {D.}~\bibnamefont {Vion}}, \bibinfo {author} {\bibfnamefont {D.}~\bibnamefont
  {Esteve}}, \bibinfo {author} {\bibfnamefont {J.~J.~L.}\ \bibnamefont
  {Morton}},\ and\ \bibinfo {author} {\bibfnamefont {P.}~\bibnamefont
  {Bertet}},\ }\bibfield  {title} {{\bibinfo {title}
  {Pulsed electron spin resonance spectroscopy in the {Purcell} regime}},\
  }\href {https://doi.org/10.1016/j.jmr.2019.106662} {\bibfield  {journal}
  {\bibinfo  {journal} {Journal of Magnetic Resonance}\ }\textbf {\bibinfo
  {volume} {310}},\ \bibinfo {pages} {106662} (\bibinfo {year}
  {2020}{\natexlab{b}})}\BibitemShut {NoStop}%
\bibitem [{\citenamefont {Sigillito}\ \emph {et~al.}(2017)\citenamefont
  {Sigillito}, \citenamefont {Tyryshkin}, \citenamefont {Schenkel},
  \citenamefont {Houck},\ and\ \citenamefont
  {Lyon}}]{sigillito_electrically_2017}%
  \BibitemOpen
  \bibfield  {author} {\bibinfo {author} {\bibfnamefont {A.~J.}\ \bibnamefont
  {Sigillito}}, \bibinfo {author} {\bibfnamefont {A.~M.}\ \bibnamefont
  {Tyryshkin}}, \bibinfo {author} {\bibfnamefont {T.}~\bibnamefont {Schenkel}},
  \bibinfo {author} {\bibfnamefont {A.~A.}\ \bibnamefont {Houck}},\ and\
  \bibinfo {author} {\bibfnamefont {S.~A.}\ \bibnamefont {Lyon}},\ }\bibfield
  {title} {\bibinfo {title} {Electrically driving nuclear spin qubits with
  microwave photonic bangap resonators},\ }\href@noop {} {\  (\bibinfo {year}
  {2017})}\BibitemShut {NoStop}%
\bibitem [{\citenamefont {Probst}\ \emph {et~al.}(2015)\citenamefont {Probst},
  \citenamefont {Rotzinger}, \citenamefont {Ustinov},\ and\ \citenamefont
  {Bushev}}]{probst_microwave_2015}%
  \BibitemOpen
  \bibfield  {author} {\bibinfo {author} {\bibfnamefont {S.}~\bibnamefont
  {Probst}}, \bibinfo {author} {\bibfnamefont {H.}~\bibnamefont {Rotzinger}},
  \bibinfo {author} {\bibfnamefont {A.~V.}\ \bibnamefont {Ustinov}},\ and\
  \bibinfo {author} {\bibfnamefont {P.~A.}\ \bibnamefont {Bushev}},\ }\bibfield
   {title} {\bibinfo {title} {Microwave multimode memory with an erbium spin
  ensemble},\ }\href {https://doi.org/10.1103/PhysRevB.92.014421} {\bibfield
  {journal} {\bibinfo  {journal} {Physical Review B}\ }\textbf {\bibinfo
  {volume} {92}},\ \bibinfo {pages} {014421} (\bibinfo {year}
  {2015})}\BibitemShut {NoStop}%
\bibitem [{\citenamefont {Kasumaj}\ and\ \citenamefont
  {Stoll}(2008)}]{kasumaj_5-_2008}%
  \BibitemOpen
  \bibfield  {author} {\bibinfo {author} {\bibfnamefont {B.}~\bibnamefont
  {Kasumaj}}\ and\ \bibinfo {author} {\bibfnamefont {S.}~\bibnamefont
  {Stoll}},\ }\bibfield  {title} {\bibinfo {title} {5- and 6-pulse electron
  spin echo envelope modulation ({ESEEM}) of multi-nuclear spin systems},\
  }\href {https://doi.org/10.1016/j.jmr.2007.11.001} {\bibfield  {journal}
  {\bibinfo  {journal} {Journal of Magnetic Resonance}\ }\textbf {\bibinfo
  {volume} {190}},\ \bibinfo {pages} {233} (\bibinfo {year}
  {2008})}\BibitemShut {NoStop}%
\bibitem [{\citenamefont {Witzel}\ \emph {et~al.}(2007)\citenamefont {Witzel},
  \citenamefont {Hu},\ and\ \citenamefont
  {Das~Sarma}}]{witzel_decoherence_2007}%
  \BibitemOpen
  \bibfield  {author} {\bibinfo {author} {\bibfnamefont {W.~M.}\ \bibnamefont
  {Witzel}}, \bibinfo {author} {\bibfnamefont {X.}~\bibnamefont {Hu}},\ and\
  \bibinfo {author} {\bibfnamefont {S.}~\bibnamefont {Das~Sarma}},\ }\bibfield
  {title} {\bibinfo {title} {Decoherence induced by anisotropic hyperfine
  interaction in {Si} spin qubits},\ }\href
  {https://doi.org/10.1103/PhysRevB.76.035212} {\bibfield  {journal} {\bibinfo
  {journal} {Physical Review B}\ }\textbf {\bibinfo {volume} {76}},\ \bibinfo
  {pages} {035212} (\bibinfo {year} {2007})}\BibitemShut {NoStop}%
\bibitem [{\citenamefont {Abe}\ \emph {et~al.}(2010)\citenamefont {Abe},
  \citenamefont {Tyryshkin}, \citenamefont {Tojo}, \citenamefont {Morton},
  \citenamefont {Witzel}, \citenamefont {Fujimoto}, \citenamefont {Ager},
  \citenamefont {Haller}, \citenamefont {Isoya}, \citenamefont {Lyon},
  \citenamefont {Thewalt},\ and\ \citenamefont {Itoh}}]{abe_electron_2010}%
  \BibitemOpen
  \bibfield  {author} {\bibinfo {author} {\bibfnamefont {E.}~\bibnamefont
  {Abe}}, \bibinfo {author} {\bibfnamefont {A.~M.}\ \bibnamefont {Tyryshkin}},
  \bibinfo {author} {\bibfnamefont {S.}~\bibnamefont {Tojo}}, \bibinfo {author}
  {\bibfnamefont {J.~J.~L.}\ \bibnamefont {Morton}}, \bibinfo {author}
  {\bibfnamefont {W.~M.}\ \bibnamefont {Witzel}}, \bibinfo {author}
  {\bibfnamefont {A.}~\bibnamefont {Fujimoto}}, \bibinfo {author}
  {\bibfnamefont {J.~W.}\ \bibnamefont {Ager}}, \bibinfo {author}
  {\bibfnamefont {E.~E.}\ \bibnamefont {Haller}}, \bibinfo {author}
  {\bibfnamefont {J.}~\bibnamefont {Isoya}}, \bibinfo {author} {\bibfnamefont
  {S.~A.}\ \bibnamefont {Lyon}}, \bibinfo {author} {\bibfnamefont {M.~L.~W.}\
  \bibnamefont {Thewalt}},\ and\ \bibinfo {author} {\bibfnamefont {K.~M.}\
  \bibnamefont {Itoh}},\ }\bibfield  {title} {\bibinfo {title} {Electron spin
  coherence of phosphorus donors in silicon: {Effect} of environmental
  nuclei},\ }\href {https://doi.org/10.1103/PhysRevB.82.121201} {\bibfield
  {journal} {\bibinfo  {journal} {Physical Review B}\ }\textbf {\bibinfo
  {volume} {82}},\ \bibinfo {pages} {121201} (\bibinfo {year}
  {2010})}\BibitemShut {NoStop}%
\bibitem [{\citenamefont {Mims}\ \emph {et~al.}(1961)\citenamefont {Mims},
  \citenamefont {Nassau},\ and\ \citenamefont {McGee}}]{mims_spectral_1961}%
  \BibitemOpen
  \bibfield  {author} {\bibinfo {author} {\bibfnamefont {W.~B.}\ \bibnamefont
  {Mims}}, \bibinfo {author} {\bibfnamefont {K.}~\bibnamefont {Nassau}},\ and\
  \bibinfo {author} {\bibfnamefont {J.~D.}\ \bibnamefont {McGee}},\ }\bibfield
  {title} {\bibinfo {title} {Spectral {Diffusion} in {Electron} {Resonance}
  {Lines}},\ }\href {https://doi.org/10.1103/PhysRev.123.2059} {\bibfield
  {journal} {\bibinfo  {journal} {Physical Review}\ }\textbf {\bibinfo {volume}
  {123}},\ \bibinfo {pages} {2059} (\bibinfo {year} {1961})}\BibitemShut
  {NoStop}%
\bibitem [{\citenamefont {Rowan}\ \emph {et~al.}(1965)\citenamefont {Rowan},
  \citenamefont {Hahn},\ and\ \citenamefont
  {Mims}}]{rowan_electron-spin-echo_1965}%
  \BibitemOpen
  \bibfield  {author} {\bibinfo {author} {\bibfnamefont {L.~G.}\ \bibnamefont
  {Rowan}}, \bibinfo {author} {\bibfnamefont {E.~L.}\ \bibnamefont {Hahn}},\
  and\ \bibinfo {author} {\bibfnamefont {W.~B.}\ \bibnamefont {Mims}},\
  }\bibfield  {title} {\bibinfo {title} {Electron-{Spin}-{Echo} {Envelope}
  {Modulation}},\ }\href {https://doi.org/10.1103/physrev.137.a61} {\bibfield
  {journal} {\bibinfo  {journal} {Physical Review}\ }\textbf {\bibinfo {volume}
  {137}},\ \bibinfo {pages} {A61} (\bibinfo {year} {1965})}\BibitemShut
  {NoStop}%
\bibitem [{\citenamefont {Mims}\ \emph {et~al.}(1990)\citenamefont {Mims},
  \citenamefont {Davis},\ and\ \citenamefont {Peisach}}]{mims_exchange_1990}%
  \BibitemOpen
  \bibfield  {author} {\bibinfo {author} {\bibfnamefont {W.~B.}\ \bibnamefont
  {Mims}}, \bibinfo {author} {\bibfnamefont {J.~L.}\ \bibnamefont {Davis}},\
  and\ \bibinfo {author} {\bibfnamefont {J.}~\bibnamefont {Peisach}},\
  }\bibfield  {title} {\bibinfo {title} {The exchange of hydrogen ions and of
  water molecules near the active site of cytochrome c},\ }\href
  {https://doi.org/10.1016/0022-2364(90)90260-G} {\bibfield  {journal}
  {\bibinfo  {journal} {Journal of Magnetic Resonance (1969)}\ }\textbf
  {\bibinfo {volume} {86}},\ \bibinfo {pages} {273} (\bibinfo {year}
  {1990})}\BibitemShut {NoStop}%
\bibitem [{\citenamefont {Childress}\ \emph {et~al.}(2006)\citenamefont
  {Childress}, \citenamefont {Dutt}, \citenamefont {Taylor}, \citenamefont
  {Zibrov}, \citenamefont {Jelezko}, \citenamefont {Wrachtrup}, \citenamefont
  {Hemmer},\ and\ \citenamefont {Lukin}}]{childress_coherent_2006}%
  \BibitemOpen
  \bibfield  {author} {\bibinfo {author} {\bibfnamefont {L.}~\bibnamefont
  {Childress}}, \bibinfo {author} {\bibfnamefont {M.~V.~G.}\ \bibnamefont
  {Dutt}}, \bibinfo {author} {\bibfnamefont {J.~M.}\ \bibnamefont {Taylor}},
  \bibinfo {author} {\bibfnamefont {A.~S.}\ \bibnamefont {Zibrov}}, \bibinfo
  {author} {\bibfnamefont {F.}~\bibnamefont {Jelezko}}, \bibinfo {author}
  {\bibfnamefont {J.}~\bibnamefont {Wrachtrup}}, \bibinfo {author}
  {\bibfnamefont {P.~R.}\ \bibnamefont {Hemmer}},\ and\ \bibinfo {author}
  {\bibfnamefont {M.~D.}\ \bibnamefont {Lukin}},\ }\bibfield  {title}
  {{\bibinfo {title} {Coherent {Dynamics} of {Coupled}
  {Electron} and {Nuclear} {Spin} {Qubits} in {Diamond}}},\ }\href
  {https://doi.org/10.1126/science.1131871} {\bibfield  {journal} {\bibinfo
  {journal} {Science}\ }\textbf {\bibinfo {volume} {314}},\ \bibinfo {pages}
  {281} (\bibinfo {year} {2006})}\BibitemShut {NoStop}%
\bibitem [{\citenamefont {Mims}(1972)}]{mims_envelope_1972}%
  \BibitemOpen
  \bibfield  {author} {\bibinfo {author} {\bibfnamefont {W.~B.}\ \bibnamefont
  {Mims}},\ }\bibfield  {title} {\bibinfo {title} {Envelope {Modulation} in
  {Spin}-{Echo} {Experiments}},\ }\href
  {https://doi.org/10.1103/physrevb.5.2409} {\bibfield  {journal} {\bibinfo
  {journal} {Physical Review B}\ }\textbf {\bibinfo {volume} {5}},\ \bibinfo
  {pages} {2409} (\bibinfo {year} {1972})}\BibitemShut {NoStop}%
\bibitem [{\citenamefont {Abragam}\ and\ \citenamefont
  {Bleaney}(2012)}]{abragam_electron_2012}%
  \BibitemOpen
  \bibfield  {author} {\bibinfo {author} {\bibfnamefont {A.}~\bibnamefont
  {Abragam}}\ and\ \bibinfo {author} {\bibfnamefont {B.}~\bibnamefont
  {Bleaney}},\ }\href@noop {} {{\emph {\bibinfo {title}
  {Electron {Paramagnetic} {Resonance} of {Transition} {Ions}}}}}\ (\bibinfo
  {publisher} {OUP Oxford},\ \bibinfo {year} {2012})\ \bibinfo {note}
  {google-Books-ID: ASNoAgAAQBAJ}\BibitemShut {NoStop}%
\bibitem [{\citenamefont {Antipin}\ \emph {et~al.}(1968)\citenamefont
  {Antipin}, \citenamefont {Katyshev}, \citenamefont {Kurkin},\ and\
  \citenamefont {Shekun}}]{antipin_a._paramagnetic_1968}%
  \BibitemOpen
  \bibfield  {author} {\bibinfo {author} {\bibfnamefont {A.}~\bibnamefont
  {Antipin}}, \bibinfo {author} {\bibfnamefont {A.}~\bibnamefont {Katyshev}},
  \bibinfo {author} {\bibfnamefont {I.}~\bibnamefont {Kurkin}},\ and\ \bibinfo
  {author} {\bibfnamefont {L.}~\bibnamefont {Shekun}},\ }\bibfield  {title}
  {\bibinfo {title} {Paramagnetic resonance and spin-lattice relaxation of
  {Er3}+ and {Tb3}+ ions in {CaWO4} crystal lattice},\ }\href@noop {}
  {\bibfield  {journal} {\bibinfo  {journal} {Sov. Phys. Solid State}\ }\textbf
  {\bibinfo {volume} {10}},\ \bibinfo {pages} {468} (\bibinfo {year}
  {1968})}\BibitemShut {NoStop}%
\bibitem [{\citenamefont {Kohn}\ and\ \citenamefont
  {Luttinger}(1955)}]{kohn_theory_1955}%
  \BibitemOpen
  \bibfield  {author} {\bibinfo {author} {\bibfnamefont {W.}~\bibnamefont
  {Kohn}}\ and\ \bibinfo {author} {\bibfnamefont {J.~M.}\ \bibnamefont
  {Luttinger}},\ }\bibfield  {title} {\bibinfo {title} {Theory of {Donor}
  {States} in {Silicon}},\ }\href {https://doi.org/10.1103/PhysRev.98.915}
  {\bibfield  {journal} {\bibinfo  {journal} {Physical Review}\ }\textbf
  {\bibinfo {volume} {98}},\ \bibinfo {pages} {915} (\bibinfo {year}
  {1955})}\BibitemShut {NoStop}%
\bibitem [{\citenamefont {Feher}(1959)}]{feher_electron_1959}%
  \BibitemOpen
  \bibfield  {author} {\bibinfo {author} {\bibfnamefont {G.}~\bibnamefont
  {Feher}},\ }\bibfield  {title} {\bibinfo {title} {Electron {Spin} {Resonance}
  {Experiments} on {Donors} in {Silicon}. {I}. {Electronic} {Structure} of
  {Donors} by the {Electron} {Nuclear} {Double} {Resonance} {Technique}},\
  }\href {https://doi.org/10.1103/PhysRev.114.1219} {\bibfield  {journal}
  {\bibinfo  {journal} {Phys. Rev.}\ }\textbf {\bibinfo {volume} {114}},\
  \bibinfo {pages} {1219} (\bibinfo {year} {1959})}\BibitemShut {NoStop}%
\bibitem [{\citenamefont {Mohammady}\ \emph {et~al.}(2010)\citenamefont
  {Mohammady}, \citenamefont {Morley},\ and\ \citenamefont
  {Monteiro}}]{mohammady_bismuth_2010}%
  \BibitemOpen
  \bibfield  {author} {\bibinfo {author} {\bibfnamefont {M.~H.}\ \bibnamefont
  {Mohammady}}, \bibinfo {author} {\bibfnamefont {G.~W.}\ \bibnamefont
  {Morley}},\ and\ \bibinfo {author} {\bibfnamefont {T.~S.}\ \bibnamefont
  {Monteiro}},\ }\bibfield  {title} {\bibinfo {title} {Bismuth {Qubits} in
  {Silicon}: {The} {Role} of {EPR} {Cancellation} {Resonances}},\ }\href
  {https://doi.org/10.1103/physrevlett.105.067602} {\bibfield  {journal}
  {\bibinfo  {journal} {Physical Review Letters}\ }\textbf {\bibinfo {volume}
  {105}},\ \bibinfo {pages} {067602} (\bibinfo {year} {2010})}\BibitemShut
  {NoStop}%
\bibitem [{\citenamefont {Hale}\ and\ \citenamefont
  {Mieher}(1969)}]{hale_shallow_1969}%
  \BibitemOpen
  \bibfield  {author} {\bibinfo {author} {\bibfnamefont {E.~B.}\ \bibnamefont
  {Hale}}\ and\ \bibinfo {author} {\bibfnamefont {R.~L.}\ \bibnamefont
  {Mieher}},\ }\bibfield  {title} {\bibinfo {title} {Shallow {Donor}
  {Electrons} in {Silicon}. {II}. {Considerations} {Regarding} the {Fermi}
  {Contact} {Interactions}},\ }\href {https://doi.org/10.1103/physrev.184.751}
  {\bibfield  {journal} {\bibinfo  {journal} {Physical Review}\ }\textbf
  {\bibinfo {volume} {184}},\ \bibinfo {pages} {751} (\bibinfo {year}
  {1969})}\BibitemShut {NoStop}%
\bibitem [{\citenamefont {Zhou}\ \emph {et~al.}(2014)\citenamefont {Zhou},
  \citenamefont {Schmitt}, \citenamefont {Bertet}, \citenamefont {Vion},
  \citenamefont {Wustmann}, \citenamefont {Shumeiko},\ and\ \citenamefont
  {Esteve}}]{zhou_high-gain_2014}%
  \BibitemOpen
  \bibfield  {author} {\bibinfo {author} {\bibfnamefont {X.}~\bibnamefont
  {Zhou}}, \bibinfo {author} {\bibfnamefont {V.}~\bibnamefont {Schmitt}},
  \bibinfo {author} {\bibfnamefont {P.}~\bibnamefont {Bertet}}, \bibinfo
  {author} {\bibfnamefont {D.}~\bibnamefont {Vion}}, \bibinfo {author}
  {\bibfnamefont {W.}~\bibnamefont {Wustmann}}, \bibinfo {author}
  {\bibfnamefont {V.}~\bibnamefont {Shumeiko}},\ and\ \bibinfo {author}
  {\bibfnamefont {D.}~\bibnamefont {Esteve}},\ }\bibfield  {title} {\bibinfo
  {title} {High-gain weakly nonlinear flux-modulated {Josephson} parametric
  amplifier using a {SQUID} array},\ }\href
  {https://doi.org/10.1103/PhysRevB.89.214517} {\bibfield  {journal} {\bibinfo
  {journal} {Phys. Rev. B}\ }\textbf {\bibinfo {volume} {89}},\ \bibinfo
  {pages} {214517} (\bibinfo {year} {2014})}\BibitemShut {NoStop}%
\bibitem [{\citenamefont {Macklin}\ \emph {et~al.}(2015)\citenamefont
  {Macklin}, \citenamefont {O’Brien}, \citenamefont {Hover}, \citenamefont
  {Schwartz}, \citenamefont {Bolkhovsky}, \citenamefont {Zhang}, \citenamefont
  {Oliver},\ and\ \citenamefont {Siddiqi}}]{macklin_nearquantum-limited_2015}%
  \BibitemOpen
  \bibfield  {author} {\bibinfo {author} {\bibfnamefont {C.}~\bibnamefont
  {Macklin}}, \bibinfo {author} {\bibfnamefont {K.}~\bibnamefont {O’Brien}},
  \bibinfo {author} {\bibfnamefont {D.}~\bibnamefont {Hover}}, \bibinfo
  {author} {\bibfnamefont {M.~E.}\ \bibnamefont {Schwartz}}, \bibinfo {author}
  {\bibfnamefont {V.}~\bibnamefont {Bolkhovsky}}, \bibinfo {author}
  {\bibfnamefont {X.}~\bibnamefont {Zhang}}, \bibinfo {author} {\bibfnamefont
  {W.~D.}\ \bibnamefont {Oliver}},\ and\ \bibinfo {author} {\bibfnamefont
  {I.}~\bibnamefont {Siddiqi}},\ }\bibfield  {title} {{\bibinfo {title} {A near–quantum-limited {Josephson} traveling-wave
  parametric amplifier}},\ }\href {https://doi.org/10.1126/science.aaa8525}
  {\bibfield  {journal} {\bibinfo  {journal} {Science}\ }\textbf {\bibinfo
  {volume} {350}},\ \bibinfo {pages} {307} (\bibinfo {year}
  {2015})}\BibitemShut {NoStop}%
\bibitem [{\citenamefont {Wolfowicz}\ \emph {et~al.}(2013)\citenamefont
  {Wolfowicz}, \citenamefont {Tyryshkin}, \citenamefont {George}, \citenamefont
  {Riemann}, \citenamefont {Abrosimov}, \citenamefont {Becker}, \citenamefont
  {Pohl}, \citenamefont {Thewalt}, \citenamefont {Lyon},\ and\ \citenamefont
  {Morton}}]{wolfowicz_atomic_2013}%
  \BibitemOpen
  \bibfield  {author} {\bibinfo {author} {\bibfnamefont {G.}~\bibnamefont
  {Wolfowicz}}, \bibinfo {author} {\bibfnamefont {A.~M.}\ \bibnamefont
  {Tyryshkin}}, \bibinfo {author} {\bibfnamefont {R.~E.}\ \bibnamefont
  {George}}, \bibinfo {author} {\bibfnamefont {H.}~\bibnamefont {Riemann}},
  \bibinfo {author} {\bibfnamefont {N.~V.}\ \bibnamefont {Abrosimov}}, \bibinfo
  {author} {\bibfnamefont {P.}~\bibnamefont {Becker}}, \bibinfo {author}
  {\bibfnamefont {H.-J.}\ \bibnamefont {Pohl}}, \bibinfo {author}
  {\bibfnamefont {M.~L.~W.}\ \bibnamefont {Thewalt}}, \bibinfo {author}
  {\bibfnamefont {S.~a.}\ \bibnamefont {Lyon}},\ and\ \bibinfo {author}
  {\bibfnamefont {J.~J.~L.}\ \bibnamefont {Morton}},\ }\bibfield  {title}
  {\bibinfo {title} {Atomic clock transitions in silicon-based spin qubits.},\
  }\href {https://doi.org/10.1038/nnano.2013.117} {\bibfield  {journal}
  {\bibinfo  {journal} {Nature Nanotechnology}\ }\textbf {\bibinfo {volume}
  {8}},\ \bibinfo {pages} {561} (\bibinfo {year} {2013})}\BibitemShut {NoStop}%
\bibitem [{\citenamefont {Probst}\ \emph {et~al.}(2019)\citenamefont {Probst},
  \citenamefont {Ranjan}, \citenamefont {Ansel}, \citenamefont {Heeres},
  \citenamefont {Albanese}, \citenamefont {Albertinale}, \citenamefont {Vion},
  \citenamefont {Esteve}, \citenamefont {Glaser}, \citenamefont {Sugny},\ and\
  \citenamefont {Bertet}}]{probst_shaped_2019}%
  \BibitemOpen
  \bibfield  {author} {\bibinfo {author} {\bibfnamefont {S.}~\bibnamefont
  {Probst}}, \bibinfo {author} {\bibfnamefont {V.}~\bibnamefont {Ranjan}},
  \bibinfo {author} {\bibfnamefont {Q.}~\bibnamefont {Ansel}}, \bibinfo
  {author} {\bibfnamefont {R.}~\bibnamefont {Heeres}}, \bibinfo {author}
  {\bibfnamefont {B.}~\bibnamefont {Albanese}}, \bibinfo {author}
  {\bibfnamefont {E.}~\bibnamefont {Albertinale}}, \bibinfo {author}
  {\bibfnamefont {D.}~\bibnamefont {Vion}}, \bibinfo {author} {\bibfnamefont
  {D.}~\bibnamefont {Esteve}}, \bibinfo {author} {\bibfnamefont {S.~J.}\
  \bibnamefont {Glaser}}, \bibinfo {author} {\bibfnamefont {D.}~\bibnamefont
  {Sugny}},\ and\ \bibinfo {author} {\bibfnamefont {P.}~\bibnamefont
  {Bertet}},\ }\bibfield  {title} {\bibinfo {title} {Shaped pulses for
  transient compensation in quantum-limited electron spin resonance
  spectroscopy},\ }\href {https://doi.org/10.1016/j.jmr.2019.04.008} {\bibfield
   {journal} {\bibinfo  {journal} {Journal of Magnetic Resonance}\ }\textbf
  {\bibinfo {volume} {303}},\ \bibinfo {pages} {42} (\bibinfo {year}
  {2019})}\BibitemShut {NoStop}%
\bibitem [{\citenamefont {Probst}\ \emph {et~al.}(2020)\citenamefont {Probst},
  \citenamefont {Zhang}, \citenamefont {Ran\v{c}i\'{c}}, \citenamefont
  {Ranjan}, \citenamefont {Le~Dantec}, \citenamefont {Zhang}, \citenamefont
  {Albanese}, \citenamefont {Doll}, \citenamefont {Liu}, \citenamefont
  {Morton}, \citenamefont {Chaneli{\`e}re}, \citenamefont {Goldner},
  \citenamefont {Vion}, \citenamefont {Esteve},\ and\ \citenamefont
  {Bertet}}]{data_repo}%
  \BibitemOpen
  \bibfield  {author} {\bibinfo {author} {\bibfnamefont {S.}~\bibnamefont
  {Probst}}, \bibinfo {author} {\bibfnamefont {G.}~\bibnamefont {Zhang}},
  \bibinfo {author} {\bibfnamefont {M.}~\bibnamefont {Ran\v{c}i\'{c}}},
  \bibinfo {author} {\bibfnamefont {V.}~\bibnamefont {Ranjan}}, \bibinfo
  {author} {\bibfnamefont {M.}~\bibnamefont {Le~Dantec}}, \bibinfo {author}
  {\bibfnamefont {Z.}~\bibnamefont {Zhang}}, \bibinfo {author} {\bibfnamefont
  {B.}~\bibnamefont {Albanese}}, \bibinfo {author} {\bibfnamefont
  {A.}~\bibnamefont {Doll}}, \bibinfo {author} {\bibfnamefont {R.~B.}\
  \bibnamefont {Liu}}, \bibinfo {author} {\bibfnamefont {J.}~\bibnamefont
  {Morton}}, \bibinfo {author} {\bibfnamefont {T.}~\bibnamefont
  {Chaneli{\`e}re}}, \bibinfo {author} {\bibfnamefont {P.}~\bibnamefont
  {Goldner}}, \bibinfo {author} {\bibfnamefont {D.}~\bibnamefont {Vion}},
  \bibinfo {author} {\bibfnamefont {D.}~\bibnamefont {Esteve}},\ and\ \bibinfo
  {author} {\bibfnamefont {P.}~\bibnamefont {Bertet}},\ }\bibfield  {title}
  {{\bibinfo {title} {Replication data for: Hyperfine
  spectroscopy in a quantum-limited spectrometer}},\ }\bibfield  {journal}
  {\bibinfo  {journal} {Harvard Dataverse}\ }\href
  {https://doi.org/10.7910/DVN/ZJ2EEX} {10.7910/DVN/ZJ2EEX} (\bibinfo {year}
  {2020})\BibitemShut {NoStop}%
\bibitem [{\citenamefont {Day}\ \emph {et~al.}(2003)\citenamefont {Day},
  \citenamefont {LeDuc}, \citenamefont {Mazin}, \citenamefont {Vayonakis},\
  and\ \citenamefont {Zmuidzinas}}]{day_broadband_2003}%
  \BibitemOpen
  \bibfield  {author} {\bibinfo {author} {\bibfnamefont {P.~K.}\ \bibnamefont
  {Day}}, \bibinfo {author} {\bibfnamefont {H.~G.}\ \bibnamefont {LeDuc}},
  \bibinfo {author} {\bibfnamefont {B.~A.}\ \bibnamefont {Mazin}}, \bibinfo
  {author} {\bibfnamefont {A.}~\bibnamefont {Vayonakis}},\ and\ \bibinfo
  {author} {\bibfnamefont {J.}~\bibnamefont {Zmuidzinas}},\ }\bibfield  {title}
  {\bibinfo {title} {A broadband superconducting detector suitable for use in
  large arrays},\ }\href {https://doi.org/10.1038/nature02037} {\bibfield
  {journal} {\bibinfo  {journal} {Nature}\ }\textbf {\bibinfo {volume} {425}},\
  \bibinfo {pages} {817} (\bibinfo {year} {2003})}\BibitemShut {NoStop}%
\bibitem [{\citenamefont {Kubo}\ \emph {et~al.}(2010)\citenamefont {Kubo},
  \citenamefont {Ong}, \citenamefont {Bertet}, \citenamefont {Vion},
  \citenamefont {Jacques}, \citenamefont {Zheng}, \citenamefont {Dréau},
  \citenamefont {Roch}, \citenamefont {Auffeves}, \citenamefont {Jelezko},
  \citenamefont {Wrachtrup}, \citenamefont {Barthe}, \citenamefont {Bergonzo},\
  and\ \citenamefont {Esteve}}]{kubo_strong_2010}%
  \BibitemOpen
  \bibfield  {author} {\bibinfo {author} {\bibfnamefont {Y.}~\bibnamefont
  {Kubo}}, \bibinfo {author} {\bibfnamefont {F.~R.}\ \bibnamefont {Ong}},
  \bibinfo {author} {\bibfnamefont {P.}~\bibnamefont {Bertet}}, \bibinfo
  {author} {\bibfnamefont {D.}~\bibnamefont {Vion}}, \bibinfo {author}
  {\bibfnamefont {V.}~\bibnamefont {Jacques}}, \bibinfo {author} {\bibfnamefont
  {D.}~\bibnamefont {Zheng}}, \bibinfo {author} {\bibfnamefont
  {A.}~\bibnamefont {Dréau}}, \bibinfo {author} {\bibfnamefont {J.-F.}\
  \bibnamefont {Roch}}, \bibinfo {author} {\bibfnamefont {A.}~\bibnamefont
  {Auffeves}}, \bibinfo {author} {\bibfnamefont {F.}~\bibnamefont {Jelezko}},
  \bibinfo {author} {\bibfnamefont {J.}~\bibnamefont {Wrachtrup}}, \bibinfo
  {author} {\bibfnamefont {M.~F.}\ \bibnamefont {Barthe}}, \bibinfo {author}
  {\bibfnamefont {P.}~\bibnamefont {Bergonzo}},\ and\ \bibinfo {author}
  {\bibfnamefont {D.}~\bibnamefont {Esteve}},\ }\bibfield  {title} {\bibinfo
  {title} {Strong {Coupling} of a {Spin} {Ensemble} to a {Superconducting}
  {Resonator}},\ }\href {https://doi.org/10.1103/PhysRevLett.105.140502}
  {\bibfield  {journal} {\bibinfo  {journal} {Physical Review Letters}\
  }\textbf {\bibinfo {volume} {105}},\ \bibinfo {pages} {140502} (\bibinfo
  {year} {2010})}\BibitemShut {NoStop}%
\bibitem [{\citenamefont {Probst}\ \emph {et~al.}(2013)\citenamefont {Probst},
  \citenamefont {Rotzinger}, \citenamefont {Wünsch}, \citenamefont {Jung},
  \citenamefont {Jerger}, \citenamefont {Siegel}, \citenamefont {Ustinov},\
  and\ \citenamefont {Bushev}}]{probst_anisotropic_2013}%
  \BibitemOpen
  \bibfield  {author} {\bibinfo {author} {\bibfnamefont {S.}~\bibnamefont
  {Probst}}, \bibinfo {author} {\bibfnamefont {H.}~\bibnamefont {Rotzinger}},
  \bibinfo {author} {\bibfnamefont {S.}~\bibnamefont {Wünsch}}, \bibinfo
  {author} {\bibfnamefont {P.}~\bibnamefont {Jung}}, \bibinfo {author}
  {\bibfnamefont {M.}~\bibnamefont {Jerger}}, \bibinfo {author} {\bibfnamefont
  {M.}~\bibnamefont {Siegel}}, \bibinfo {author} {\bibfnamefont {A.~V.}\
  \bibnamefont {Ustinov}},\ and\ \bibinfo {author} {\bibfnamefont {P.~A.}\
  \bibnamefont {Bushev}},\ }\bibfield  {title} {\bibinfo {title} {Anisotropic
  {Rare}-{Earth} {Spin} {Ensemble} {Strongly} {Coupled} to a {Superconducting}
  {Resonator}},\ }\href {https://doi.org/10.1103/PhysRevLett.110.157001}
  {\bibfield  {journal} {\bibinfo  {journal} {Physical Review Letters}\
  }\textbf {\bibinfo {volume} {110}},\ \bibinfo {pages} {157001} (\bibinfo
  {year} {2013})}\BibitemShut {NoStop}%
\bibitem [{\citenamefont {Guillot-Noël}\ \emph {et~al.}(2007)\citenamefont
  {Guillot-Noël}, \citenamefont {Vezin}, \citenamefont {Goldner},
  \citenamefont {Beaudoux}, \citenamefont {Vincent}, \citenamefont {Lejay},\
  and\ \citenamefont {Lorgeré}}]{guillot-noel_direct_2007}%
  \BibitemOpen
  \bibfield  {author} {\bibinfo {author} {\bibfnamefont {O.}~\bibnamefont
  {Guillot-Noël}}, \bibinfo {author} {\bibfnamefont {H.}~\bibnamefont
  {Vezin}}, \bibinfo {author} {\bibfnamefont {P.}~\bibnamefont {Goldner}},
  \bibinfo {author} {\bibfnamefont {F.}~\bibnamefont {Beaudoux}}, \bibinfo
  {author} {\bibfnamefont {J.}~\bibnamefont {Vincent}}, \bibinfo {author}
  {\bibfnamefont {J.}~\bibnamefont {Lejay}},\ and\ \bibinfo {author}
  {\bibfnamefont {I.}~\bibnamefont {Lorgeré}},\ }\bibfield  {title} {\bibinfo
  {title} {Direct observation of rare-earth-host interactions in er:y2sio5},\
  }\href {https://doi.org/10.1103/PhysRevB.76.180408} {\bibfield  {journal}
  {\bibinfo  {journal} {Physical Review B}\ }\textbf {\bibinfo {volume} {76}},\
  \bibinfo {pages} {180408} (\bibinfo {year} {2007})}\BibitemShut {NoStop}%
\bibitem [{\citenamefont {Car}\ \emph {et~al.}(2018)\citenamefont {Car},
  \citenamefont {Veissier}, \citenamefont {Louchet-Chauvet}, \citenamefont
  {Le~Gouët},\ and\ \citenamefont {Chanelière}}]{car_selective_2018}%
  \BibitemOpen
  \bibfield  {author} {\bibinfo {author} {\bibfnamefont {B.}~\bibnamefont
  {Car}}, \bibinfo {author} {\bibfnamefont {L.}~\bibnamefont {Veissier}},
  \bibinfo {author} {\bibfnamefont {A.}~\bibnamefont {Louchet-Chauvet}},
  \bibinfo {author} {\bibfnamefont {J.-L.}\ \bibnamefont {Le~Gouët}},\ and\
  \bibinfo {author} {\bibfnamefont {T.}~\bibnamefont {Chanelière}},\
  }\bibfield  {title} {\bibinfo {title} {Selective {Optical} {Addressing} of
  {Nuclear} {Spins} through {Superhyperfine} {Interaction} in {Rare}-{Earth}
  {Doped} {Solids}},\ }\href {https://doi.org/10.1103/PhysRevLett.120.197401}
  {\bibfield  {journal} {\bibinfo  {journal} {Physical Review Letters}\
  }\textbf {\bibinfo {volume} {120}},\ \bibinfo {pages} {197401} (\bibinfo
  {year} {2018})}\BibitemShut {NoStop}%
\bibitem [{\citenamefont {Byrd}\ \emph {et~al.}(1995)\citenamefont {Byrd},
  \citenamefont {Lu}, \citenamefont {Nocedal},\ and\ \citenamefont
  {Zhu}}]{byrd_limited_1995}%
  \BibitemOpen
  \bibfield  {author} {\bibinfo {author} {\bibfnamefont {R.}~\bibnamefont
  {Byrd}}, \bibinfo {author} {\bibfnamefont {P.}~\bibnamefont {Lu}}, \bibinfo
  {author} {\bibfnamefont {J.}~\bibnamefont {Nocedal}},\ and\ \bibinfo {author}
  {\bibfnamefont {C.}~\bibnamefont {Zhu}},\ }\bibfield  {title} {\bibinfo
  {title} {A {Limited} {Memory} {Algorithm} for {Bound} {Constrained}
  {Optimization}},\ }\href {https://doi.org/10.1137/0916069} {\bibfield
  {journal} {\bibinfo  {journal} {SIAM Journal on Scientific Computing}\
  }\textbf {\bibinfo {volume} {16}},\ \bibinfo {pages} {1190} (\bibinfo {year}
  {1995})}\BibitemShut {NoStop}%
\bibitem [{\citenamefont {Mansir}\ \emph {et~al.}(2018)\citenamefont {Mansir},
  \citenamefont {Conti}, \citenamefont {Zeng}, \citenamefont {Pla},
  \citenamefont {Bertet}, \citenamefont {Swift}, \citenamefont {Van~de Walle},
  \citenamefont {Thewalt}, \citenamefont {Sklenard}, \citenamefont {Niquet},\
  and\ \citenamefont {Morton}}]{mansir_linear_2018}%
  \BibitemOpen
  \bibfield  {author} {\bibinfo {author} {\bibfnamefont {J.}~\bibnamefont
  {Mansir}}, \bibinfo {author} {\bibfnamefont {P.}~\bibnamefont {Conti}},
  \bibinfo {author} {\bibfnamefont {Z.}~\bibnamefont {Zeng}}, \bibinfo {author}
  {\bibfnamefont {J.}~\bibnamefont {Pla}}, \bibinfo {author} {\bibfnamefont
  {P.}~\bibnamefont {Bertet}}, \bibinfo {author} {\bibfnamefont
  {M.}~\bibnamefont {Swift}}, \bibinfo {author} {\bibfnamefont
  {C.}~\bibnamefont {Van~de Walle}}, \bibinfo {author} {\bibfnamefont
  {M.}~\bibnamefont {Thewalt}}, \bibinfo {author} {\bibfnamefont
  {B.}~\bibnamefont {Sklenard}}, \bibinfo {author} {\bibfnamefont
  {Y.}~\bibnamefont {Niquet}},\ and\ \bibinfo {author} {\bibfnamefont
  {J.}~\bibnamefont {Morton}},\ }\bibfield  {title} {\bibinfo {title} {Linear
  {Hyperfine} {Tuning} of {Donor} {Spins} in {Silicon} {Using} {Hydrostatic}
  {Strain}},\ }\href {https://doi.org/10.1103/PhysRevLett.120.167701}
  {\bibfield  {journal} {\bibinfo  {journal} {Physical Review Letters}\
  }\textbf {\bibinfo {volume} {120}},\ \bibinfo {pages} {167701} (\bibinfo
  {year} {2018})}\BibitemShut {NoStop}%
\bibitem [{\citenamefont {Pla}\ \emph {et~al.}(2018)\citenamefont {Pla},
  \citenamefont {Bienfait}, \citenamefont {Pica}, \citenamefont {Mansir},
  \citenamefont {Mohiyaddin}, \citenamefont {Zeng}, \citenamefont {Niquet},
  \citenamefont {Morello}, \citenamefont {Schenkel}, \citenamefont {Morton},\
  and\ \citenamefont {Bertet}}]{pla_strain-induced_2018}%
  \BibitemOpen
  \bibfield  {author} {\bibinfo {author} {\bibfnamefont {J.}~\bibnamefont
  {Pla}}, \bibinfo {author} {\bibfnamefont {A.}~\bibnamefont {Bienfait}},
  \bibinfo {author} {\bibfnamefont {G.}~\bibnamefont {Pica}}, \bibinfo {author}
  {\bibfnamefont {J.}~\bibnamefont {Mansir}}, \bibinfo {author} {\bibfnamefont
  {F.}~\bibnamefont {Mohiyaddin}}, \bibinfo {author} {\bibfnamefont
  {Z.}~\bibnamefont {Zeng}}, \bibinfo {author} {\bibfnamefont {Y.}~\bibnamefont
  {Niquet}}, \bibinfo {author} {\bibfnamefont {A.}~\bibnamefont {Morello}},
  \bibinfo {author} {\bibfnamefont {T.}~\bibnamefont {Schenkel}}, \bibinfo
  {author} {\bibfnamefont {J.}~\bibnamefont {Morton}},\ and\ \bibinfo {author}
  {\bibfnamefont {P.}~\bibnamefont {Bertet}},\ }\bibfield  {title} {\bibinfo
  {title} {Strain-{Induced} {Spin}-{Resonance} {Shifts} in {Silicon}
  {Devices}},\ }\href {https://doi.org/10.1103/PhysRevApplied.9.044014}
  {\bibfield  {journal} {\bibinfo  {journal} {Physical Review Applied}\
  }\textbf {\bibinfo {volume} {9}},\ \bibinfo {pages} {044014} (\bibinfo {year}
  {2018})}\BibitemShut {NoStop}%
\bibitem [{\citenamefont {Graaf}\ \emph {et~al.}(2012)\citenamefont {Graaf},
  \citenamefont {Danilov}, \citenamefont {Adamyan}, \citenamefont {Bauch},\
  and\ \citenamefont {Kubatkin}}]{graaf_magnetic_2012}%
  \BibitemOpen
  \bibfield  {author} {\bibinfo {author} {\bibfnamefont {S.~E.~d.}\
  \bibnamefont {Graaf}}, \bibinfo {author} {\bibfnamefont {A.~V.}\ \bibnamefont
  {Danilov}}, \bibinfo {author} {\bibfnamefont {A.}~\bibnamefont {Adamyan}},
  \bibinfo {author} {\bibfnamefont {T.}~\bibnamefont {Bauch}},\ and\ \bibinfo
  {author} {\bibfnamefont {S.~E.}\ \bibnamefont {Kubatkin}},\ }\bibfield
  {title} {\bibinfo {title} {Magnetic field resilient superconducting fractal
  resonators for coupling to free spins},\ }\href
  {https://doi.org/10.1063/1.4769208} {\bibfield  {journal} {\bibinfo
  {journal} {Journal of Applied Physics}\ }\textbf {\bibinfo {volume} {112}},\
  \bibinfo {pages} {123905} (\bibinfo {year} {2012})}\BibitemShut {NoStop}%
\bibitem [{\citenamefont {Samkharadze}\ \emph {et~al.}(2016)\citenamefont
  {Samkharadze}, \citenamefont {Bruno}, \citenamefont {Scarlino}, \citenamefont
  {Zheng}, \citenamefont {DiVincenzo}, \citenamefont {DiCarlo},\ and\
  \citenamefont {Vandersypen}}]{samkharadze_high-kinetic-inductance_2016}%
  \BibitemOpen
  \bibfield  {author} {\bibinfo {author} {\bibfnamefont {N.}~\bibnamefont
  {Samkharadze}}, \bibinfo {author} {\bibfnamefont {A.}~\bibnamefont {Bruno}},
  \bibinfo {author} {\bibfnamefont {P.}~\bibnamefont {Scarlino}}, \bibinfo
  {author} {\bibfnamefont {G.}~\bibnamefont {Zheng}}, \bibinfo {author}
  {\bibfnamefont {D.}~\bibnamefont {DiVincenzo}}, \bibinfo {author}
  {\bibfnamefont {L.}~\bibnamefont {DiCarlo}},\ and\ \bibinfo {author}
  {\bibfnamefont {L.}~\bibnamefont {Vandersypen}},\ }\bibfield  {title}
  {{\bibinfo {title} {High-{Kinetic}-{Inductance}
  {Superconducting} {Nanowire} {Resonators} for {Circuit} {QED} in a {Magnetic}
  {Field}}},\ }\href {https://doi.org/10.1103/PhysRevApplied.5.044004}
  {\bibfield  {journal} {\bibinfo  {journal} {Physical Review Applied}\
  }\textbf {\bibinfo {volume} {5}},\ \bibinfo {pages} {044004} (\bibinfo {year}
  {2016})}\BibitemShut {NoStop}%
\bibitem [{\citenamefont {Mahashabde}\ \emph {et~al.}(2020)\citenamefont
  {Mahashabde}, \citenamefont {Otto}, \citenamefont {Montemurro}, \citenamefont
  {de~Graaf}, \citenamefont {Kubatkin},\ and\ \citenamefont
  {Danilov}}]{mahashabde_fast_2020}%
  \BibitemOpen
  \bibfield  {author} {\bibinfo {author} {\bibfnamefont {S.}~\bibnamefont
  {Mahashabde}}, \bibinfo {author} {\bibfnamefont {E.}~\bibnamefont {Otto}},
  \bibinfo {author} {\bibfnamefont {D.}~\bibnamefont {Montemurro}}, \bibinfo
  {author} {\bibfnamefont {S.}~\bibnamefont {de~Graaf}}, \bibinfo {author}
  {\bibfnamefont {S.}~\bibnamefont {Kubatkin}},\ and\ \bibinfo {author}
  {\bibfnamefont {A.}~\bibnamefont {Danilov}},\ }\bibfield  {title} {\bibinfo
  {title} {Fast tunable high {Q}-factor superconducting microwave resonators},\
  }\href {http://arxiv.org/abs/2003.11068} {\bibfield  {journal} {\bibinfo
  {journal} {arXiv:2003.11068 [cond-mat, physics:physics]}\ } (\bibinfo {year}
  {2020})},\ \bibinfo {note} {arXiv: 2003.11068}\BibitemShut {NoStop}%
\end{thebibliography}
\end{document}



\title{Hyperfine spectroscopy in a quantum-limited spectrometer}

\author{S.~Probst$^{1}$, G.-L. Zhang$^{2}$, M. Ran\v{c}i\'{c}$^{1}$, V. Ranjan$^{1}$, M. Le Dantec$^{1}$, Z. Zhang$^{3}$ B. Albanese$^{1}$, A. Doll$^{4}$, R.-B. Liu$^{2}$, J.J.L. Morton$^{5}$, T. Chaneli{\`e}re$^{6}$, P. Goldner$^{3}$, D. Vion$^{1}$, D. Esteve$^{1}$, P. Bertet$^{1}$}

\affiliation{$^{1}$Quantronics group, SPEC, CEA, CNRS, Universit\'e Paris-Saclay, CEA Saclay 91191 Gif-sur-Yvette Cedex, France}

\affiliation{$^{2}$Department of Physics and The Hong Kong Institute of Quantum Information Science and Technology, The Chinese University of Hong Kong, Shatin, New Territories, Hong Kong, China}

\affiliation{$^{3}$Chimie ParisTech, PSL University, CNRS, Institut de Recherche de Chimie Paris, 75005 Paris, France}

\affiliation{$^{4}$Laboratory of nanomagnetism and oxides, SPEC, CEA, CNRS, Universit\'e Paris-Saclay, CEA Saclay 91191 Gif-sur-Yvette Cedex, France}

\affiliation{$^{5}$London Centre for Nanotechnology, University College London, London WC1H 0AH, United Kingdom}

\affiliation{$^{6}$Univ. Grenoble Alpes, CNRS, Grenoble INP, Institut N\'eel, 38000 Grenoble, France}

\date{\today}

\pacs{07.57.Pt,76.30.-v,85.25.-j}

\maketitle

\section*{Section S1 - Details of the Model for ESEEM of Bi:Si}
In this Section, we give the details of the Si:Bi Hamiltonian, as a supplementary note to Section~III~B of the main text.
\\

The Hamiltonian of the bismuth donor spins and the $^{29}$Si nuclear spins is
\begin{equation}
H= H_{\rm Bi}+H_{\rm Si}+H_{{\rm hf}},
\end{equation}
where $ H_{\rm Bi}$ is the Hamiltonian for the Bismuth electron and nuclear spin, $H_{\rm Si}$ is  the Zeeman energy and the dipole-dipole interaction of the $^{29}$Si nuclear spins, and $H_{{\rm hf}}$ is  the hyperfine interaction between the Bismuth electron spin and the $^{29}$Si nuclear spins.

The Hamiltonian of the Bi center spins is~\citep{Ma2015} 
\begin{align}
        H_{\rm Bi} & = g_{\rm e} \beta_{\rm e} S_{0,z} + A_\text{Bi} \mathbf{S}_0\cdot \mathbf{I}_{{0}} ,
        \label{eq:H_of_Bi}
\end{align}
where the notations are the same as in main text. The electron gyromagnetic ratio $ g_{\rm e}\beta_{\rm e}= 1.76\times 10^{11}\,\mathrm{S^{-1}T^{-1}}$  and the Fermi contact hyperfine coupling $A_\text{Bi}/2\pi = 1.4754\,\mathrm{GHz}$. Note that the Zeeman energy of the nuclear spin is neglected (as explained in the main text).

The internal Hamiltonian of the $^{29}$Si nuclear spin bath is
$$
H_{\rm Si}=\omega_{I}\sum_{j}I_{j,z}+\sum_{i<j}{\mathbf I}_{i}\cdot \frac{\mu_0g_{\rm n}^2\beta_n^2}{4\pi r_{ij}^3} \left(1-\frac{3{\mathbf r}_{ij}{\mathbf r}_{ij}}{r_{ij}^2}\right)\cdot{\mathbf I}_j,
$$
where $\omega_{I}=g_{\rm n}\beta_nB_0$ denotes the Larmor frequency of the $^{29}$Si nuclear spin with the gyromagnetic ratio $g_{\rm n}\beta_n=-5.319 \times 10^7 {\rm s}^{-1}{\rm T}^{-1}$, the spin operator ${\mathbf I}_j$ denotes the nuclear spin at position ${\mathbf r}_j$, and ${\mathbf r}_{ij}={\mathbf r}_i-{\mathbf r}_j$.

The hyperfine interaction between the electron spin and the $^{29}$Si nuclear spins is
\begin{equation}
H_{{\rm hf}}=\sum_j {\mathbf S}_0\cdot {\bar{\mathbf A}}_j\cdot{\mathbf I}_j \equiv \sum_j H_{{\rm hf},j},
\end{equation}
where the hyperfine coupling tensor includes both the Fermi contact term $A_{{\rm cf},j}$ (a scalar) and the dipolar interaction ${\bar{\mathbf A}}_{{\rm dd},j}$, i.e., ${\bar{\mathbf A}}_j=A_{{\rm cf},j}{\mathbf 1}+{\bar{\mathbf A}}_{{\rm dd},j}$. 
The dipolar hyperfine coupling tensor is
$${\bar{\mathbf A}}_{{\rm dd},j}=\frac{\mu_0 g_{\rm e}\beta_{\rm e}g_{\rm n}\beta_{\rm n}}{4\pi r_{j}^3} \left(1-\frac{3{\mathbf r}_j{\mathbf r}_j}{r_{j}^2}\right)
\equiv A_{{\rm dd},j}\left(1-\frac{3{\mathbf r}_j{\mathbf r}_j}{r_{j}^2}\right),$$
decaying with a cubic power of the distance. 
The Fermi contact interaction is proportional to the electron density $|\psi({\mathbf r}_j)|^2$ of the Bismuth donor at the nuclear spin position ${\mathbf r}_j$ \citep{Feher1959, Kohn1957, Sousa2003}, 
$$
A_{{\rm cf},j}=\frac{2}{3}\mu_0 g_{\rm e}\beta_{\rm e}g_{\rm n}\beta_{\rm n}|\psi({\mathbf r}_j)|^2.
$$
The electron density
\begin{align}
|\psi({\mathbf r}_j)|^2 = \frac{2}{3}\eta |f(\mathbf{r}_j)|^2\left[\cos(k_0x_j) + \cos(k_0y_j) + \cos(k_0z_j)\right]^2 ,
\nonumber
\end{align}
where  $k_0 = 0.85\frac{2\pi}{a_{\text{Si}}}$ (with $a_{\text{Si}} = 0.543\,\mathrm{nm}$) is the wavenumber of the conduction band minimum,
 $\eta \approx 180$ is the charge density on each site, and the envelope function is taken as the Kohn-Luttinger wave function form~\citep{Sousa2003}
\begin{equation}
    f_j(\mathbf{r}) = \frac{1}{\sqrt{\pi(sa)^2(sb)}}\exp\left(-\sqrt{\frac{z^2}{(sb)^2} + \frac{x^2+y^2}{(sa)^2}}\right), \nonumber
\end{equation}
with $a = 2.51\,\mathrm{nm}$ and $b=1.44\,\mathrm{nm}$ being the characteristic lengths for hydrogenic impurities in Si and the scaling factor $s=0.64$ 
for bismuth~\citep{Hale1969}.
The hyperfine coupling strength $|{\bar{\mathbf A}}_j|$ is mostly $<<$~MHz.

\section*{Section S2 - Applying the fictitious spin-1/2 model to Bi:Si coupled to $^{29}$Si spins}
In this section, we provide detailed justifications for applying the fictitious model in Section~II~D of main text  to Bi:Si coupled to $^{29}$Si spins.
\\

The eigenstates $|\pm,m\rangle$ of the Bismuth Hamiltonian $ H_{\rm Bi}$ in Eq.~(\ref{eq:H_of_Bi}) are
\begin{equation}
|\pm,m\rangle = \cos \frac{\theta_m}{2} |\pm \frac{1}{2},m \mp \frac{1}{2} \rangle \pm \sin \frac{\theta_m}{2} |\mp \frac{1}{2},m \pm \frac{1}{2} \rangle,
\label{eq:eigenstates}
\end{equation}
with $
\tan \theta_m = \frac{\sqrt{25-m^2}}{m+(1+\delta){{g_{\rm e}\beta_{\rm e}B_0}}/{A_{\rm Bi}}}$, and the corresponding eigenenergies are
\begin{align}
E_m^\pm =&  -\frac{A_{\rm Bi}}{4}
\pm \frac{A_{\rm Bi}}{2} \sqrt{\left(m+\frac{{g_{\rm e}\beta_{\rm e}B_0}}{A_{\rm Bi}}\right)^2 + 25 - m^2}.
\label{eq:BiSiE}
\end{align}
For the experimental condition $\left|{g_{\rm e}\beta_{\rm e}B_0}\right|\ll \left|A_{\rm Bi}\right|$, the eigenenergies in Eq.~(\ref{eq:BiSiE}) can be approximated as
$$
E_m^{\pm} \approx -\frac{A_{\rm Bi}}{2} \pm \frac{5A_{\rm Bi}}{2} \pm\frac{m {g_{\rm e}\beta_{\rm e}B_0}}{10},
$$
and the level splitting 
$$
E_m^{\pm}-E_{m-1}^{\pm} \approx  \pm\frac{{g_{\rm e}\beta_{\rm e}B_0}}{10}.
$$
For a field about 1~Gauss (a typical value in our experiments), $(2\pi)^{-1} {g_{\rm e}\beta_{\rm e}B_0}/10\sim 300$~kHz.


The hyperfine coupling to the $^{29}$Si nuclear spins can be written in the basis of the eigenstates of $H_{\rm Bi}$.
The non-vanishing matrix elements of the electron spin operators are
\begin{subequations}
\begin{align}
 \left\langle\pm,m\left|S_{0,z}\right|\pm,m\right\rangle & = \pm\frac{1}{2}\cos\theta_m\equiv \pm\frac{1}{2}\alpha_m, \label{eq_alpha_m}\\ 
\left\langle +,m\left|S_{0,x}\right|+,m-1\right\rangle &=+\frac{1}{2}\cos\frac{\theta_m}{2}\sin\frac{\theta_{m-1}}{2}, \\
\left\langle -,m\left|S_{0,x}\right|-,m-1\right\rangle &=-\frac{1}{2}\sin\frac{\theta_m}{2}\cos\frac{\theta_{m-1}}{2}. 
\end{align}
\end{subequations}
The diagonal matrix elements
cause the frequency shift due to the
coupling to the nuclear spins,which is $\sim |{\bar{\mathbf A}}_j|/2$, and the off-diagonal elements cause the
coupling between neighboring energy levels $\sim |{\bar{\mathbf A}}_j|/4$. 
For a nuclear spin with relatively weak hyperfine coupling, namely,
\begin{equation}
|{\bar{\mathbf A}}_j|/2\ll |{g_{\rm e}\beta_{\rm e}B_0}|/10,
\label{eq:CCE_condition}
\end{equation}
the hyperfine-induced mixing between different energy levels of the Bismuth center can be neglected. Then the interaction Hamiltonian becomes
\begin{equation}
H_{{\rm hf}}\approx \sum_{\eta=\pm}\sum_{m}\eta\alpha_m|\eta,m\rangle\langle\eta,m|\otimes   \sum_j{\mathbf e}_z\cdot \bar{\mathbf A}_j\cdot {\mathbf I}_j,
\label{eq:puredephasing}
\end{equation}
where  
\begin{equation}
    \alpha_m = \frac{{g_{\rm e}\beta_{\rm e}B_0} + mA_\text{Bi}}{\sqrt{A_\text{Bi}^2 (25-m^2) + ({g_{\rm e}\beta_{\rm e}B_0}+mA_\text{Bi})^2}},
\end{equation} 
as defined in Eq.~(\ref{eq_alpha_m}). This Hamiltonian is diagonal in the Bismuth center eigenstates, corresponding to the pure dephasing model (or secular approximation).

As will be demonstrated in Sec.~\ref{sec:strong}, the contribution of a relatively strongly coupled $^{29}$Si nuclear spin ($|{\bar{\mathbf A}}_j|/2 \gtrsim |{g_{\rm e}\beta_{\rm e}B_0}|/10$) to the ESEEM signal in our experiments is negligible.
Thus the pure dephasing model in Eq.~(\ref{eq:puredephasing}) is justified.

The microwave pulses in general can induce many transitions in the Bi center as along as the transitions are within the bandwidth of the microwave cavity ({164~kHz for the high-Q resonator}). Since the interference between the transitions whose frequencies are not near-degenerate would cause oscillation much faster than the time resolution of the experiments and the transitions with near-degenerate frequencies do not share an eigenstate of the Bi center (and therefore has no interference), we can reduce the Bi center to independent two-level systems (``fictitious spin-1/2'') coupled to the Si spin bath.  

The ``fictitious spin-1/2'' for  the transition 
$\lvert +,m \rangle\leftrightarrow \lvert -,m-1 \rangle$ has the Hamiltonian
\begin{align}
H_m\equiv |+,m\rangle \langle +,m|\otimes H_m^{+}+|-,m-1\rangle\langle -,m-1|\otimes H_m^{-},
\label{eq:Hm}
\end{align}
with
\begin{subequations}
\begin{align}
H_m^{+} &= E_m^{+}+\frac{1}{2} \alpha_m h_z + H_{\mathrm{Si}}, \\
H_m^{-} &=  E_{m-1}^{-}-\frac{1}{2} \alpha_{m-1} h_z + H_{\mathrm{Si}},
\end{align}
\end{subequations}
where the Overhauser field along the $z$ direction is 
$$
h_z\equiv  \sum_j {\mathbf e}_z\cdot \bar{\mathbf A}_j\cdot {\mathbf I}_j.
$$
The transition frequency or effective Larmor frequency of the fictitious spin-1/2
$$\omega_S =E^+_m-E^-_{m-1}\approx 5A_{\rm Bi}+\frac{(2m-1){g_{\rm e}\beta_{\rm e}B_0}}{10}.
$$

Using a fictitious spin-1/2 to represent the transition  $\lvert -,m-1 \rangle\leftrightarrow \lvert +, m \rangle$ and define $\delta_m \equiv (\alpha_m - \alpha_{m-1})/2$ and 
$\bar{\alpha}_m \equiv (\alpha_m + \alpha_{m-1})/2 $, we obtain the Hamiltonian
\begin{align}
H_m= \omega_SS_z+\bar{\alpha}_mS_zh_z+\left(\frac{\delta_m}{2}h_z+H_{\rm Si}\right) ,
\end{align}
which becomes Eq. 2 of main text if the bath has only one nuclear spin.
\\

For a single nuclear spin (denoted as spin-$j$), the conditional Hamiltonian is
\begin{equation}
H_m^{\pm} = (\omega_{I} + \frac{\delta_m}{2}A_{j,zz})I_{j,z} + \frac{\delta_m}{2}A_{j,zx} I_{j,x} \pm \frac{\bar{\alpha}_m}{2}(A_{j,zz}I_{j,z} + A_{j,zx} I_{j,x})
\equiv \bar{H}_m \pm V_m
.
\label{eq:effective_Hm}
\end{equation}
Note that we have  dropped the $\omega_S$-term (by working in the rotating reference frame) and for convenience chosen the coordinate system such that $A_{j,zy}=0$.
In the basis of the eigenstates of $ \bar{H}_m$, the effective Hamiltonian can be written as
\begin{equation}
H_m^{\pm} = \tilde{\omega}_{I}I_{j,z} \pm \frac12({A}I_{j,z} + {B}I_{j,x}), 
\end{equation}
where 
\begin{align}
\tilde{\omega}_{I} & = \sqrt{(\omega_{I} + \delta_mA_{j,zz}/2) ^2+ (\delta_mA_{j,zx}/2)^2}, \nonumber\\
{A} & = \bar{\alpha}_m(A_{j,zz}\cos\theta + A_{j,zx} \sin\theta), \nonumber \\
{B} & = \bar{\alpha}_m(A_{j,zx} \cos\theta - A_{j,zz}\sin\theta),\nonumber
\end{align}
with $\theta = \arcsin\left(\delta_mA_{j,zx}/2\tilde{\omega}_{I}\right)$.

\section*{Section S3 - Analytical solutions of ESEEM envelopes for the fictitious spin-1/2 model}
In this Section, we provide details on the analytical solutions for the 2p-, 3p- and 5p-ESEEM~\citep{Kasumaj2008}, 
as a supplementary note on how the curves in Figs. 9-11 of Sec. VB in main text are calculated.

\subsection*{Section S3.1 - Formula}

In the $n$-pulse ESEEM experiment, a sequence of $n$ pulses $R_j\in\{R^{x}_{\pi/2},R^y_{\pi/2}, R^{x}_{\pi},R^{y}_{\pi}\}$ are applied at $t_j$ ($j=1,2,\ldots n$), and then the signal $V_{n{\rm p}}$ is measured at the echo time $t$. In this Section, we take the pulses as ideal, i.e.,
$$R^{x/y}_{\theta}=\exp\left(-i\theta S_{x/y}\right).$$ 
With the pulse sequences described in Sec. VB of main text, the evolution for the 2p-, 3p-, and 5p-ESEEM is in turn
\begin{subequations}
\label{U2p3p5p}
\begin{align}
& U_{\rm 2p}=e^{-iH_m\tau}R^y_{\pi}e^{-iH_m\tau}R^x_{\pi/2},  \\
& U_{\rm 3p}=e^{-iH_m\tau}R^x_{\pi/2}e^{-iH_m T}R^x_{\pi/2}e^{-iH_m\tau}R^x_{\pi/2}, \\
& U_{\rm 5p}=e^{-iH_m\tau_2}R^y_{\pi}e^{-iH_m\tau_2}R^y_{\pi/2}e^{-iH_m T}R^y_{\pi/2}e^{-iH_m\tau_1}R^x_{\pi/2}e^{-iH_m\tau_1}R^x_{\pi/2}.
\end{align}
\end{subequations}
We assume the Bi center is initially in the state $\rho_{S}=|-,m-1\rangle\langle -,m-1|$ and the nuclear spin bath is in the maximally mixed state $\rho_{\rm B}$. The spin coherence at the echo time is
\begin{align}
V_{\rm 2p/3p/5p} =2{\rm Tr}\left[S_{0,x} U_{\rm 2p/3p/5p} \left(\rho_S\otimes\rho_{\rm B}\right) U^{\dag}_{\rm 2p/3p/5p}\right].
\end{align}

\subsection*{Section S3.2 - Exact formula for a single nuclear spin}
\label{exact_single}

For 2p-ESEEM, the modulation amplitude due to the $j$-th $^{29}$Si spin is 
\begin{equation}
V_{2 \mathrm{p},j}(\tau)=1-\frac{k_j}{4}\left[2-2 \cos \left(\omega_{\uparrow} \tau\right)-2 \cos \left(\omega_{\downarrow} \tau\right)
+\cos \left(\omega_{\uparrow}\tau-\omega_{\downarrow} \tau\right)+\cos \left(\omega_{\uparrow}\tau+\omega_{\downarrow} \tau\right)\right],
\end{equation}
where $k_j=\left(\frac{{B} \tilde{\omega}_{I}}{\omega_{\uparrow} \omega_{\downarrow}}\right)^{2}$, and (the same as Eq. 6 of main text)
\begin{align}
\omega_\uparrow \equiv \sqrt{(\tilde{\omega}_{I}+\frac{{A}}{2})^{2}+\frac{{B}^{2}}{4}},  \ \ \ \ 
\omega_\downarrow \equiv \sqrt{(\tilde{\omega}_{I}-\frac{{A}}{2})^{2}+\frac{{B}^{2}}{4}}. \nonumber
\end{align}
\\

For the 3p-ESEEM, the modulation amplitude is given by
\begin{align}
V_{3 \mathrm{p},j}(\tau, T) &= \frac{1}{2}\left[V_{3 \mathrm{p},j}^{\uparrow}(\tau, T) + V_{3 \mathrm{p},j}^{\downarrow}(\tau, T) \right] \nonumber \\
&= 1-\left\{\frac{k_j}{4}\left[1-\cos (\omega_{\downarrow} \tau)\right]\left[1-\cos (\omega_{\uparrow}T+\omega_{\uparrow}\tau)\right]
+\left[\omega_{\uparrow} \leftrightarrow  \omega_{\downarrow}\right]\right\}.
\end{align}
where 
$$V_{3 \mathrm{p},j}^{\uparrow}(\tau, T) = 1-\frac{k_j}{2}\left[1-\cos (\omega_{\downarrow} \tau)\right]\left[1-\cos (\omega_{\uparrow}T+\omega_{\uparrow}\tau)\right]$$ 
and $V_{3 \mathrm{p},j}^{\downarrow}(\tau, T)$ is obtained by exchanging $\uparrow$ and $\downarrow$. 
\\

For the 5p-ESEEM, the modulation amplitude is
\begin{equation}
V_{{\rm 5p},j}(\tau_1, \tau_2, T) = \frac{1}{4}\left(V_{{\rm 5p},j}^{\uparrow,+} - V_{{\rm 5p},j}^{\uparrow, -} + V_{{\rm 5p},j}^{\downarrow, +} - V_{{\rm 5p},j}^{\downarrow, -}\right),
\end{equation}
where 
\begin{align} V_{{\rm 5p},j}^{\uparrow, {\pm}} = & V_{2 \mathrm{p},j}\left(\tau_{1}\right) V_{2 \mathrm{p},j}\left(\tau_{2}\right)\pm
b_{{\rm 5p},j} \left[4 k_j^2 C_{j}^{\uparrow} -2 k_j^{2} \cos \phi_{\downarrow, -} \cos \left(T \omega_{\uparrow}+\phi_{\uparrow, +}\right)
\right.\nonumber \\
&\left. - 4 k_j \cos ^{4} \eta_j \cos \left(T \omega_{\uparrow}+\phi_{\uparrow, +}+\phi_{\downarrow, +}\right) -4 k_j \sin ^{4} \eta_j \cos \left(T \omega_{\uparrow}+\phi_{\uparrow, +}-\phi_{\downarrow, +}\right)\right], 
\end{align}
and $V_{{\rm 5p},j}^{\downarrow, {\pm}}$ is obtained from the expression above by exchanging $\uparrow$ and $\downarrow$, with 
\begin{align}
\phi_{\uparrow/\downarrow, \pm} &=\left(\tau_{1} \pm \tau_{2}\right)\omega_{\uparrow/\downarrow} / 2, \nonumber \\
\cos^2\eta_j & = \left[\tilde{\omega}_{I}^{2} - \frac{1}{4} (\omega_\uparrow - \omega_\downarrow)^2\right]/\left(\omega_\uparrow\omega_\downarrow\right),\nonumber \\
\sin^2\eta_j & = \left[ \frac{1}{4} (\omega_\uparrow + \omega_\downarrow)^2-\tilde{\omega}_{I}^{2} \right]/\left(\omega_\uparrow\omega_\downarrow\right), \nonumber \\
 C_{j}^{\uparrow} &=\cos \left(\frac{\tau_{1} \omega_{\uparrow}}{2}\right) \cos \left(\frac{\tau_{2} \omega_{\uparrow}}{2}\right) \sin \left(\frac{\tau_{1} \omega_{\downarrow}}{2}\right) \sin \left(\frac{\tau_{2} \omega_{\downarrow}}{2}\right), \nonumber \\
b_{{\rm 5p},j} &=\sin \left(\frac{\tau_{1} \omega_{\uparrow}}{2}\right) \sin \left(\frac{\tau_{2} \omega_{\uparrow}}{2}\right) \sin \left(\frac{\tau_{1} \omega_{\downarrow}}{2}\right) \sin \left(\frac{\tau_{2} \omega_{\downarrow}}{2}\right).
\nonumber 
\end{align}

\subsection*{Section S3.3 - Contributions of multiple nuclear spins}
For multiple nuclear spins, if they are taken as independent (with interactions between the nuclear spins neglected), 
the ESEEM signal is obtained by applying the product rule for each pathway followed by average over different pathways (\citep{Kasumaj2008}). For the 2p/3p/5p-ESEEM signals are in turn
\begin{subequations}
\label{V2p3p5p}
\begin{align}
V_{\mathrm{2p}} & = \prod_j V_{2 \mathrm{p},j}, \\
V_{\mathrm{3p}} & = \frac{1}{2}\left(\prod_j V_{3 \mathrm{p},j}^{\uparrow} + \prod_j V_{3 \mathrm{p},j}^{\downarrow} \right), \\
V_{\mathrm{5p}} &= \frac{1}{4}\left(\prod_j V_{{\rm 5p},j}^{\uparrow,+} - \prod_j V_{{\rm 5p},j}^{\uparrow, -} +\prod_j  V_{{\rm 5p},j}^{\downarrow, +} - \prod_j V_{{\rm 5p},j}^{\downarrow, -}\right).
\end{align}
\end{subequations}

Above we have considered the ESEEM signal due to the transition $\lvert -,m-1 \rangle\leftrightarrow \lvert +,m\rangle$.
{The transitions $\lvert -,m \rangle\leftrightarrow \lvert +,m-1 \rangle$ can be considered similarly.} We use  $V^{(m,\pm)}_{\rm 2p/3p/5p}$ to denote the contribution to the 2p/3p/5p-ESEEM signal by the transition $\lvert -,m \rangle\leftrightarrow \lvert +,m\pm 1\rangle$ [with a superscript index $(m,\pm)$ attached to the signals in Eq.~(\ref{V2p3p5p})]. 

\subsection*{Section S3.4 - Weighting factors of different Bi spin transitions}

To take into account the contributions of all possible transitions, we assume the Bi center spin is initially randomly populated in the lower manifold of the eigenstates, i.e.,
$$
\rho_{\rm Bi}=\sum_{m=-4}^{+4} P_m |-,m\rangle\langle -,m|,
$$
with the probabilities $\sum_m P_m=1$. 
The coherence of different transitions is summed up as 
\begin{align}
V_{\rm 2p/3p/5p}=\sum_{m,\pm }W_{m,\pm} V^{(m,\pm)}_{\rm 2p/3p/5p},
\label{V_weighted}
\end{align}
with the weighting factor $W_{m,\pm}$ accounting for the initial probabilities $P_m$ and the different amplitudes for different transitions given the microwave pulse spectra and the distribution of the hyperfine coupling $A_{\rm Bi}$ due to the strain in the Si layer.

\begin{figure}[]%
\centering
\includegraphics[width=\columnwidth]{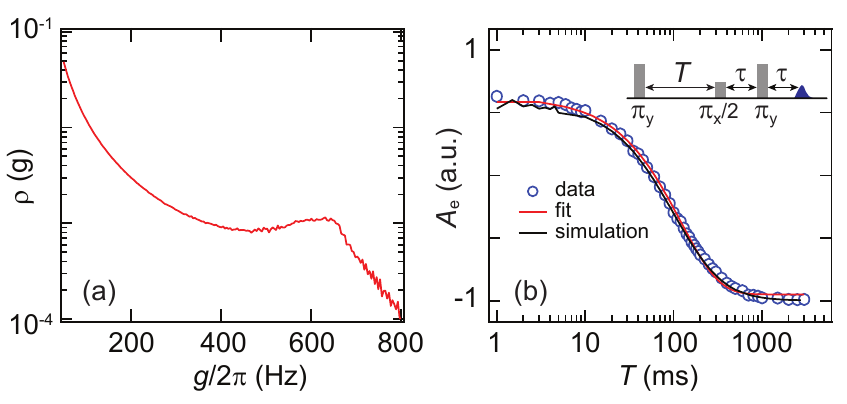}%
\caption{Inhomogeneous coupling strength. (a) The normalized density of spins $\rho (g)$ versus the spin-photon coupling strength for the transition $|-,4 \rangle \rightarrow |+,5 \rangle$. (b) Inversion recovery of spin polarization measured (in symbols) at a repetition rate of 0.2~Hz. The fit yields a $T_1 \approx 120$~ms. The simulation incorporating 18~transitions and $\rho(g)$ is consistent with the data for only a suitable pulse amplitude. }%
\label{fig:T1}%
\end{figure}

To determine the relative transition weights, we resort to a complete model of the experiment, based on the physical parameters, namely the control pulse amplitude and temporal profile, the Rabi frequencies distribution, and the repetition time $\Gamma_{\rm rep}^{-1}$. We use a simulation code that was purposely written. 
For each class of fictitious spin-1/2's with given Larmor frequency $\omega_S$ and Rabi frequency $\Omega_R$, the program integrates the Bloch equations to compute the time dependence of the spin density matrix $\rho(t)$. The initial conditions take into account the Rabi-frequency-dependent spin relaxation $T_1$ because of the Purcell effect, by taking $\rho(0) =|x\rangle\langle x|$ (the spin polarized along the $x$-axis). The simulation results are furthermore averaged over the strain-induced distribution of the $^{209}$Bi hyperfine coupling $\sigma_A(A_{\rm Bi})$, which results in the distribution of the transition frequency $\sigma_S(\omega_{S})$, and over the distribution of the Rabi frequency $\sigma_R(\Omega_R)$. The Bi hyperfine coupling distribution $\sigma_A(A_{\rm Bi})$ is taken to be flat since the inhomogeneous broadening {($\sim$50~MHz)} is two orders of magnitude larger than the cavity bandwidth {($\sim$160~kHz)}.  The Rabi frequency distribution $\sigma_R(\Omega_R)$ is computed using finite-element modelling of the AC field spatial profile generated by running a constant current through the resonator inductance wire. The AC field map and the resulting distribution $\sigma_R(\Omega_R)$ are shown in Fig.~\ref{fig:T1}. To calibrate the AC pulse amplitude, we rely on the fact that due to the AC field inhomogeneity and to the Purcell spin relaxation, the measured $T_1$ in an inversion recovery sequence is in fact amplitude-dependent. We then adjust the pulse amplitude in the simulation so that the simulated inversion recovery sequence reproduces the same relaxation curve as measured.
The weighting factors thus derived from simulating the experimental data are given in Table.~\ref{tab:weights}.

\begin{table}[t]
\caption{The relative transition weights $W_{m,\pm}$ from the simulation.}
\vskip 0.2cm
	\begin{tabular}{|c|c|c|c|c|c|c|c|c|c|}
		\hline
		m   & 4      & 3      & 2      & 1      & 0      & -1     & -2     & -3     & -4     \\ \hline
		$\lvert -, m\rangle \leftrightarrow\lvert +, m+1\rangle$ & 0.1075 & 0.0977 & 0.0909 & 0.0766 & 0.0586 & 0.0326 & 0.0204 & 0.0086 & 0.007  \\ \hline
		$\lvert -, m\rangle \leftrightarrow\lvert +, m-1\rangle$ & 0.007  & 0.0086 & 0.0204 & 0.0326 & 0.0586 & 0.0766 & 0.0909 & 0.0977 & 0.1075 \\ \hline
	\end{tabular}
		\label{tab:weights}
\end{table}

\subsection*{Section S3.5 - Comparison with experiments}

For comparison with experimental data, the curves in Figs. 9-11 of main text are obtained by first calculating 
$V_{\rm{2p/3p/5p}}$ using Eq.~(\ref{V_weighted}) for each nuclear spin configuration (with random positions of the $^{29}$Si nuclei) and then averaging over different nuclear spin spatial configurations.

\section*{Section S4 - Justification of independent bath spin approximation}

In the calculations above, the nuclear spins in the bath are taken as independent, i.e., the interactions between the $^{29}$Si spins are neglected. 
In this Section, we will provide a justification of the approximation by calculating the Bi spin coherence with the cluster correlation expansion 
(CCE)~\citep{Yang2008, Yang2009, Zhao2012}. The leading order of expansion (CCE-1) corresponds to the independent bath spin approximation used in the previous Section. The numerical calculation shows that the CCE-1 is a good approximation for the timescale relevant to the experimental studies of the ESEEM.


We check the contributions of various orders of the nuclear spin correlations to the ESEEM signals using the fictitious spin-1/2 model in Eq.~(\ref{eq:Hm}) and the CCE method~\citep{Yang2008, Yang2009, Zhao2012}. The interactions between the $^{29}$Si nuclear spins are included in $H_{\rm Si}$. 
The CCE method allows us to consider in a recursive way the dynamics of the nuclear spin bath due to correlations of different 
sizes (CCE-$n$ for $n$-spin irreducible correlation). 

\begin{figure}[h]
	\centering
	\includegraphics[width=\linewidth]{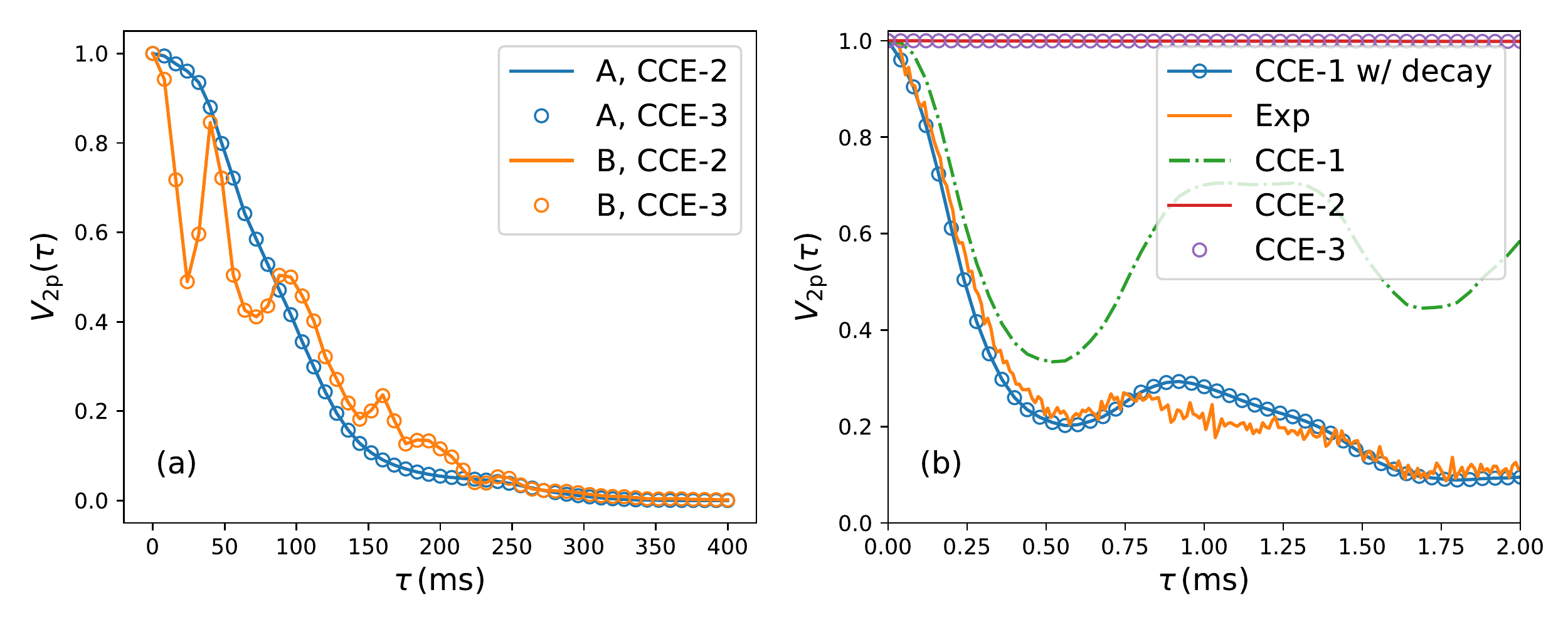}
	\caption{\label{fig:T2Compare} CCE calculation of the Hahn echo signal of a Bi donor spin coupled to a $^{29}$Si nuclear spin bath. The CCE-2 and CCE-3 results have excluded the CCE-1 effects.  (a) Decoherence for two different random bath configurations (denoted as ``A'' and ``B''), calculated up to CCE-2 or CCE-3 (both with CCE-1 contributions excluded). $B_0=1$~G. The number of bath spins $N =2000$. The negligible difference between CCE-2 and CCE-3 indicates that the CCE-2 has already converged. 
(b) Comparison of the CCE results with the experimental ESEEM signal for $B=1$~G. For the ``CCE-1 w/ decay'', an overall exponential decay is included as explained in the main text.  }
\end{figure}

Considering the 2p-ESEEM for example, the spin coherence is expanded as
\begin{equation}
	V_{\rm 2p} = \prod_C \tilde{V}_{\rm 2p}^C,
\end{equation}
with $\tilde{V}_{\rm 2p}^C$ defined as the irreduible correlation of cluster $C$ excluding the irreducible correlations of all sub-clusters, i.e., 
\begin{equation}
	\tilde{V}_{\rm 2p}^C\equiv \frac{V_{\rm 2p}^C}{\prod_{C'\subset C}\tilde{V}_{\rm 2p}^{C'}},
\end{equation}
where $V_{\rm 2p}^C$ is the center spin coherence under coupling to the cluster $C$ in the spin bath (the bath spins outside the cluster dropped). 

 As shown in Fig.~\ref{fig:T2Compare}, in the timescale relevant to the ESEEM signals in the experiments, the echo is affected mainly by the single-spin dynamics (CCE-1) and the contributions from pair dynamics (CCE-2) and higher order correlations in the bath are negligible.  The decoherence due to CCE-2 and CCE-3 occurs at timescales of $\sim 100$~ms [Fig.~\ref{fig:T2Compare}(a)], which are much longer than the experimental time regime ($\sim 1$~ms). 

Thus it is well justified to take the nuclear spins in the bath as independent of each other.



\section*{Section S5 - Effect of a strongly coupled Si-29 nuclear spin}
\label{sec:strong}

In this Section, we show that the strongly coupled $^{29}$Si nuclear spins have negligible contributions to the ESEEM signals. Therefore the calculated ESEEM signals presented in Figs.~9-11 in the main text are those from $^{29}$Si with hyperfine couplings $<20$~kHz.

In the calculations, we have assumed the pure dephasing model (the secular approximation) in which the transitions between different eigenstates of the Bi center spin due to the hyperfine coupling to the $^{29}$ nuclear spins are taken as negligible. This approximation is well justified if the hyperfine coupling is much less than the energy splitting between different Bi center spin states [Eq.~(\ref{eq:CCE_condition})].  When the coupling is strong the secular and the CCE-1 approximation become invalid. Using simulations that take into account exactly the effect of the strongly coupled nuclear spin, we show that a strongly coupled spin has negligible effects on the ESEEM signal in the timescale considered in the experiments. A strongly coupled spin contributes only fast oscillations in the signal, which would vanish if we take into account the inhomogeneous broadening effects. In addition, the influence of the strongly coupled spins on the distant weakly coupled spins are also negligible. 

For a $^{29}$Si with hyperfine coupling $\gtrsim 200$~kHz, the state mixing due to the hyperfine coupling enables many transitions and the interference between these transitions will cause rather complicated and fast oscillations in the spin echo signal. This is seen in Figs.~\ref{fig:3pOneSi29_FFT_F101kHz}. However, in such a strong coupling case, the ESEEM frequencies depend sensitively on the local Overhauser field on the Bi electron spin. Since the Overhauser field has a large inhomogeneous broadening ($\sim 0.5$~MHz), the ensemble average over the nuclear spin state thermal distribution leads to a rapid decay of the signal (decay in $<1\ \mu$s). We estimate that about $10\%$ Bi centers have one or more $^{29}$Si with coupling $>200$~kHz in the proximity, which would contribute to a fast initial decay of the total echo signal in $< 1\ \mu$s by about $10\%$. In the experiments, the echo signal is measured at times much greater than $\mu$s (the first time point is $\sim 1\,\mathrm{ms}$).  As shown in Figs.~\ref{fig:3p_Compare_Exact_CCE1_F2kHz} and \ref{fig:3pOneSi29_FFT_F101kHz}, the ESEEM amplitude of a nuclear spin with coupling strength between $20$~kHz and $200$~kHz is much less than $1\%$. And for nuclear spins with the relatively weak hyperfine coupling $(2\pi)^{-1}|{\bar{\mathbf A}}_j|<100$~kHz, the pure dephasing model produces results with negligible errors of the modulation frequencies from the exact solution (Figs.~\ref{fig:3p_Compare_Exact_CCE1_F2kHz}-\ref{fig:CCE_Exact_Cutoff}). Furthermore, the systematic numerical studies (Fig.~\ref{fig:3p_OneClose_Compare_FFT}) show that a nearby Si nuclear spin with coupling $<200$~kHz has negligible effects on the ESEEM due to other distant nuclear spins.
 
Considering these different contributions of Si nuclear spins of different hyperfine couplings, it is justified to assume the pure-dephasing model and consider only the contributions of those Si nuclear spins that have couplings weaker than a certain cut-off (chosen as $|{\bar{\mathbf A}}_j|\le 20$~kHz).

\subsection*{Section S5.1 - CCE-1 for the multi-level central system}
\label{subsec: CCE-1}

\begin{figure}[h]
    \centering
    \includegraphics[width=0.9\linewidth]{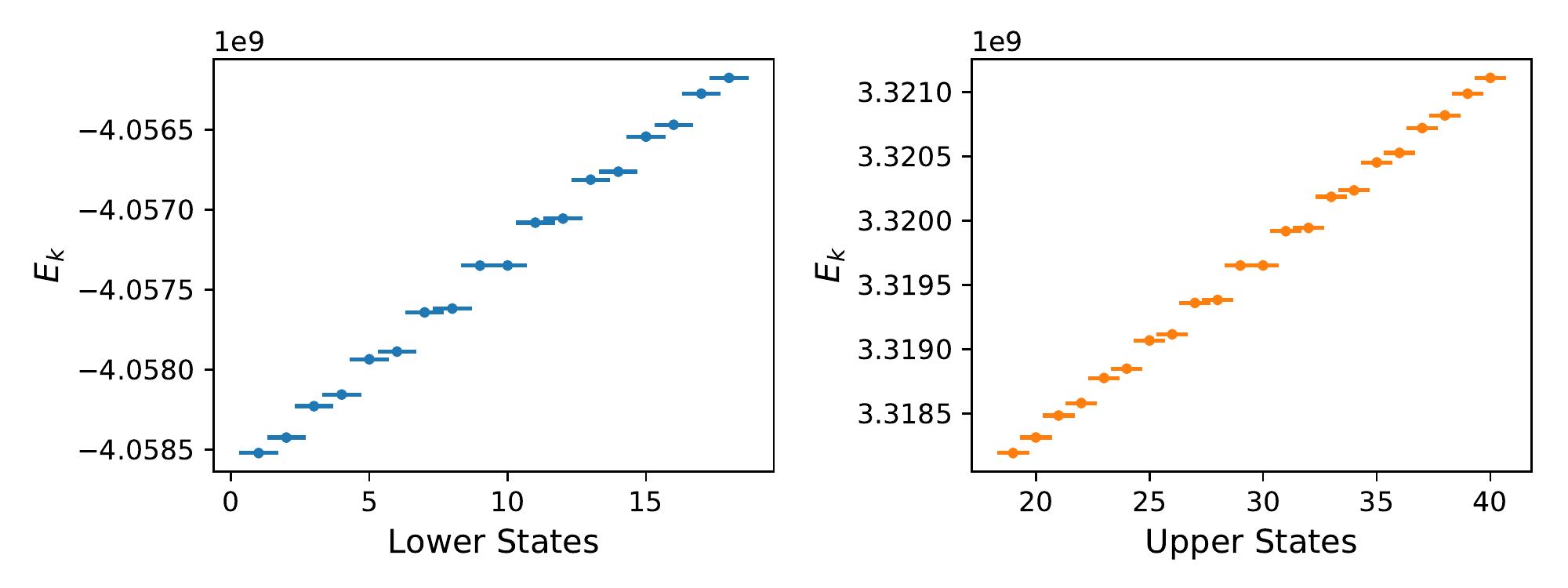}
    \caption{Eigenenergies of a hybrid spin system composed of  the Bi donor electron spin ${\mathbf S}_0$, the $^{209}$Bi nuclear spin ${\mathbf I}_0$, and a ``strongly coupled'' $^{29}$Si nuclear spin at position ${\mathbf r}_1 = (1.8, 0.4, -0.9)$~nm. The Fermi-contact hyperfine coupling to the Si spin is $A_{{\rm fc},1}/(2\pi)\approx  250$~kHz. The magnetic field along the $z$-axis $B_{0}=1$~G. The $x$ axis denotes the state label for the upper and lower eigenstates (the $\pm$ manifolds).}
    \label{fig:EngSi0}
\end{figure}

Strong hyperfine interaction would cause the mixing between the Bismuth center eigenstates. To consider the effect of a strongly coupled $^{29}$Si nuclear spin, we take the electron spin ${\mathbf S}_0$, the $^{209}$Bi nuclear spin ${\mathbf I}_0$, and the $j_0$ ``strongly coupled'' $^{29}$Si nuclear spins (denoted as ${\mathbf I}_j$ for $1\le j\le j_0$) as a hybrid center spin system. The hybrid center system can be diagonalized, with $2\times 10\times 2^{j_0}$ eigenstates, separated into two manifold $\{|\pm,m\rangle\}$.  The corresponding eigenenergies of the hybrid center spin system is denoted as $E^{\pm}_m$. 
See Fig.~\ref{fig:EngSi0} for an example of the eigenenergies of a hybrid spin system with one strongly coupled $^{29}$Si nuclear spin.

The Hamiltonian of the hybrid center system plus the rest $N-j_0$ weakly coupled $^{29}$Si nuclear spins can be written as
\begin{equation}
H = \sum_{\eta=\pm}\sum_m E^{\eta}_{m}\lvert \eta,m\rangle \langle \eta,m\rvert + \sum_{j=j_0+1}^N \left(\mathbf{S}_0\cdot \bar{\mathbf A}_j \cdot \mathbf{I}_j + g_n\beta_n \mathbf{B}_0\cdot \mathbf{I}_j\right),
\label{eq:hybrid_H}
\end{equation}
where the dipole-dipole interaction between the nuclear spins in the bath is neglected because it is weak and would have effects only on higher order CCE.
By assumption, the hyperfine couplings $|{\bar{\mathbf A}}_j|$ (for $j>j_0$) are much smaller than the energy differences among the eigenstates of the central system (otherwise they would have been absorbed into the hybrid central system). Therefore, we have the pure dephasing model with the bath Hamiltonians conditioned on the states of the central system (for the conciseness, we label $(\eta,m)$ as $k$)
\begin{equation}
H^{(k)}\equiv H'_{\rm Si}+\langle k|H_{{\rm hf}}|k\rangle
=H'_{\rm Si}+\eta \alpha_m\sum_j {\mathbf A}_j\cdot{\mathbf I}_j,
\end{equation}
where the Si nuclear spin bath Hamiltonian $H'_{\rm Si}$ excludes the strongly coupled $^{29}$Si nuclear spins ${\mathbf I}_j$ (for $j\le j_0$) and neglects the 
dipolar interactions between the Si nuclear spins.

In the echo experiment, a sequence of $n$ pulses 
$$R_j\in\{R^x_{\pi/2},R^x_{\pi},R^y_{\pi/2},R^y_{\pi}\}$$ 
are applied at $t_j$ ($j=1,2,\ldots n$), and then the signal $V_{\rm 3p}$ is measured at time $t$.  The $\pi/2$ and $\pi$ control pulses are taken as $R^{x/y}_{\pi/2}\equiv e^{-i\left(\Omega_{\rm R} S_{0,x/y}+ H_{\rm Bi}\right)\tau_{\rm p}/2}$ and $R^{x/y}_{\pi}=e^{-i\left(2 \Omega_{\rm R} S_{0,x/y}+ H_{\rm Bi}\right)\tau_{\rm p}/2}$, where $\tau_{\rm p}$ is the duration of a $\pi$ pulse and $\Omega_{\rm R}$ the Rabi frequency. Note that the pulse durations are assumed much shorter than the timescales of dynamics of the $^{29}$Si bath spins and thus the rotation transform does not include $H'_{\rm Si}$. If  the system starts with a certain state $|k_0\rangle\otimes |J\rangle$ at $t=0$ (in which $|k_0\rangle$ denotes a certain eigenstate $|-,m\rangle$ and $|J\rangle$ a certain bath state), the state at time $t$ is
$$
\sum_{k_1,k_2,\ldots,k_n} C_{k_0,k_1,\ldots k_n}e^{-i\phi_{k_1,k_2,\ldots k_n}}|k_n\rangle \left|J_{k_1,k_2,\ldots k_n}\right\rangle,
$$
where $|k_j\rangle$ denotes an eigenstate $|\pm,m\rangle$, the phase $$\phi_{\mathbf k}=E_{k_1}t_1+E_{k_2}(t_2-t_1)+\cdots+E_{k_n}(t-t_n),$$ the coefficients of the eigenstates 
$$
C_{k_0,{\mathbf k}}\equiv \langle k_n|R_n|k_{n-1}\rangle\cdots \langle k_2|R_2|k_1\rangle \langle k_1|R_1|k_0\rangle, 
$$
and 
the bath state
$$
\left|J_{\mathbf k}\right\rangle
\equiv e^{-iH^{(k_n)}(t-t_{n})}\cdots e^{-iH^{(k_2)}(t_2-t_1)} e^{-iH^{(k_1)}t_1}|J\rangle,
$$
with the shorthand notation ${\mathbf k}=(k_1,k_2,\ldots,k_n)$. The echo signal is
\begin{align}
V_{\rm 2p/3p/5p} & \equiv \langle S_{0,x} (t)\rangle 
 =\sum_{{\mathbf k},{\mathbf k}'} C_{k_0,{\mathbf k}'}^*C_{k_0,{\mathbf k}}e^{i\phi_{{\mathbf k}'}-i\phi_{\mathbf k}} \langle k'_n|S_{0,x}|k_n\rangle \left\langle J_{{\mathbf k}'} \right.\left|J_{\mathbf k}\right\rangle.
\end{align}
In general, the  nuclear spin state overlap $\left\langle J_{{\mathbf k}'} \right.\left|J_{\mathbf k}\right\rangle$ is computed using the CCE~\citep{Yang2008, Yang2009, Zhao2012}, and in our current case for the Hamiltonian in Eq.~(\ref{eq:hybrid_H}), CCE-1 gives the exact solution.
The result of $V_{\rm 2p/3p/5p}$ is further averaged over different initial states $|k_0\rangle$ and $|J\rangle$ in the thermal distribution.

\begin{table}[t]
\caption{Absolute values of the matrix elements of $S_{0,x}$ in the basis $\{\lvert \eta, m\rangle\}$ for $B_{0} =1$~G.}
\vskip 0.2cm
\scriptsize
	\begin{tabular}{|c||ccccccccc|ccccccccccc|} \hline
		$\lvert -, 4\rangle$  & 0    & 0.14 & 0    & 0    & 0    & 0    & 0    & 0    & 0    & 0    & 0    & 0    & 0    & 0    & 0    & 0    & 0    & 0.07 & 0    & 0.47 \\
		$\lvert -, 3\rangle$  & 0.14 & 0    & 0.19 & 0    & 0    & 0    & 0    & 0    & 0    & 0    & 0    & 0    & 0    & 0    & 0    & 0    & 0.12 & 0    & 0.42 & 0    \\
		$\lvert -, 2\rangle$  & 0    & 0.19 & 0    & 0.21 & 0    & 0    & 0    & 0    & 0    & 0    & 0    & 0    & 0    & 0    & 0    & 0.17 & 0    & 0.37 & 0    & 0    \\
		$\lvert -, 1\rangle$  & 0    & 0    & 0.21 & 0    & 0.22 & 0    & 0    & 0    & 0    & 0    & 0    & 0    & 0    & 0    & 0.22 & 0    & 0.32 & 0    & 0    & 0    \\
		$\lvert -, 0\rangle$  & 0    & 0    & 0    & 0.22 & 0    & 0.22 & 0    & 0    & 0    & 0    & 0    & 0    & 0    & 0.27 & 0    & 0.27 & 0    & 0    & 0    & 0    \\
		$\lvert -, -1\rangle$  & 0    & 0    & 0    & 0    & 0.22 & 0    & 0.21 & 0    & 0    & 0    & 0    & 0    & 0.32 & 0    & 0.22 & 0    & 0    & 0    & 0    & 0    \\
		$\lvert -, -2\rangle$  & 0    & 0    & 0    & 0    & 0    & 0.21 & 0    & 0.19 & 0    & 0    & 0    & 0.37 & 0    & 0.17 & 0    & 0    & 0    & 0    & 0    & 0    \\
		$\lvert -, -3\rangle$  & 0    & 0    & 0    & 0    & 0    & 0    & 0.19 & 0    & 0.14 & 0    & 0.42 & 0    & 0.12 & 0    & 0    & 0    & 0    & 0    & 0    & 0    \\
		$\lvert -, -4\rangle$  & 0    & 0    & 0    & 0    & 0    & 0    & 0    & 0.14 & 0    & 0.47 & 0    & 0.07 & 0    & 0    & 0    & 0    & 0    & 0    & 0    & 0    \\ \hline
		$\lvert +, -5\rangle$ & 0    & 0    & 0    & 0    & 0    & 0    & 0    & 0    & 0.47 & 0    & 0.16 & 0    & 0    & 0    & 0    & 0    & 0    & 0    & 0    & 0    \\
		$\lvert +, -4\rangle$ & 0    & 0    & 0    & 0    & 0    & 0    & 0    & 0.42 & 0    & 0.16 & 0    & 0.21 & 0    & 0    & 0    & 0    & 0    & 0    & 0    & 0    \\
		$\lvert +, -3\rangle$ & 0    & 0    & 0    & 0    & 0    & 0    & 0.37 & 0    & 0.07 & 0    & 0.21 & 0    & 0.24 & 0    & 0    & 0    & 0    & 0    & 0    & 0    \\
		$\lvert +, -2\rangle$ & 0    & 0    & 0    & 0    & 0    & 0.32 & 0    & 0.12 & 0    & 0    & 0    & 0.24 & 0    & 0.26 & 0    & 0    & 0    & 0    & 0    & 0    \\
		$\lvert +, -1\rangle$ & 0    & 0    & 0    & 0    & 0.27 & 0    & 0.17 & 0    & 0    & 0    & 0    & 0    & 0.26 & 0    & 0.27 & 0    & 0    & 0    & 0    & 0    \\
		$\lvert +, 0\rangle$ & 0    & 0    & 0    & 0.22 & 0    & 0.22 & 0    & 0    & 0    & 0    & 0    & 0    & 0    & 0.27 & 0    & 0.27 & 0    & 0    & 0    & 0    \\
		$\lvert +, 1\rangle$ & 0    & 0    & 0.17 & 0    & 0.27 & 0    & 0    & 0    & 0    & 0    & 0    & 0    & 0    & 0    & 0.27 & 0    & 0.26 & 0    & 0    & 0    \\
		$\lvert +, 2\rangle$ & 0    & 0.12 & 0    & 0.32 & 0    & 0    & 0    & 0    & 0    & 0    & 0    & 0    & 0    & 0    & 0    & 0.26 & 0    & 0.24 & 0    & 0    \\
		$\lvert +, 3\rangle$ & 0.07 & 0    & 0.37 & 0    & 0    & 0    & 0    & 0    & 0    & 0    & 0    & 0    & 0    & 0    & 0    & 0    & 0.24 & 0    & 0.21 & 0    \\
		$\lvert +, 4\rangle$ & 0    & 0.42 & 0    & 0    & 0    & 0    & 0    & 0    & 0    & 0    & 0    & 0    & 0    & 0    & 0    & 0    & 0    & 0.21 & 0    & 0.16 \\
		$\lvert +, 5\rangle$ & 0.47 & 0    & 0    & 0    & 0    & 0    & 0    & 0    & 0    & 0    & 0    & 0    & 0    & 0    & 0    & 0    & 0    & 0    & 0.16 & 0\\   \hline
	\end{tabular}
		\label{tab:BiSxe}
\end{table}

Without loss of generality we consider the 3p-ESEEM experiment with the magnetic field $B_{0} =1$~G. For the 3p-ESEEM, the CCE-1 result is
\begin{align}
V_{\rm 3p} = \sum_{k_1,k_2,k_3,k_1',k_2',k_3'}  & C_{k_0,k'_1, k'_2, k'_3}^\ast C_{k_0,k_1, k_2, k_3}\exp\left(i(\phi^{(k'_1, k'_2, k'_3)}-\phi^{(k_1, k_2, k_3)})\right) \nonumber \\
&\langle k'_3\rvert S_{0,x}\lvert k_3\rangle \left| \prod_{j=j_0+1}^N \mathrm{Tr}\left[ \frac{1}{2}U_j^{(k'_1, k'_2, k'_3)\dagger} U_j^{(k_1, k_2, k_3)}\right] \right|,
\label{eq:final_ESEEM}
\end{align}
where  $U_j^{(k_1, k_2, k_3)}$ is the evolution operator of the spin ${\mathbf I}_j$ for the center spin pathway $k_0\rightarrow k_1\rightarrow k_2\rightarrow k_3$. 

For the hybrid center that contains only the Bi electron and nuclear spins (no $^{29}$Si spins), the  nonzero elements of the $S_{0,x}$ operator are
\begin{equation}
\begin{aligned}
\left\langle +,m\left|S_{0,x}\right|-,m-1\right\rangle &=+\frac{1}{2}\cos\frac{\theta_m}{2}\cos\frac{\theta_{m-1}}{2},\\
\left\langle -,m\left|S_{0,x}\right|+,m-1\right\rangle &=-\frac{1}{2}\sin\frac{\theta_m}{2}\sin\frac{\theta_{m-1}}{2},\\
\left\langle +,m\left|S_{0,x}\right|+,m-1\right\rangle &=+\frac{1}{2}\cos\frac{\theta_m}{2}\sin\frac{\theta_{m-1}}{2},\\
\left\langle -,m\left|S_{0,x}\right|-,m-1\right\rangle &=-\frac{1}{2}\sin\frac{\theta_m}{2}\cos\frac{\theta_{m-1}}{2}.
\end{aligned}
\nonumber
\end{equation}
See numerical values in Table.~\ref{tab:BiSxe}.

\begin{figure}[t]
	\centering
	\includegraphics[width=0.9\linewidth]{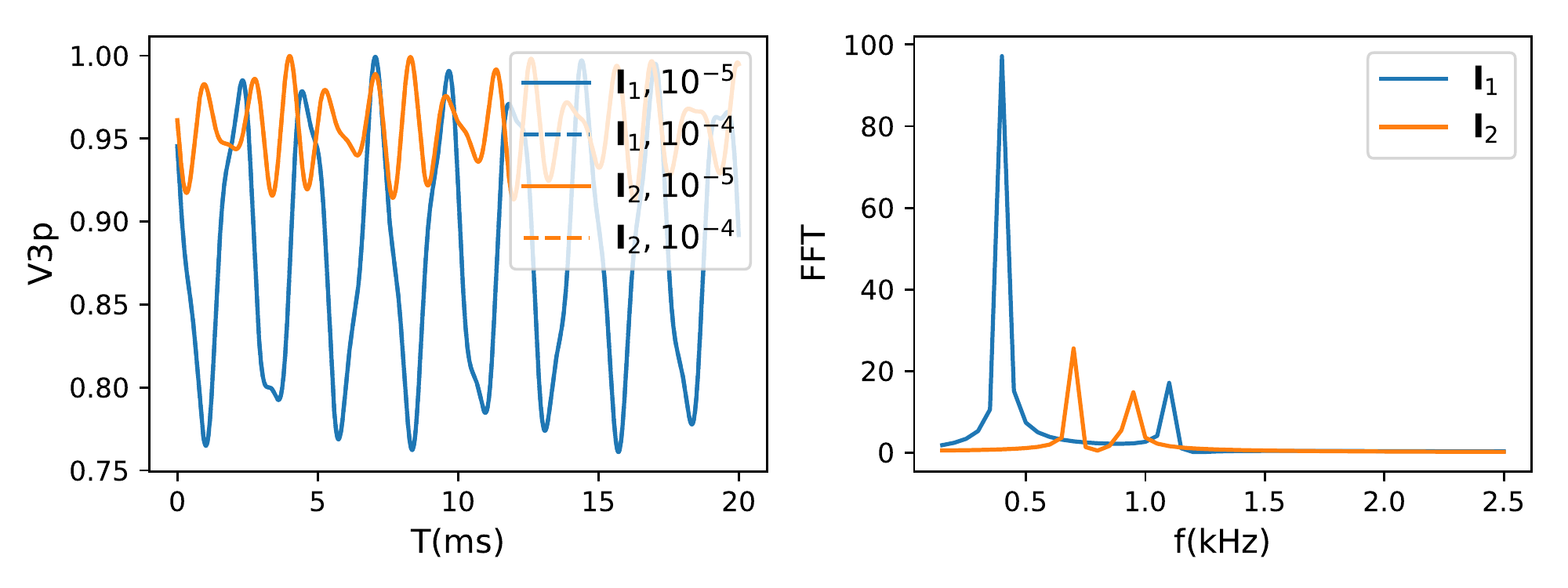}
	\caption{Exact solution of the 3p-ESEEM of a Bi donor spin coupled to only one $^{29}$Si nuclear spin. The Fermi contact and dipolar hyperfine couplings are $A_{{\rm cf},1} = 2.4$~kHz and $A_{{\rm dd},1} = 0.6$~kHz in the first case (labeled as ${\mathbf I}_1$), and  $A_{{\rm cf},2} = 0.8$~kHz and $A_{{\rm dd},2} = 0.5$~kHz in the second case (labeled as ${\mathbf I}_2$). The external field $B_{0} = 1$~G.  The left panel shows the ESEEM, and the right panel shows the Fourier transform of the ESEEM. The cutoff thresholds $|C_{{\mathbf k}'}^*C_{{\mathbf k}}|<10^{-5}$ and $10^{-4}$ for a pathway to be dropped are indicated in the left panel. The two choices of cutoff thresholds produce nearly identical results.}
	\label{fig:3pTwoSi29_Separate}
\end{figure}

Not all of the pathways have contributions to the ESEEM signal because of the inhomogeneous broadening and selection rules. For instance, considering a hybrid center spin system that contains only the Bi electron and nuclear spins, the phase difference accumulated for $k_j=(+,m)$ and $k'_j = (+,m')$ during time from $t_{j-1}$ to $t_j$ is
\begin{equation}
\label{eq:inhomo_eff}
    \phi_{k'_j} - \phi_{k_j} = (E_{m'}^{+}-E_m^{+})\cdot (t_j-t_{j-1}) \approx \frac{m'-m}{10} ({g_{\rm e}\beta_{\rm e}B_0} + h_z)\cdot (t_j-t_{j-1}),
\end{equation}
where $h_z$ is the Overhauser field from the bath spins. For the relevant timescales ($\sim 1$~ms) and inhomogeneous broadening of $h_z$ (which is $\sim 0.5$~MHz), the ensemble averaged phase factor would vanish unless $k_j = k'_j$. In the numerical simulation, a pathway $({\mathbf k}',{\mathbf k})$ is dropped if (1) the echo condition
\begin{align}
\phi_{\mathbf k}- \phi_{{\mathbf k}'}=0,
\label{eq:echo_condition}
\end{align}
is not satisfied, or  (2) the amplitude is too small, e.g., $ |C_{{\mathbf k}'}^*C_{{\mathbf k}}|< 10^{-4}$ (see Fig.~\ref{fig:3pTwoSi29_Separate}).

\subsection*{Section S5.2 - Exact simulation for one $^{29}$Si spin}
\label{subsec: pathway_model}


Taking the special case of Eq.~(\ref{eq:final_ESEEM}) for $j_0=1$ and $N=1$, we can obtain the exact solution of the ESEEM due to a single $^{29}$Si spin. The result for
a certain initial state $|k_0\rangle$ is
\begin{equation}
V_{\rm 3p} = \sum_{k_1,k_2,k_3,k'_1,k'_2,k'_3} C_{k'_0, k'_1, k'_2, k_3}^\ast C_{k_0, k_1, k_2, k_3} \langle k'_3\rvert S_{0,x}\lvert k_3\rangle \exp\left(i\phi^{(k'_1, k'_2, k'_3)}-i\phi^{(k_1, k_2, k_3)}\right),
\label{eq:exact-1spin}
\end{equation}
where $\phi^{(k_1, k_2, k_3)}\equiv (E_{k_1} + E_{k_3})\tau +E_{k_2}T$ is the phase accumulated for the pathway $k_0\rightarrow k_1\rightarrow k_2\rightarrow k_3$.

In numerical simulations we neglect the pathways that have negligible probabilities $C_{k'_0, k'_1, k'_2, k_3}^\ast C_{k_0, k_1, k_2, k_3} |$. For instance, Fig.~\ref{fig:3pTwoSi29_Separate} compares the results with pathways neglected when $|C_{{\mathbf k}'}^*C_{{\mathbf k}}|<10^{-5}$ or $10^{-4}$. Actually, for these two cases shown
in the figure, there are only $4$ main contributing pathways, since most pathways have no contributions in the ESEEM signal after taking into account of the inhomogeneous broadening effects (see Eq.~\ref{eq:inhomo_eff}).

\subsection*{Section S5.3 - Justification of CCE-1 approximation for weakly coupled bath spins}
\label{subsec: cutoff}

\begin{figure}[h]
    \centering
    \includegraphics[width=0.8\linewidth]{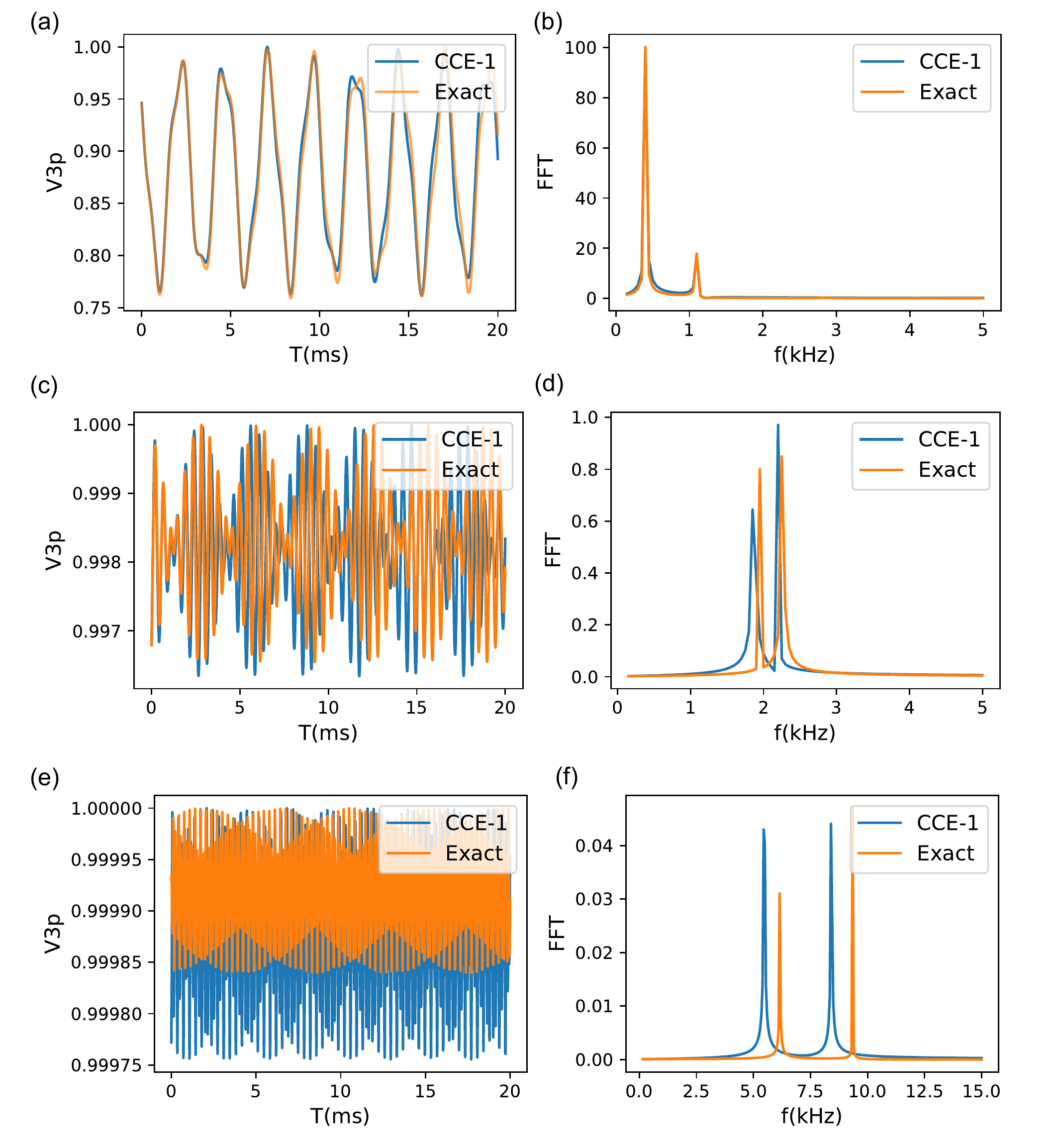}
  \caption{Comparison between exact solution [Eq.~(\ref{eq:exact-1spin})] and the CCE-1 approximation [Eq.~(\ref{eq:final_ESEEM}) with $j_0=0$ and $N=1$] of the ESEEM of a Bi donor spin coupled to one single $^{29}$Si spin with various hyperfine coupling strength. 
(a,c,e) are the ESEEM and (b,d,f) are the Fourier transform (the ESEEM spectra). 
The hyperfine coupling is about $2.4$~kHz, $13.4$~kHz, and $45.5$~kHz corresponding to (a/b), (c/d), and (e/f). The external field $B_{0} = 1$~G. The microwave pulse is chosen resonant with the transition $6\leftrightarrow 13$ and of duration $1$~$\mu$s for the $\pi$-pulse.
}
    \label{fig:3p_Compare_Exact_CCE1_F2kHz}
\end{figure}


Now we compare the exact solution and CCE-1 approximation for a single $^{29}$Si bath spins ($N=1$). We use Eq.~(\ref{eq:exact-1spin}) for the exact solution and Eq.~(\ref{eq:final_ESEEM}) with $j_0=0$ and $N=1$ for the CCE-1 approximation.  The pathways of the Bi spin that do not satisfy the echo condition [Eq.~(\ref{eq:echo_condition})] are not included in the simulation. 

\begin{figure}[t]
    \centering
    \includegraphics[width=0.6\linewidth]{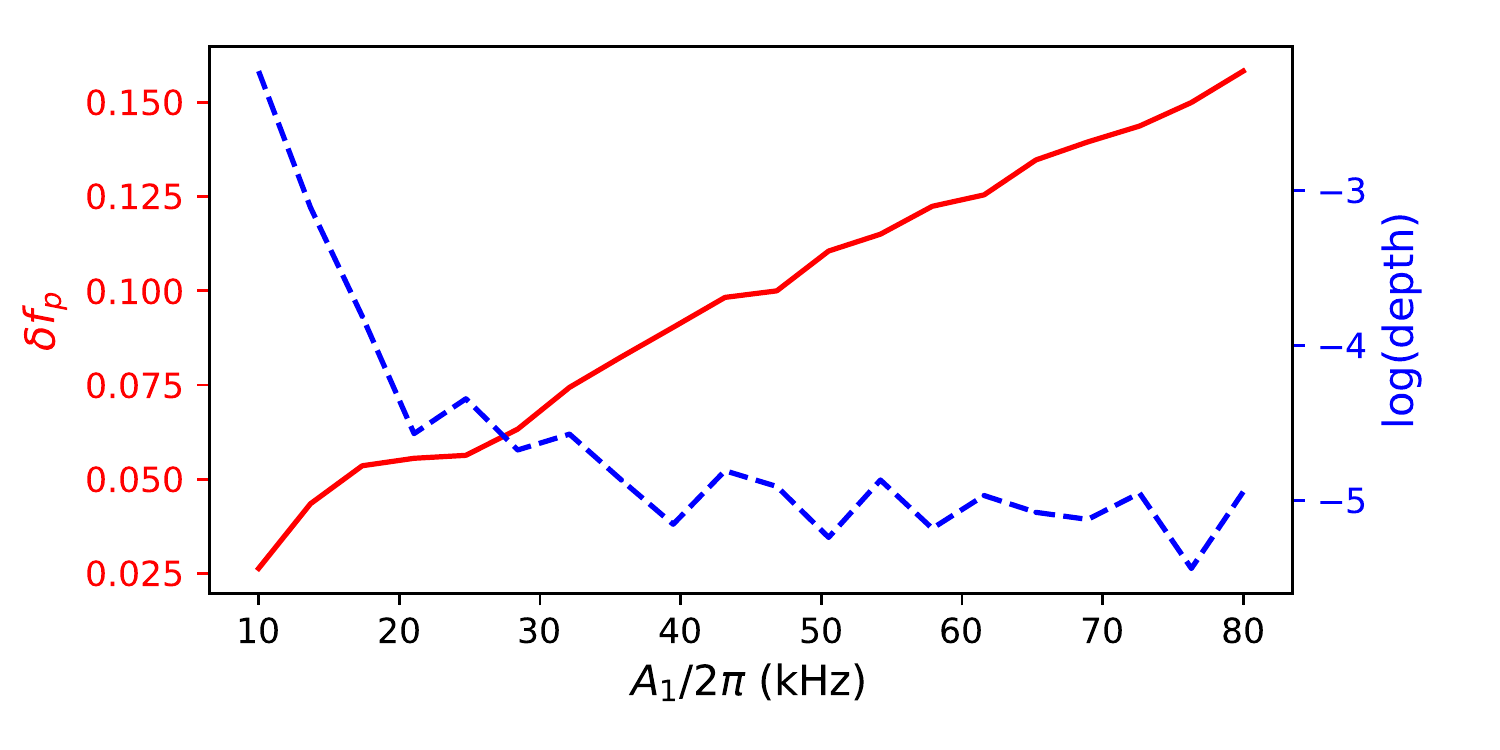}
    \caption{Left axis: The relative frequency deviation ($\delta f_p = |f_\text{CCE}-f_\text{Exact}|/f_\text{Exact}$) of the CCE-1 approximation from the exact results for as a function of the hyperfine coupling $A_{1}$ (which refers to Fermi contact coupling in this figure).  For CCE-1 to be a good approximation, we can set the cutoff to be $20$~kHz, for an error up to $5\%$. Right axis: The dependence of the modulation depth on $A_1$. The microwave pulse is chosen resonant with the transition $6\leftrightarrow 13$ and of duration $1$~$\mu$s for the $\pi$-pulse. The external field is $B_{0}=1$~G.}
    \label{fig:CCE_Exact_Cutoff}
\end{figure}

We expect that when the hyperfine coupling is comparable to the level splitting of the Bismuth donor spin within each manifold, the secular and the CCE-1 approximation become invalid (Eq.~\ref{eq:CCE_condition}).  Figure~\ref{fig:3p_Compare_Exact_CCE1_F2kHz} compares the exact solution and the CCE-1 approximation for various hyperfine coupling strengths. As expected, the CCE-1 agrees well with the exact solution for relatively weak coupling. The deviation of the ESEEM frequency calculated by the CCE-1 from the exact solution is shown in Fig.~\ref{fig:CCE_Exact_Cutoff} as a function of the hyperfine coupling strength. For coupling weaker than $20$~kHz, the error is less than $5\%$. Furthermore, the ESEEM depth (shown in Fig.~\ref{fig:CCE_Exact_Cutoff}) becomes negligible for coupling greater than $20$~kHz for the field.  (It should be noted that if the echo condition in Eq.~(\ref{eq:echo_condition}) is applied, strong hyperfine coupling would lead to mixing between Bi spin states, causing fast oscillations with amplitude increasing with the coupling strength. Such fast oscillations, however, decay rapidly to zero due to inhomogeneous broadening. See Sec.~\ref{Sec:Strong_exact} below for more discussions.)

Considering both the ESEEM frequency calculation precision and the modulation depth, a nuclear spin with hyperfine coupling weaker than $20$~kHz can be well approximated  by the CCE-1.

\subsection*{Section S5.4 - ESEEM due to a strongly coupled $^{29}$Si spin}
\label{Sec:Strong_exact}

\begin{figure}[h]
    \centering
    \includegraphics[width=0.9\linewidth]{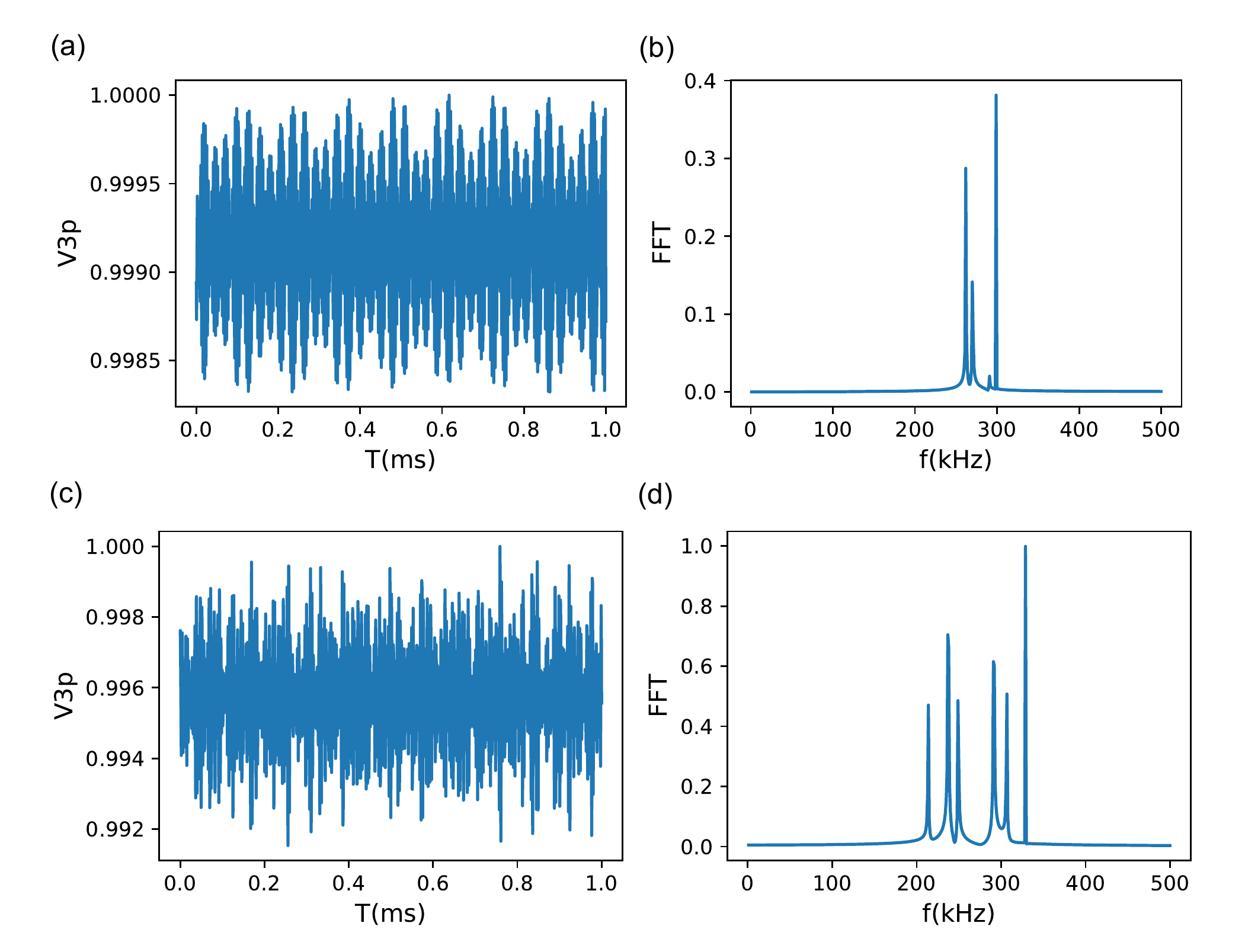}
    \caption{Exact simulation of the 3p-ESEEM due to a strongly coupled $^{29}$Si.  (a) and (c) are the ESEEM and (b) and (d) are the Fourier transform. The locations and the hyperfine couplings are ${\mathbf r}_1=(13, 6, -7)\times a_{\rm Si}/4 $,  $A_{{\rm,cf},1}\approx 101.8$~kHz, and $A_{{\rm dd},1}\approx 1.6$~kHz for (a) and (b), and ${\mathbf r}_1=(8,5,-9)\times a_{\rm Si}/4 $, $A_{{\rm cf},1} \approx 207.1$~kHz, and $A_{{\rm dd},1}\approx 2.8$~kHz for (c) and (d). The external field $B_{0} = 1$~G. All pathways with $\left|C_{k_0,{\mathbf k}'}^*C_{k_0,{\mathbf k}}\right|>10^{-4}$ are taken into account, without imposing the echo condition Eq.~(\ref{eq:echo_condition}).} 
    \label{fig:3pOneSi29_FFT_F101kHz}
\end{figure}

Figure.~\ref{fig:3pOneSi29_FFT_F101kHz} show the exact simulation of the 3p-ESEEM due to a strongly coupled $^{29}$Si spin. Two coupling strengths ($\sim 100$~kHz and $\sim 200$~kHz) are considered. The strong hyperfine coupling induces mixing between the Bi spin states (violating the pure dephasing approximation). To show the effect of state mixing, we do not impose the echo condition Eq.~(\ref{eq:echo_condition}), and therefore need to include about 100 pathways to produce converged results. The interference between different pathways (due to the mixing among Bi spin states) induce fast and complicated modulations. The Fourier transform shows that the ESEEM frequencies are spread around $280$~kHz, which is approximately the splitting between Bi levels in absence of Si spins. The fact that modulation frequency is near $280$~kHz confirms that the ESEEM is mostly due to the mixing of the Bi spin states. On the contrary, in the case of weakly coupled $^{29}$Si spins where the mixing between different Bi levels is negligible,  the ESEEM is due to the mixing between different $^{29}$Si nuclear spin states, so there the ESEEM has frequencies essentially determined by the $^{29}$Si nuclear spin Larmor frequency (see Fig.~\ref{fig:3p_Compare_Exact_CCE1_F2kHz}). 
 
\begin{figure}[t]
    \centering
    \includegraphics[width=0.9\linewidth]{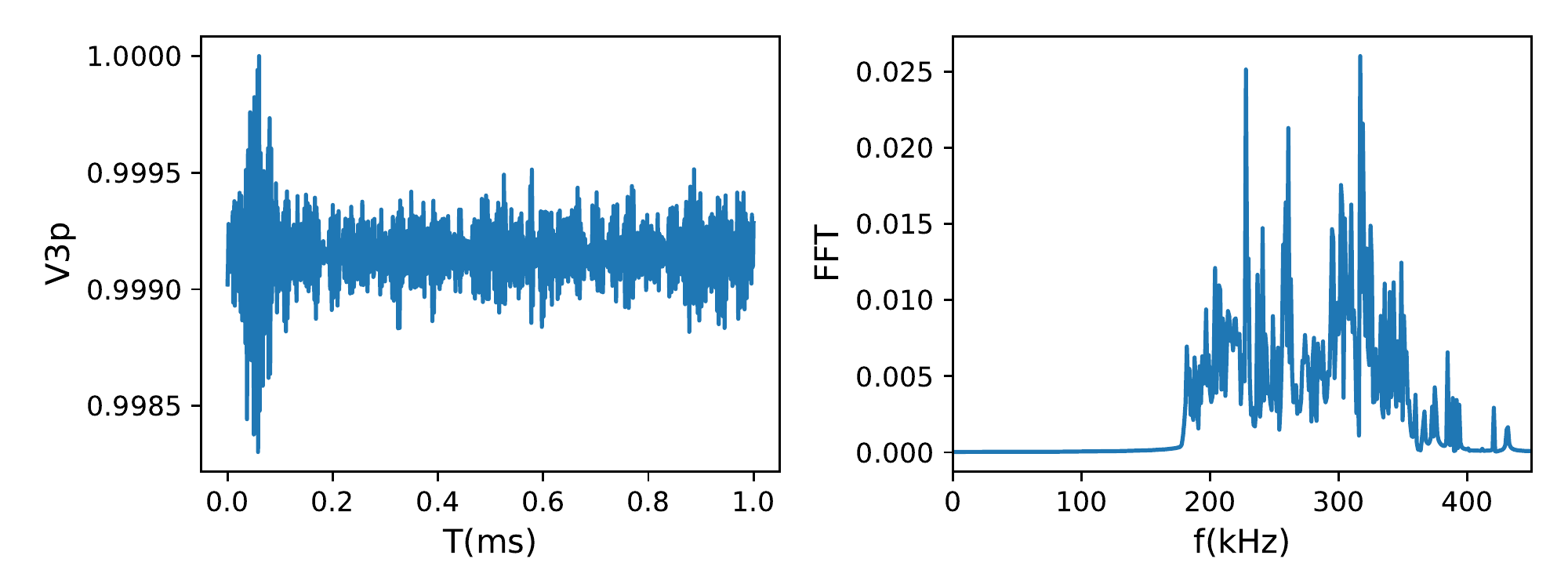}
    \caption{Ensemble average of (a) 3p-ESEEM over 100 samples of nuclear spin states in a thermal distribution and (b) the spectrum.  The locations and the hyperfine couplings are ${\mathbf r}_1=(8,5,-9)\times a_{\rm Si}/4 $, $A_{{\rm cf},1} \approx 207.1$~kHz, and $A_{{\rm dd},1}\approx 2.8$~kHz. The external field $B_{0} = 1$~G. All pathways with $\left|C_{k_0,{\mathbf k}'}^*C_{k_0,{\mathbf k}}\right|>10^{-4}$ are taken into account.
}
    \label{fig:3pOneClose_Ave_50}
\end{figure}

The rapid oscillations due to interference between different Bi spin states, however, have their frequencies sensitively depending on the external field and the Overhauser field due to the random configurations of the nuclear spin baths (the inhomogeneous broadening). Therefore, the ensemble average over the distribution of the Overhauser field $h_z$ (which has a broadening of about 0.5~MHz) would lead to rapid decay of the modulation (in $\mu$s timescales). This is indeed shown in 
Fig.~\ref{fig:3pOneClose_Ave_50}. Thus, the contributions of the strongly coupled nuclear spins are not observable in the timescales relevant in the experiments ($\sim$ms).



\subsection*{Section S5.5 - Influence of a strongly coupled $^{29}$Si spin on weakly coupled $^{29}$Si spins}

\begin{figure}[t]
    \centering
    \includegraphics[width=0.9\linewidth]{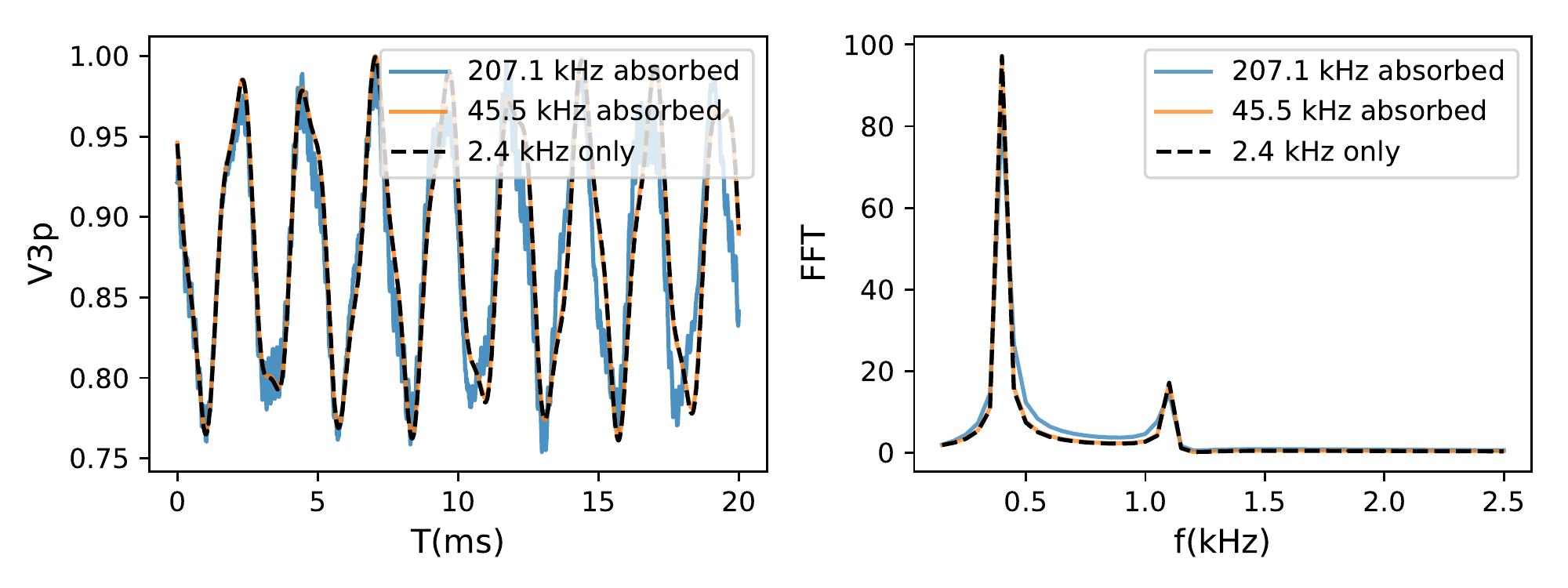}
    \caption{The ESEEM of a weakly coupled $^{29}$Si nuclear spin with or without another nuclear spin strongly coupled to the Bi donor spin.
The Fermi contact hyperfine coupling $A_{{\rm fc},2}\approx 2.4$~kHz for the weakly coupled nuclear spin, and $A_{{\rm fc},1}\approx 45.5$ or $207.1$~kHz for the strongly coupled nuclear spin. The external field  $B_{0} = 1$~G. }
    \label{fig:3p_OneClose_Compare_FFT}
\end{figure}

In general, a strongly coupled nuclear spin may affect the ESEEM due to a weakly coupled nuclear spin. This is because the Bi spin states can be mixed by the strong coupling.
The state mixing renormalizes the effective couplings between the weakly coupled nuclear spin and the ``ficticious'' spin-1/2 of the Bi donor [in particular, $\alpha_m$ and $\alpha_{m-1}$ in Eq.~(\ref{eq:effective_Hm})] and hence affects the ESEEM. To exam such an effect, we calculate the ESEEM due to a weakly coupled nuclear spin ${\mathbf I}_2$ in the presence or in the absence of a strongly coupled nuclear spin ${\mathbf I}_1$. The weakly coupled nuclear spin is considered using the CCE-1 and the strongly coupled one, if taken into account, is exactly considered by absorbing it into the hybrid center spin system  [corresponding to Eq.~(\ref{eq:final_ESEEM}) with $j_0=1$ and $N=2$].

Figure~\ref{fig:3p_OneClose_Compare_FFT}  shows that while the strongly coupled spin causes some fast, small-amplitude modulations, it has negligible effect on the ESEEM due to the weakly coupled nuclear spin. The weak effect is understandable considering that the matrix elements  $\alpha_{m}$ and $\alpha_{m-1}$ in Eq.~(\ref{eq:effective_Hm}) are only slightly ($<1/10$) modified by the hyperfine coupling to the spin ${\mathbf I}_1$ even when $A_{{\rm fc},1}$ is as large as $200$~kHz.

\section*{Section S6 - Phase cycling scheme}
\label{sec:phasecycling}

In the tables below is provided the phase cycling scheme for the 3- and 5-pulse ESEEM measurements.

\begin{table}[h]
\caption{Phase cycling table for 3 pulse ESEEM with CPMG and offset removal.}
\centering
\begin{tabular}{ c c c c | c c }
  \multicolumn{4}{c|}{3 pulse ESEEM}& \multicolumn{2}{c}{CPMG} \\
  $\pi/2$ & $\pi/2$ & $\pi/2$ & Det. & $\pi$ & Det.\\ \hline 
	+x & +x & +x & +y & +y & +y\\
	-x & +x & +x & -y & +y & -y\\
	-x & -x & +x & +y & +y & +y\\
	+x & -x & +x & -y & +y & -y\\
\end{tabular}

\label{tab:3phasecycle}
\end{table}

\begin{table}[h]
\caption{Phase cycling table for 5 pulse ESEEM with CPMG and offset removal.}
\centering
\begin{tabular}{ c c c c c  c | c c }
  \multicolumn{6}{c|}{5 pulse ESEEM}& \multicolumn{2}{c}{CPMG} \\
  $\pi/2$ & $\pi$ & $\pi/2$ & $\pi/2$ & $\pi$ & Det. & $\pi$ & Det.\\ \hline 
 	+x & +x & +y & +y & +y & +y & +y & +y\\
	-x & -x & -y & +y & +y & +y & +y & +y\\
	-x & -x & -y & -y & -y & -y & +y & -y\\  
	+x & +x & +y & -y & -y & -y & +y & -y\\  
	+x & -x & -y & -y & +y & +y & +y & +y\\
	-x & +x & +y & -y & +y & +y & +y & +y\\
	-x & +x & +y & +y & -y & -y & +y & -y\\
	+x & -x & -y & +y & -y & -y & +y & -y\\ 
\end{tabular}
\label{tab:5phasecycle}
\end{table}